\documentclass[final,3p,times]{elsarticle}

\usepackage{lineno,hyperref}
\modulolinenumbers[5]

\usepackage{amssymb}
\usepackage{amsmath}
\usepackage{graphics}
\usepackage{graphicx}
\usepackage{subfigure}
\usepackage{color}

\numberwithin{equation}{section}

\journal{Physics Reports}
\begin{document}

\begin{frontmatter}

\title{Quantum Chaos and Thermalization in Isolated Systems of Interacting Particles}

\author[BRESCIA]{F. Borgonovi}
\ead{fausto.borgonovi@unicatt.it}
\address[BRESCIA]{Dipartimento di Matematica e Fisica and Interdisciplinary Laboratories for Advanced Materials Physics,
Universit\'a Cattolica, via Musei 41, 25121 Brescia, and INFN,
Sezione di Pavia, Italy}

\author[BUAP,MSU]{F.~M. Izrailev}
\ead{felix.izrailev@gmail.com}
\address[BUAP]{Instituto de F{\'i}sica, Universidad Aut\'onoma de Puebla,
Apt. Postal J-48, Puebla, Pue., 72570, Mexico}


\author[YESHIVA]{L.~F. Santos}
\ead{lsantos2@yu.edu}
\address[YESHIVA]{Department of Physics, Yeshiva University, 245 Lexington Ave, New York, NY 10016, USA}

\author[MSU]{V.~G. Zelevinsky\corref{CorAuth}}
\ead{Zelevins@nscl.msu.edu}
\cortext[CorAuth]{Corresponding author}
\address[MSU]{NSCL and Department of Physics and Astronomy, Michigan State
University, East Lansing, MI 48824-1321, USA}

\begin{abstract}

This review is devoted to the problem of thermalization in a small isolated conglomerate of interacting constituents.
A variety of physically important systems of intensive current interest belong to this category: complex atoms,
molecules (including biological molecules), nuclei, small devices of condensed matter and quantum optics on nano- and micro-scale, cold atoms in optical lattices, ion traps. Physical implementations of quantum computers, where there are many interacting qubits, also fall into this group. Statistical regularities come into play through inter-particle interactions, which
have two fundamental components: mean field, that along with external conditions, forms the regular component
of the dynamics, and residual interactions responsible for the complex structure of the actual stationary states.
At sufficiently high level density, the stationary states become exceedingly complicated superpositions
of simple quasiparticle excitations. At this stage, regularities typical of quantum chaos emerge and bring in signatures of thermalization. We describe all the stages and the results of the processes leading to thermalization, using analytical
and massive numerical examples for realistic atomic, nuclear, and spin systems, as well as for models with
random parameters. The structure of stationary states, strength functions of simple configurations, and concepts
of entropy and temperature in application to isolated mesoscopic systems are discussed in detail.
We conclude with a schematic discussion of the time evolution of such systems to equilibrium.

\end{abstract}


\end{frontmatter}

\tableofcontents

\section{Introduction}
The goal of this review is two-fold. First, we discuss the problem of the emergence of
thermalization in isolated quantum systems, as caused by the interaction between particles
(or quasi-particles). Second, we follow an approach that allows one to link thermalization to
quantum chaos, the latter arising when the inter-particle interactions are sufficiently strong. This approach
has been developed during the last two decades in applications to nuclear and atomic physics,
as well as for appropriate random matrix models. Here, we show how this approach can be
extended to various models of interacting fermions, bosons, and spins, being therefore
relevant for quantum dots, nuclear magnetic resonance platforms, and the more recent
experiments with cold atoms and molecules in optical lattices as well as trapped ions. Our
review can be considered as a complementary viewpoint on the subject of thermalization, to
which many papers have been recently devoted (see, for example, the reviews \cite
{ZelevinskyRep1996,Kota2001,Polkovnikov2011RMP,Huse2015,Rigol-review}).

Thermalization of isolated many-body quantum systems is a subject that belongs to the branch
of science referred to as mesoscopic physics, the broad field living in between the
macroscopic and microscopic worlds. The analysis of the mesoscopic systems deals both with the
physics of complexity on a relatively small scale and the physics of individual quantum states,
which can be studied theoretically and  experimentally. The wealth of ideas coming from these
seemingly opposite directions is extremely rich, instructive and promising in applications
covering nuclear, atomic, and molecular  physics; condensed matter on micro- and nano-scale;
quantum informatics. It can also spread its achievements all the way to biophysics.

In a  generic thermalization process,  accompanied by the growth of complexity appropriately defined through
entropy-like quantities, the system acquires typical features of statistical equilibrium.
 We treat the onset of
 thermalization in an isolated quantum system as the crossover from a time-periodic (regular) dynamics to a behavior described by the standard methods of statistical mechanics.
 The key
point of this approach is to consider systems of a finite number of particles on a finite
time scale. The problems of the thermodynamic limit as well as infinite time scale, which are
widely discussed in the literature, are only touched upon in the last section of this review.
In particular we show that this approach is applicable to integrable systems  of few
particles.

In classical Hamiltonian mechanics (again, apart from the thermodynamic limit), it is well
understood that the mechanism of statistical behavior is based on the so-called deterministic
chaos. This term stresses that the phenomenon of classical chaos occurs in strictly
deterministic systems without any  {\it a priori} randomness.
Deterministic classical chaos emerges due to the non-linearity of the equations of motion and is related 
to the high sensitivity of the motion with respect to  small changes of initial
conditions. The theory of classical chaos is nowadays well developed and quite often serves
as an extreme limit to be compared with its quantum analog, if such limit exists. In the
early days of the theory of quantum chaos, the main studies were performed for quantum systems
with a well defined classical limit. The simplest example is a billiard-like system where
classical chaos has been proven rigorously, based on the ergodic theory. Nowadays, the notion
of quantum chaos is used in a much broader context, due to underlying generic relations
between quantum mechanics and classical wave mechanics. In both cases the employed tools of
analysis are very similar and one can relate the dynamics  to the properties of their energy
spectra and eigenstates.

With great success, the notion of quantum chaos has been  extended to realistic many-body
systems without any randomness and with no classical limit, as well as to models with some
intrinsic randomness or disorder in their structure. The presence of randomness itself does
not  guarantee  the onset of strong statistical properties $-$ this depends on the disorder
strength. The current understanding of quantum chaos is broader and clearly different from
its original meaning. However, since the underlying mechanism of quantum chaoticity is wave
chaos, the methods developed in the study of quantum chaos are found to be useful in various
fields of classical physics, such as  acoustics, optics, and electromagnetism.

With respect to our claim that quantum chaos provides the mechanism driving thermalization in
quantum systems, we would like to comment on the folkloric statement that {\it quantum chaos
does not exist}. The notion of deterministic chaos was originally related to the
exponential divergence of classical trajectories started at close initial conditions in the
phase space. In quantum mechanics there are no trajectories and no precisely defined initial
conditions in the phase space of canonically conjugate dynamical variables; therefore,
quantum chaos was claimed  not to be {\it real} chaos. However, as argued by Chirikov in Ref.~\cite{c97},
the world is governed by quantum mechanics. Therefore, if we accept that
quantum mechanics is {\it real} then we have to accept that  classical chaos is just a mathematical approximation based on the
well developed ergodic theory. One of the main topics of this review is the discussion
of manifestations of quantum chaos in specific properties of the energy spectra and structure
of eigenstates.

Another controversial question arises when considering, as in our review, quantum isolated systems. The
energy spectrum of bound systems  is discrete and the dynamics is
quasi-periodic, formally implying the absence of any chaos. However, the quasi-periodicity is  related
to the limit $t \rightarrow \infty$, which should be treated as non-physical. Even though the
time scale on which chaotic behavior emerges is finite, it can be much larger than any
physical time scale. This situation has been termed by Chirikov \cite{c97} as {\it linear chaos} to
stress that the mechanism of quantum chaos is different from that of classical chaos.
The physical question is to establish the time scale on which quantum chaos comes into play. As
discussed by Chirikov, the mechanism leading to quantum chaos is due to a large number of
independent frequencies and random phases that are present in the evolution of a wave packet.
Since this evolution is mainly  defined by the structure of the eigenstates involved in the
dynamics, the problem of quantum chaos is naturally reduced to the emergence of chaotic
eigenstates, which depend on  physical parameters and initial conditions.

Since the mechanism of quantum chaos is directly related to the chaotic structure of the
eigenstates, our approach quantifies chaos in terms of chaotic eigenstates, rather than
in terms of local spectral statistics. The latter is a tool to distinguish between integrable
and non-integrable systems, but our interest is in the onset of relaxation rather than
integrability, so we focus on the conditions allowing to treat the eigenstates as random (or
pseudo-random) superpositions of a large number of components. To speak of the structure of
eigenstates, one has, first, to define the basis in which we consider the system evolution.
The habitual objection is that the components of the  eigenstates are specific to the choice
of the basis, therefore  an approach based on the eigenstates rather than on the
basis-invariant spectral statistics might be not appropriate. However the basis representation is
quite often well defined and chosen by the  physics of the problem. A typical example is the
concept of Anderson localization where the eigenstates are exponentially localized in the
configuration representation so that the basis is naturally chosen by physics. In many
physical systems, such as atoms, molecules and nuclei, the mean field basis, being singled
out by the physical picture and by the reaction of the system to external perturbations, is
deeply ingrained in the description of numerous processes, including that of thermalization.

The  concept of the mean field is a key ingredient of our approach to many-body problems.
The use of the mean field allows one to choose the most representative basis where the
structure of the eigenstates is analyzed and transparently related to the dynamics of interactions. In this
approach the total Hamiltonian $H=H_0 +V$ of an isolated system is split into two parts. The
part $H_0$ (mean field) corresponds to non-interacting particles or quasi-particles, and
the rest, $V$, describes the {\it residual} interaction between them. If the residual
interaction, even without random parameters, is sufficiently strong, the eigenstates of
the total Hamiltonian $H$ can  be treated as  chaotic superpositions of many-body
unperturbed states of $H_0$. The interaction becomes effectively stronger with the growth of
the excitation energy and the related combinatorial growth of the level density. Therefore,
in an isolated system, thermalization can emerge for sufficiently high energy when the interaction effectively mixes the simple states.

In this review we demonstrate how this approach works for various Hamiltonians, either
described by random matrices or corresponding to realistic physical systems, such as heavy
nuclei, complex atoms, trapped ions, cold atoms or molecules in optical lattices and
interacting spins. An important concept  is the so-called {\it energy shell}
to which the eigenstates are compared. The partial filling of this shell is associated with
the many-body localization in the energy representation. Contrary, when the eigenstates fill
completely the energy shell, this  typically indicates  maximal quantum chaos. We demonstrate
that the onset of chaos and thermalization in various models can  be predicted by simple
analytical estimates of the width and filling of the energy shell.

The notion of chaotic eigenstates plays a key role in  the statistical description of
isolated quantum systems. As mentioned by Landau and Lifshitz in their {\sl Statistical
Mechanics}, the full description of quantum statistical mechanics can be done on the level of
individual states, not only by means of the Gibbs distribution. However, this statement does
not answer the important questions of {\it when} or {\it under what conditions} this fact
holds in practice, especially for mesoscopic systems. The answer is: when an individual eigenstate can be treated as a very
complicated superposition of many components, in our words - in the case of chaotic
eigenstates. Thus, in order to speak about thermalization one needs to know the conditions
required for the onset of chaotic eigenstates.

Characteristic manifestations of the
appearance of quantum statistics in small (mesoscopic) systems are given in Refs.
\cite{ZelevinskyRep1996} and \cite{GGF99} and  where the Fermi-Dirac distributions have been found for
individual eigenstates of heavy atom and nuclei. An actual (and practical) problem is to find
the conditions under which the eigenstates can be, indeed, treated as random (or
pseudo-random) ones. Our approach suggests a tool to derive these conditions and we show how
to apply it to various specific cases.

We start our review with a detailed discussion of the general setup for the Hamiltonians in
the mean-field representation (Section 2.1). Few basic models are introduced to be used
subsequently. As mentioned above, these models  are characterized by a Hamiltonian $H=
H_0+V$, where the first term, $H_0$, describes the non-interacting constituents, particles or
quasi-particles, while the interaction between them is embedded into the $V$ term. This
approach was originally used by Wigner in his attempts to describe generic properties of
heavy nuclei. The basic idea was that due to the very complex interactions inside the
nucleus, the many-body matrix elements corresponding to the interaction between nucleons could 
be modeled as random entries, thus allowing to introduce random matrices. In the first models
of such random matrices, Wigner suggested to take $H_0$ as a diagonal matrix with equally
spaced elements and  $V$ as a banded random matrix. In this way we come to what is
 nowadays known as Wigner banded random matrices (WBRM).

We introduce the  WBRM model in Section 2.2 in a more general form where the {\it unperturbed}
part $H_0$ can also have random diagonal elements with a constant spacing on average. In this
way, one can relate $H_0$ to an integrable Hamiltonian reflected by the Poisson distribution
for the nearest energy  level spacings. As for $V$, it can be treated as a residual
interaction that cannot be embedded into $H_0$ due to its random structure. Such matrices can
be considered as a generalization of full random matrices, described by the well developed
random matrix theory (RMT). The full random matrices can be treated as an extension of WBRM, when the
mean-field part is neglected and the band size of $V$ spreads to the total size of the
matrix.

Concerning  full random matrices, it is known that their total level density has a
nonphysical semicircle form. This and other facts led to the conclusion that in order to have
a closer relation to realistic many-body systems, one needs to introduce single-particle
states. This was realized by fixing a finite number of quantum particles distributed over a
number of single-particle levels. The off-diagonal matrix elements carry on an
important feature of the dynamics, namely the rank of the interactions. In many applications one
can assume that the inter-particle interaction is two-body. In this case the rank
equals two and the corresponding ensemble of such matrices consisting of both the mean-field
part $H_0$ and the random two-body interactions $V$ has been termed two-body random ensemble
(TBRE) as discussed in Sections 2.3. These matrices are not fully random since the two-body type
of interaction leads to the emergence of a large number of zero matrix elements and to
correlations between the non-zero elements. Still, many properties of the spectra, such as
the level spacing distribution, are quite close to those known for full random matrices.

In order to show that the general approach, initially derived in terms of WBRM and TBRE
models, can be effectively applied to realistic physical systems, in Sec. 2.4 we introduce
two one-dimensional (1D) models of interacting spins $1/2$. Since one of our goals is to
compare global statistical properties emerging in integrable and non-integrable models, we
consider an integrable spin-1/2 model and a non-integrable one. In the same section we also
discuss another physical system, namely, the  shell model, widely used in nuclear physics.

In Section 3  we consider the concept of  the {\it strength function} (SF) introduced long ago in
nuclear physics in order to characterize the main properties of the relaxation process inside
heavy nuclei. Relaxation occurs in nuclei after the initial
excitation of a given many-body state of $H_0$, and develops due to the interaction $V$
between nucleons. The SF is the projection of a particular basis state, defined by the
unperturbed Hamiltonian $H_0$, onto the exact stationary states of $H$. Defined in the energy
representation, the width of the SF, in general, characterizes the relaxation time for the
initial state to spread among other basis states of $H_0$. The mathematical definition of the
SF is given in Section 3.1, together with the discussion of how it can be computed. The SF
can be associated with the local density of states (LDOS) used in solid state physics. We
give examples of the SF obtained numerically for  the nuclear shell model and for the spin-1/2
models.

In addition to the width of the SF, another important feature is its shape. In Section 3.2 we
reproduce the ``standard model'' of the Breit-Wigner (BW) shape (or, mathematically referred
to as Lorentzian) typical of situations where the complicated states admixed to the initial
excitation are of approximately the same degree of complexity; this corresponds to the Fermi
golden rule limit but is not necessarily related to the weakness of interaction and perturbation theory. In typical
physical systems, however, due to the finite energy range of inter-particle interactions, the
tails of the SF deviate from the Lorentzian. This fact is demonstrated with numerical data
obtained for few nuclei in the framework of the shell model. Furthermore we show that the
shape of SF  typically  changes from Breit-Wigner to Gaussian as the interaction increases.
This crossover is thoroughly studied  for  WBRM and spin systems. This result is directly
relevant to the onset of chaos and thermalization. The Gaussian shape of the SF implies
chaotic states that fully occupy the energy shell defined by $V$, consequently leading to a
fast statistical relaxation.

In Section 3.3 we use a nuclear physics example of the situation where the interaction $V$ is
very strong and one is  beyond the Fermi golden rule regime. Such a situation arises for
heavy nuclei and leads to the so-called {\it chaotic enhancement of perturbation}. This
effect leads to an impressive  result of the large parity violation that was
observed experimentally. We show how the enhancement of perturbation, that is directly due to the chaotic compound eigenstates in the region of  high level density, helps to explain the
experimental results.

For a long time the onset of quantum chaos  was mainly linked to specific properties of
fluctuations in the energy spectrum. These fluctuations were thoroughly studied for full
random matrices and the results have served as a reference for the comparison with  the
spectral properties of realistic physical systems. The quantities associated with the
eigenvalues, such as the level spacing distribution $P(s)$, can be instrumental in the
interpretation of results obtained with realistic many-body systems in experiments with poor
resolution. Important complementary  information is contained in the structure of eigenstates, which
justifies our focus on  the problem of the emergence of chaotic eigenstates as a function of
the model parameters. In Section 4.1 we  briefly mention the reasons for shifting our interest
from spectral statistics to chaotic eigenstates and discuss pioneer studies on this subject.

While some of the paradigmatic models for one-body chaos, such as billiard-type systems, are
characterized by eigenstates spread all over the unperturbed basis, this does not
typically occur when many-body systems are considered. This is due to the finite rank of the inter-particle interactions  in realistic systems. As a result, the total Hamiltonian $H$ of real systems cannot be associated
with full random matrices. Despite having a large number of uncorrelated components,
many-body eigenstates typically span over a finite region of the Hilbert space. One of the
first studies of such chaotic eigenstates was  performed by Chirikov in 1985 when he analyzed experimental data for highly excited states of the cerium atom \cite{c85}. In Section 4.1 we compare the
chaotic structure of the eigenstates of the cerium atom  obtained by a direct {\it ab-initio} computation with the results of the TBRE model. We also discuss the chaotic eigenstates of heavy nuclei, when the dimension of the many-body basis defined by the orbitals of
non-interacting nucleons turns out to be extremely large.

The notion of chaotic eigenstates plays a key role in  the statistical description of
isolated quantum systems. As mentioned by Landau and Lifshitz in their {\sl Statistical
Mechanics}, the full description of quantum statistical mechanics can be done on the level of
individual states, not only by means of the Gibbs distribution. However, this statement does
not answer the important questions of {\it when} or {\it under what conditions} this fact
holds in practice, especially for mesoscopic systems. The answer is: when an individual eigenstate can be treated as a very
complicated superposition of many components, in our words - in the case of chaotic
eigenstates. Thus, in order to speak about thermalization one needs to know the conditions
required for the onset of chaotic eigenstates.

Due to the importance of the study of individual eigenstates, we discuss in Sections 4.2-4.4 their main 
statistical characteristics. We start with the {\it  Shannon, or information, entropy}. It quantifies the effective
number of components of a given eigenstate in a specific basis and can serve as a measure
of chaos or complexity of individual eigenstates. The value of this entropy characterizes the
dynamical interrelationship between the basis vectors of the  representation used and the
eigenbasis of the full Hamiltonian. Similarly to the studies of localization in disordered
systems, the effective number of components of an eigenstate can be associated with its
localization length in the chosen basis (in this case, in the many-body basis of the
unperturbed Hamiltonian $H_0$). As a paradigmatic example we consider in Section 4.2
 the $sd$-shell model of
nuclear physics and study  how  the localization length depends on energy. In Section 4.3 the
statistical moments of the component distribution are discussed. Specifically,  the second moment defines the  {\it inverse
participation ratio}, which is commonly studied  in solid state physics. In Section 4.4 we
present results for the {\it invariant entropy}, which, contrary to the Shannon entropy, is basis
independent. We show how this quantity can be effectively used for the description of phase
transitions or crossovers occurring in systems described by chaotic eigenstates. We
demonstrate the effectiveness of the invariant entropy by employing it to the nucleus
$^{24}Mg$ where there is an interplay between isoscalar and isovector pairing.

Section 4.5 is dedicated to the important question of the relationship between chaotic
eigenstates and the energy shell. The latter defines the maximal energy region that can be
occupied by the eigenstates with respect to the unperturbed energy spectrum of $H_0$. The
width of the energy shell is entirely determined by the interaction term $V$. With the use of the
WBRM and spin models, we show how the width of the energy shell can be estimated, and how  it
gets filled as the interaction strength increases. We argue that, for a finite number of
particles distributed over a finite number of single-particle levels, the crossover to
chaotic eigenstates  filling the energy shell is accompanied by the change of the
strength function from the Breit-Wigner to the Gaussian-like shape. This is used  to
obtain approximate estimates for the onset of fully chaotic states and thermalization. In
fact, such a crossover from localized to extended eigenstates with respect to the energy
shell can be associated with many-body delocalization, a subject still under investigation.

In Section 5 we discuss how thermalization can occur in isolated finite systems of interacting
particles. We start (Section 5.1) by noting that the idea of thermalization in systems isolated
from a heat bath is not usually treated in standard textbooks on statistical mechanics. There
are many possible definitions of the effective temperature in such cases. We show that, in a
finite system with a self-consistent mean field and chaotic eigenfunctions, various
thermometers used for defining the effective temperature do agree. In a sense, the system
behaves as its own heat bath and basic notions of statistical mechanics can still be
employed. For example, one of the general properties of thermalization in conventional
statistical mechanics of non-interacting identical particles is the emergence of the standard
Fermi-Dirac and Bose-Einstein distributions. From these distributions one can define the
single-particle temperature. In Section 5.2 we demonstrate the emergence of the Fermi-Dirac
distribution in the realistic nuclear shell model. An essential part of this discussion
supported by numerical data covers the onset of Fermi-Dirac distribution seen at the level of
individual compound states. We conclude Section 5.2 with the comparison of the single-particle
temperature to the thermodynamic temperature defined through the  density of states. Under
the conditions of validity of a single-particle thermometer, the two temperatures essentially
coincide, thus indicating that, indeed, one can speak of genuine thermalization in the
absence of a heat bath. We also take the opportunity to briefly discuss the meaning of a
thermometer.

Section 5.3 discusses the relevance and interrelation of the standard canonical
distribution. For this, the TBRE model with a given  number of fermions occupying a finite
number of single-particle levels is used. We start with the definition of the occupancies
distribution directly involving  the structure of the chaotic eigenstates. The key point here
is that when the eigenstates consist of a very large number of statistically independent
components, the occupation number distribution can be evaluated in terms of the shape of the
eigenstates written in the unperturbed basis. Thus, we do not need to know the eigenstates
themselves, $-$ instead it is sufficient to know  the $F$-{\it function}, that is the envelope
of the distribution of the projections of exact eigenstates of the total Hamiltonian $H$ onto
the unperturbed states of the  Hamiltonian $H_0$. When computing the $F$-function, we do not
need to average over few eigenstates, but can instead use the "moving window"
method for a single eigenstate. The analysis of the $F-$function and the strength function
allows one to identify  chaotic eigenstates. In this section  we also show how the occupation
number distribution obtained from an {\it isolated} eigenstate is related to the canonical
distribution.

The explanation of how the Fermi-Dirac distribution emerges in the TBRE model is given in
Section 5.4. If the inter-particle interaction is weak, one cannot speak of the occupation
number distribution as a smooth function of energy. When the perturbation strength exceeds
some critical value, the occupation number distribution has a form close to the Fermi-Dirac
function allowing for the introduction of temperature in a consistent way. Rough analytical
estimates allow one to describe this distribution in terms of the model parameters.

Different definitions of temperature are discussed in Section 5.5. We introduce the canonical
temperature and show how, in the TBRE model, this temperature reproduces the energy
dependence of the temperature defined by the global level density. We also provide a detailed
analysis of different temperatures for a model of two interacting spins. It is shown that for
high values of spins, when the classical limit is approached, the eigenstates turn out to be
quite chaotic in some energy region. Thus, even for two interacting particles one can speak
of chaotic states and thermalization at sufficiently high energy.

In Section 6 we come to the description of  dynamical properties of isolated systems by relating
them to the strength function and the structure of the eigenstates. This subject is now under
close scrutiny due to the experiments with trapped ions and with interacting particles (fermions and bosons) in optical lattices. We start with the definition of the survival probability (Section
6.1), which is simply the Fourier transform of the strength function. It is commonly assumed
that in quantum systems the survival probability decreases exponentially in time, apart from
a very short initial quadratic decay. The exponential behavior  is a direct
consequence of the Breit-Wigner form of the strength function. If the strength function has a
Gaussian form, as happens at a relatively strong inter-particle interaction, the Gaussian
decay can last for a very long time  prior to the typical restoration of  the exponential
behavior. With the use of the TBRE model we  show how these two regimes  emerge depending on the interaction $V$. We also present analytical and numerical arguments describing the
restoration of the exponential behavior after a long Gaussian decay. In the very long time limit,
the time dependence of the decay agrees with the power law defined by the lower boundary
of the energy spectrum.

In Section 6.2 we show numerical and analytical results for the survival probability  for the
spin  models. The same transition in the strength function from a Breit-Wigner to a
Gaussian form as a function of the perturbation strength is observed and studied for both
chaotic and integrable systems. We also analyze how the dynamics depends on the choice of the
initial state. This is particularly important for experiments with optical lattices and trapped ions, with a relative freedom in the preparation of the system. We also show that on a certain
time scale the dynamics in integrable and  non-integrable models may be very  similar. Finally,
we briefly discuss situations  where the decay of the survival probability is faster than
Gaussian.

In Section 6.3 the relaxation process is related  to the notion of the energy shell by studying
the spreading of wave packets. In contrast to  one-body chaos, the spread of probability in
the many-body unperturbed basis is not diffusive  but it can be described by the so-called
{\it cascade dynamics} on the Cayley tree. The data for the TBRE model demonstrate how the
dynamics depend on whether the strength function has a Breit-Wigner or a Gaussian form. With
the use of the cascade model, one can obtain simple expressions for dynamical quantities,
such as the Shannon entropy of time-dependent  wave functions. In some practical limits, the
increase of the  Shannon entropy has a simple form, being linearly proportional to time up to
the saturation due to finite size effects. This linear dependence is a generic
property of the dynamics corresponding to eigenstates fully extended in the energy shell. We
also discuss how the evolution of the participation ratio and of the occupation number
distribution depend  on the inter-particle interaction.

In Section 6.4 we show that the linear increase of the Shannon entropy with time can be used to
detect the onset of strong chaos and statistical relaxation. To demonstrate this we consider
a one-dimensional model of interacting bosons. The dynamics is periodic in time for a
relatively weak interaction, and when the interaction strength exceeds some critical value, the
Shannon entropy clearly displays a linear time dependence, in correspondence with the
analytical predictions. The crossover from the time-periodic dependence of the Shannon
entropy to the  linear time dependence corresponds to the onset of the so-called
Tonks-Girardeau regime which was  predicted analytically long ago.

In Sec. 7 we briefly discuss the relevance of our approach to recent studies of the onset of
thermal equilibrium in view of the so-called {\it eigenstate thermalization hypothesis}. We
argue that this hypothesis, essentially equivalent to the classical statement of Landau and Lifshitz,
is a natural consequence of the onset of thermalization, but it does not define the
conditions for it to occur. Our approach suggests the way to identify these conditions.
\section{Hamiltonians}
\subsection{Basic ideas}
In this review we consider various models of isolated systems of interacting particles described by a Hamiltonian that usually can be separated into two parts,
\begin{equation}
H=H_0+V,                          \label{1}
\end{equation}
with $H_0$ describing a finite number of {\sl non-interacting} particles or quasi-particles kept together by some field, external or self-consistent. Frequently, the part $H_0$ is treated as the unperturbed Hamiltonian and $V$ is considered as {\sl residual interaction}, usually of two-body type. The strength of the interaction can be regulated driving the system from the perturbative regime to that of strong interaction and quantum chaos. By the latter we mean specific properties of spectra and eigenstates, that allow us to develop a statistical approach. As we demonstrate below, the main property of quantum chaos can be attributed to a chaotic structure of eigenstates in the unperturbed basis of $H_0$.

In what follows we discuss both the dynamical systems  without any random parameters as well
as  the systems where the matrix elements of the Hamiltonian (\ref{1}) may include random entries or can be completely random and uncorrelated. In such applications we study {\it ensembles} of related Hamiltonians which embody different realizations of randomness. It is worth to stress that the randomness of the perturbation $V$ by itself does not guarantee the emergence of chaos and the validity of a statistical description. On the other hand, realistic interactions without random elements naturally produce chaotic features in a region of a sufficiently high level density. The two-body nature of typical interactions leads to important restrictions,
such as the band-like structure of the Hamiltonian matrix and the presence of many vanishing and repeated matrix elements, which we will discuss later on.

The separation of the total Hamiltonian
(\ref{1}) into two parts is common in the description of many-body systems, including complex
atoms and nuclei, where the starting point is the {\sl mean field approximation}. The mean field can be introduced phenomenologically or it can be derived self-consistently with the aid of variational principles. In mesoscopic systems such as cold atoms in traps, $H_{0}$ is created by external fields. In these cases, the mean field generates a ``natural" basis, where the most regular part of the dynamics is absorbed in $H_0$ thus defining the single-particle states (quasi-particles) with their quantum numbers and symmetries. In contrast, the residual interaction $V$ may include the correlations between particles and fluctuational terms which cannot be embedded into the mean field. In this way, the system acquires both collective motion on the background of  the mean field, and {\sl chaotic} dynamical features. Although the choice of the mean field is not uniquely defined, one can expect that if the most regular features of a system are well described by $H_0$, the main results are not sensitive to the details coming from the specific choice of the mean field.

Many ideas for a treatment of systems with strongly chaotic properties came from the {\sl random matrix theory} (RMT), formulated in 1951-1963 by the works of Wigner, Dyson and others (see, for example, the collection of main contributions edited by Porter \cite{porter65}). The original intention of Wigner was to use random matrices in order to describe statistical properties of complex nuclei and nuclear reactions.

Initially the applications of the RMT were limited to atomic and nuclear physics. According to the suggestion by N. Bohr \cite{bohr36} to consider nuclear reactions induced by
slow neutrons as proceeding in two separated stages (the first one related to the excitation of a
nucleus caused by the primary interaction of the captured neutron, and the second one as the decay process after a relatively long internal evolution), the multiple narrow neutron resonances
correspond to {\sl compound states} with a long lifetime that allows for intrinsic equilibration and actual independence of those two stages. The processes governed by very complicated interactions in compound nuclei can be described only statistically, and here the random matrices provide an appropriate and powerful instrument. Indeed, if one assumes an extremely complex character of strong interactions between the nucleons in the excited nucleus at high level density, one can imagine that in the considered energy region the Hamiltonian matrix is so complicated that its matrix elements can be treated as random entries. This interaction keeps memory only of constants of motion (energy, angular moment, parity, isospin in nuclei) creating exceedingly entangled quasistationary wave functions, a process that can be compared to thermal equilibration in macroscopic bodies. In each class of states with the fixed values of
exactly conserved quantities, the components of the many-body stationary states in the mean-field basis are practically uncorrelated, and their statistical distribution is typically close to the normal one. Through multiple avoided level crossings, the spectral repulsion forms aperiodic crystal structure of  the levels similar to what is known in simple billiard-like systems which reveal one-body classical chaos in an appropriate classical limit. The difference is in the mechanism of chaotization $-$ here it is governed by the interparticle interactions rather than by the violation of the billiard symmetry. The complicated interactions effectively heat the system, so that
 the idea of many-body chaos reveals its connection to {\sl statistical thermalization} that will be one of the  main topics of this review article.

Instead of a dynamical description of nuclear reactions, Wigner focused the attention on the statistical aspects of nuclear spectra revealed by the neutron scattering data
\cite{wigner1951,wigner1955,wigner1957,wigner1958}. This idea is similar to the approach used in the modern theory of dynamical chaos emerging {\sl in classical systems} due to a local instability of  motion. As is now well understood, the statistical method, that seems to be an approximation to the ``genuine" deterministic description, is in fact the {\sl only adequate way} to treat chaotic systems. The link between the dynamical (with a deterministic time evolution)
behavior of a large number of constituents of a quantum system and the statistical approach
can be justified by the negligible role of tiny correlations between the ``true" matrix elements of the Hamiltonian and the numerous components of the generic initial wave function.

At a first glance, such an approach to the spectra of real systems looks misleading since every system has its own specific set of quantum levels forming its density of states. However, the point is that {\it locally} the fluctuating properties of energy spectra may have universal properties independently of the global evolution of the energy spectrum. At the excitation energy where the level density is so high that the mixing of the mean-field states by residual interactions becomes effectively strong, the wave functions expressed in the mean-field basis are inevitably extremely complicated superpositions and their statistical properties acquire {\sl universal features} determined mostly by the fundamental symmetry laws. This is the starting point of the RMT. At the time of early works by Wigner \cite{wigner1951,wigner1955,wigner1957,wigner1958}, this assumption was far from being obvious. Later on, various experiments with heavy nuclei (see Refs. in \cite{Brody81,Bohigas84}),
complex atoms \cite{rosenzweig60,camarda83} and many-electron molecules \cite{haller83,AFIIK84} confirmed the predictive power of the RMT, even if the chaotic RMT limit is not always fully reached in reality.

\subsection{Wigner Band Random Matrices}

In the first attempt to establish the relation between statistical properties of complex quantum systems and random matrix models \cite{wigner1951,wigner1955,wigner1957,wigner1958}, Lane, Thomas and Wigner introduced an ensemble of {\sl banded matrices} for the description of conservative systems like atomic nuclei \cite{lane55}. Assuming time-reversal invariant dynamics, an ensemble of real symmetric infinite Hamiltonian matrices was considered,
\begin{equation}
H_{mn}=\epsilon_n\delta _{mn}+V_{mn}, \,\,\,\, \epsilon _n=nD, \quad  V_{mn}\,=\,V_{nm},      \label{2}
\end{equation}
The diagonal part was modeled with an equidistant spectrum (``{\sl picket fence}"), $\epsilon_{n+1}-\epsilon_n=D=1/\rho_0$, where $\rho_0 $ is the level density of the ``unperturbed" Hamiltonian $H_0=\epsilon_{n}\delta_{mn}$. In Refs. \cite{wigner1951,wigner1955,wigner1957,wigner1958} the absolute values of the off-diagonal matrix elements were taken equal, $V_{mn}=\pm v$, while the signs were assumed to be random and statistically independent within the band of width $2b$ around the main diagonal. In a more general case thoroughly studied in Ref.~\cite{FCIC96}, the diagonal elements $\epsilon_n$ are random entries, the matrix elements $V_{mn}$ are distributed randomly with $\langle V_{mn}\rangle=0$ and $\langle V^2_{mn}\rangle=v^2$ for $|m-n|<b$, while  $V_{mn}=0$ outside the band (here and below the angular brackets stand for the ensemble average). The assumption of random character of the ``perturbation" $V$ was a pioneering step in the statistical description of complex quantum systems. In his seminal paper of 1955, Wigner wrote that the considered quantum-mechanical systems {\it ``are assumed to be so complicated that statistical consideration can be applied to them}".

In the model (\ref{2}), the unperturbed density $\rho_0$ is constant which simplifies the analytical treatment. In modern presentations of banded random matrices (BRM), it is typically assumed that the unperturbed spectrum of $H_0$ corresponds to the {\sl Poisson level statistics} \cite{gurevich57}, so that the eigenvalues are random entries with the constant {\it average} density $\rho_0$, instead of the picket fence spectrum used by Wigner. The new element here is the presence of large fluctuations in the unperturbed spectrum.

Two important features of the BRM should be stressed. The first point is that the band-like appearance of a matrix is not invariant with respect to orthogonal transformations of the basis. The special basis diagonalizing $H_0$ should be thought of as corresponding to the mean-field representation. In this way it is assumed the existence of a physically singled out basis in which the treatment of the {\sl total Hamiltonian} is preferential. The second point is that the banded structure  reflects a finite range of interaction in the energy representation that may emerge from the physical selection rules.

After Wigner's pioneering work, the BRM were almost forgotten (curiously enough, by Wigner himself \cite{wigner1951,wigner1955,wigner1957,wigner1958}), apparently because of their mathematical inconvenience, namely the absence of invariance with respect to basis rotations. Due to this, attention was paid mainly to full random matrices for which a fairly complete mathematical analysis has been developed \cite{porter65,mehta67,Brody81}. However, in real physical applications full random Hamiltonian matrices can be only used to describe the {\sl local statistical properties} of spectra and not the global ones. For this reason, such matrices were criticized by Dyson \cite{Dyson62a,Dyson62b,Dyson62c} because of the ``unphysical" semicircle law of the total level density.

\subsection{Two-body random interaction}
In order to use the random matrix approach with a more realistic level density, an ensemble of random matrices was suggested in Refs.~\cite{French1970,French1971,Bohigas1971,Bohigas1971a,mon75} that takes into account the $n$-body nature of interaction between the particles (for details and other references, see \cite{Brody81}). Since in the majority of physical applications the main contribution is due to two-body interactions ($n=2$), this kind of random matrices, known as {\it two-body random interaction} (TBRI) matrices, has been studied in great detail. The ensemble of such matrices, referred to as the two-body random ensemble (TBRE), is defined in the secondary quantized form as
\begin{equation}
H=\sum \epsilon _s\,a_s^{\dagger }a_s + \frac 12\sum V_{s_1s_2s_3s_4}\,a_{s_1}^
{\dagger }a_{s_2}^{\dagger}a_{s_3}a_{s_4} .                              \label{3}
\end{equation}
Here the term $H_0=\sum\epsilon _s\,a_s^{\dagger }a_s$ corresponds to non-interacting particles and $V$ absorbs the two-body interaction. In a more general context, $H_0$ can be treated as a regular one-body part of the total Hamiltonian written in the mean-field basis, and $V$ represents the {\it residual interaction} which, due to its very complicated structure cannot
be embedded into the mean field. The entries $\epsilon_s$ represent single-particle (or quasi-particle) energies corresponding to single-particle states $\left| s\right\rangle $,
while $a_{s}^{\dagger }$ and $ a_{s}$ are particle creation and annihilation operators for fermions or bosons. These operators define the many-particle basis $\left| k;n\right\rangle=a_{s_1}^{\dagger }\,.\,\,.\,\,.\,a_{s_n}^{\dagger }\left| 0\right\rangle $ of non-interacting particles, where the symbol $k$ labels the whole $n$-body configuration. In this basis $H_0$ is diagonal with eigenvalues $E_{k}=\sum\epsilon_{s}$ defined  by the single-particle levels occupied in the many-body state $\left|k;n\right\rangle$.

The matrix elements $V_{s_1s_2s_3s_4}$ of the perturbation $V$ describe a two-body
process with indices $s_1,s_2,s_3,s_4$ indicating initial $ (s_3,s_4)$ and final $(s_1,s_2)$
single-particle states connected by this interaction. It is convenient to reorder this basis
according to the growth of unperturbed energies $ E_k$ with an increase of the index $ k=1,\,.\,.\,.\,,\,{\cal N}\,$. The size ${\cal N}$ of the basis for $H$ depends on the statistics of the particles and practical truncation of the single-particle space. For instance, in the case of Fermi-statistics, any single-particle state can be occupied by one particle only, and
\begin{equation}
{\cal N}=\frac{N_s!}{N_p!(N_s-N_p)!},                                       \label{4}
\end{equation}
where $N_s$ is the number of single-particle levels and $N_p < N_s$ is the number of particles
occupying these levels. The total number of many-body states increases very fast with the particle number and the number of available orbitals.

\begin{figure}[htb]
\centering
\includegraphics[angle=-90,width=0.5\textwidth]{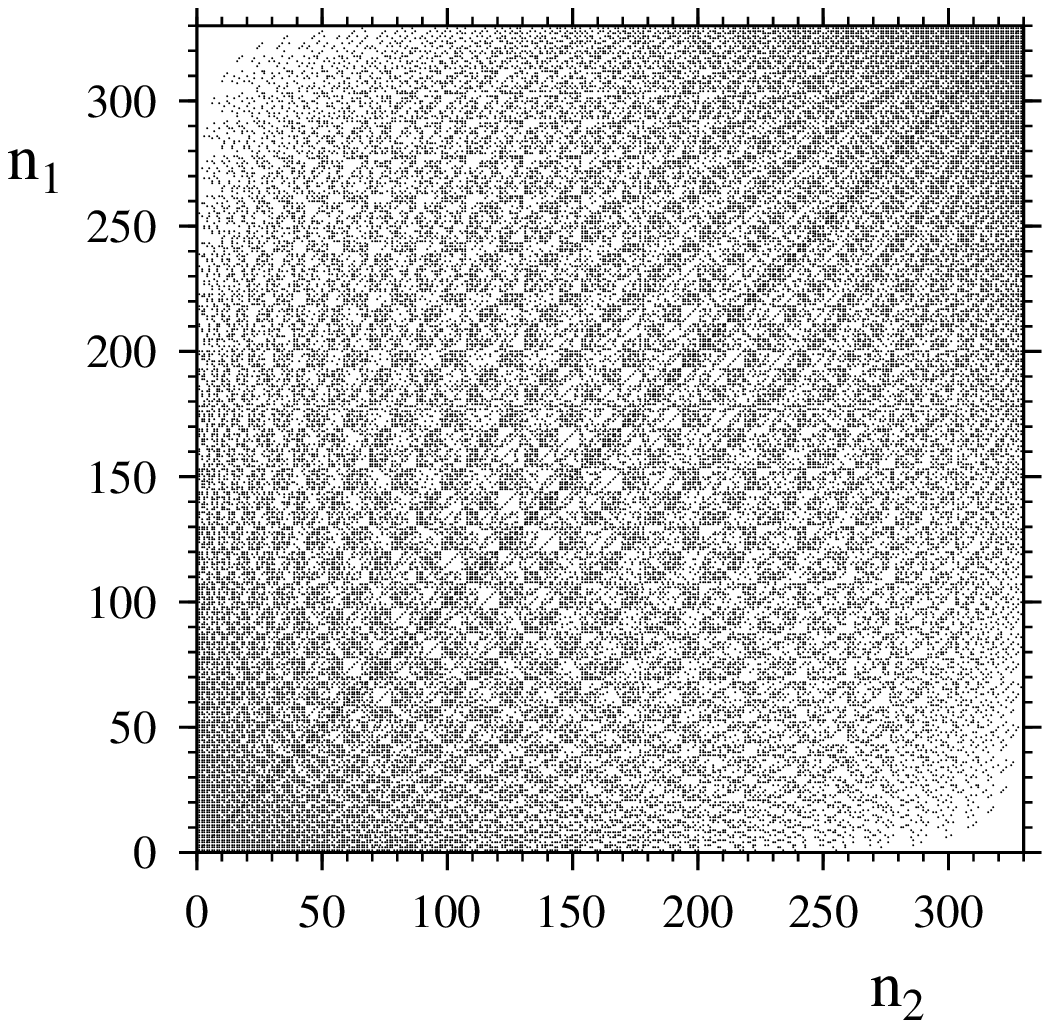}
\caption{Sparsity of the Hamiltonian matrix $H_{n_1,n_2}$ for $N_p=4$ particles and $N_s=11$ single-particle levels. Black points are non-zero matrix elements of two-body interaction (after \cite{IVarenna}).}
\label{matrix99}
\end{figure}

In the TBRI model all matrix elements $V_{s_1s_2s_3s_4}$ are assumed to be random independent
variables. However, due to the two-body nature of the interaction for $N_s \gg N_p \gg 1$, the matrix $H_{kk'}$ turns out to be band-like, with many vanishing elements inside the band, see example in Fig.~\ref{matrix99}. The total number $K$ of non-zero elements $H_{nm}$ in each line of $H$ can be estimated as $K \approx \frac{1}{4}N_p^2 N_s^2$ which is much less than its size ${\cal N}$. Moreover, many non-vanishing matrix elements turn out to be correlated, even in the case of complete randomness of the interaction matrix elements $V_{s_1s_2s_3s_4}$. This is a consequence of the $n$-body character of interactions with $n\ll N_{p}$: the interaction matrix elements are the same for any configuration of spectator $(N_p-n)$ particles occupying single-particle states not involved in the given matrix element $V$. As shown in Ref.~\cite{Flambaum1996a}, this is important when analyzing the statistical properties of some observables. Apart from the sparsity and intrinsic correlations in TBRI matrices, another difference from the Wigner BRM with the sharp band boundary is that the amplitudes of the matrix elements $H_{kl}$ decrease smoothly away from the diagonal. It should be stressed that all these peculiarities are quite typical for physical systems such as complex atoms and nuclei (see, for example, \cite{FGGK94,ZelevinskyRep1996}).

Quite specific properties emerge if the Hamiltonian reveals additional symmetries so that the Hilbert space can be  decomposed into separate subspaces of states, and the dynamics within each subspace is either regular or chaotic. For instance,  an unusual result was observed in a simple simulation \cite{johnson98} for few fermions occupying a single level with a large total angular momentum quantum number $j$ and interacting through all types of two-body matrix elements of random magnitude but restricted by rotational invariance. The new aspect here is the interrelation between non-overlapping classes of states due to the dynamics driven by the same Hamiltonian. The unexpected result is a clear predominance of ground states of total spin $J=0$. Statistical considerations \cite{ZVPhysrep04} qualitatively explain this by assuming that the wave functions are randomized and prefer maximal or minimal values of total spin (precursor of ferromagnetic or anti-ferromagnetic order). It is interesting that the observed effect seems to be different from the spin glass system \cite{oitmaa01}, where the ground state spin on average grows as the square root of the number $N_{p}$ of interacting spins. Other regular collective effects also appear with significant probability in such systems with random interactions \cite{HZ10,abramkina11}, where a quantitative theory is still absent.

One can also use random interactions in order to study possible landscapes arising in the sectors with different values of random parameters. This was done in the interacting boson models \cite{bijker}, where it was possible to delineate the parameter space areas corresponding to different symmetries of the system, and in the nuclear shell model \cite{HZ10}, where the random interactions allowed one to find out the sectors of the random parameter space responsible for the predominance of prolate deformation of the mean field.

\subsection{Realistic models}

In order to demonstrate our approach to the problem of thermalization, below we consider few realistic models of interacting spins-$1/2$, as well as the nuclear shell model widely used in nuclear physics. Contrary to TBRI, these models are deterministic since they have no random entries. However, under some conditions their main properties can be compared with those described by random TBRI Hamiltonians.

\subsubsection{Spin-1/2 models}

Spin-1/2 models describe systems experimentally studied with nuclear magnetic resonance, optical lattices and trapped ions. They also model real magnetic compounds and quantum computers.
Two models of interacting spins 1/2 are considered here. One is completely integrable (analytically solvable), and the other is non-integrable \cite{Santos2012PRL,Santos2012PRE}.

The Hamiltonian for the integrable case (Model 1) has only nearest-neighbor (NN) interaction:
\begin{equation}
H_1 = H_0 + \mu V_1 ,\qquad
H_0 = \sum_{i=1}^{L-1} J \left(S_i^x S_{i+1}^x + S_i^y S_{i+1}^y \right) , \qquad
 V_1 = \sum_{i=1}^{L-1} J  S_i^z S_{i+1}^z,                   \label{5}
\end{equation}
where $H_0$ corresponds to the unperturbed part of the Hamiltonian and determines the mean field basis, and $\mu$ is the strength of the perturbation.
Above and below, $L$ is the number of sites, and $S^{x,y,z}_i$ are the spin operators at site $i$.  We assume both $J$ and $\mu$ positive, thus favoring anti-ferromagnetic order. The coupling parameter $J$ determines the energy scale and will be set to 1.

The unperturbed part $H_0$, known as the flip-flop term, is responsible for moving the excitations through the chain. It is integrable and can be mapped onto a system of noninteracting spinless fermions~\cite{Jordan28} or hardcore bosons \cite{Holstein40}. The Hamiltonian remains integrable with the addition of the Ising interaction $V_1$, no matter how large the anisotropy parameter $\mu $ is. The total Hamiltonian $H_1$, referred to as the XXZ Hamiltonian, can be solved with the Bethe ansatz \cite{Bethe1931}.

Model 2 is described by the Hamiltonian,
\begin{equation}
H_2 = H_1 + \lambda V_2 ,\qquad
V_2 = \sum_{i=1}^{L-2} J\left[ \left( S_i^x S_{i+2}^x + S_i^y S_{i+2}^y
\right) + \mu S_i^z S_{i+2}^z  \right] ,                      \label{6}
\end{equation}
that includes both nearest and next-nearest-neighbor (NNN) couplings.  Here the mean-field is defined by $H_1$ which is the same XXZ Hamiltonian as in Model 1. The parameter $\lambda \geq 0$ refers to the relative strength of the NNN exchange determined by the perturbation $V_2$ and the NN couplings characterized by $H_1$. The dynamics generated by this Hamiltonian becomes chaotic when the strengths of the NN and NNN couplings are comparable. In particular, for finite systems, there is a threshold value of $\lambda$ above which the level spacing distribution becomes of the Wigner-Dyson form~\cite{Hsu1993,Kudo2005,Santos2009JMP}. The threshold value decreases as $L$ increases suggesting that, in the thermodynamic limit, the Wigner-Dyson distribution might be achieved with an infinitesimally small integrability breaking term~\cite{Santos2010PRE,Torres2014PRE}.

Both Model 1 and Model 2 conserve the total spin in the $z$ direction, ${\cal S}^z_{{\rm tot}}$.
The Zeeman splittings, caused by a static magnetic field in the $z$ direction, are the same for all sites and are not shown in the Hamiltonians above. We note that the presence of random Zeeman splitting~\cite{Avishai2002,SantosEscobar2004,Brown2008} or even a single Zeeman splitting different from the others~\cite{Santos2004,Gubin2012,Torres2014PRE} can also led Model 1 to show a Wigner-Dyson distribution. A comparison between the latter case and Model 2 is presented in Ref.~\cite{Torres2014PRE}.

\begin{figure}[htb]
\centering
\includegraphics[width=0.65\textwidth]{Fig02_long_aMatrix.eps}
\caption{(Color online.) Absolute values of the matrix elements of Model 1, eq. (\ref{5}), left panel, and
Model 2, eq. (\ref{6}), with $\lambda =0.5$ (right panel) for $L=12$ and $\mu=0.5$. The mean-field basis is ordered in energy. Only even states are considered. Light color indicates large values (after \cite{Santos2012PRE}).}
\label{structure}
\end{figure}

In Fig.~\ref{structure}
the structure of the Hamiltonian matrices is shown for both models in the basis of $H_0$ for Model 1 and in the basis of $H_1$ for Model 2, for typical values of control parameters. The global structure of these Hamiltonians is similar to that emerging for the TBRI Hamiltonian (\ref{3}), although some peculiarities of the Models 1 and 2 can be seen. We observe a more regular structure of the Hamiltonian for the integrable Model 1 in comparison with the non-integrable Model 2. Some kind of symmetries can be detected for Model 1, which can be treated as a fingerprint of the integrability of this model. In spite of this, many properties of the eigenstates, as well as of the dynamics for these two models look quite similar and can be compared with those of the TBRI model (\ref{3}) with random two-body interaction.

\subsubsection{Nuclear shell model}

Another example of a model describing physical systems with global features similar to TBRI is the nuclear shell model (frequently called {\sl configuration interaction}). Various versions of this model have similar global features. Here the nucleons are supposed to move in the mean field (for example, that of the isotropic harmonic oscillator with added spin-orbit coupling or another potential well) and interact through two-body interactions allowed by the global conservation laws (angular momentum, parity, isospin, $J^{\Pi}T$). The orbital space is truncated; one of the best examples is provided by the $sd$-model that includes, both for protons and for neutrons, only three $j$-levels, $d_{5/2}, d_{3/2}$ and $s_{1/2}$, whose energies are taken as parameters. Supposedly, this space should describe all isotopes between doubly-magic oxygen, $^{16}_{8}$O$_{8}$, and calcium, $^{40}_{20}$Ca$_{20}$. Only 63 two-body matrix elements conserving the constants of motion are possible here; they are taken from experimental data and renormalized to account for the truncation of the space \cite{ZelevinskyRep1996}. Thousands of observables corresponding to low-lying nuclear states and transition rates between them are well described, and this characterizes the reliability of the model as a powerful working tool \cite{Brown2001}. Varying the input, one can extract the matrix elements mainly responsible for specific observable trends \cite{HZ10}. Similar constructions are used in atomic physics, quantum chemistry, and for cold atoms interacting in traps \cite{Alcaraz87,Karbach1997,Reimann2002}.

In models of this type the full large-scale diagonalization of Hamiltonian matrices in each $(J^{\Pi}T)$ sector provides much more eigenstates than resolved experimentally, especially at high energy where the level density increases exponentially. Then we have at our disposal a good model of interacting fermions {\sl without any random parameters}. With an increase of excitation energy and level density, the stationary states become exceedingly complicated combinations of simple single-particle excitations. Then the fully deterministic system reveals the signatures of quantum chaos, which, therefore, becomes a generic property of the conglomerates of interacting particles. These signatures are close to those of band random matrix ensembles even if the distributions of many-body matrix elements are not Gaussian.

\section{Strength Functions}

\subsection{Definitions}

The original Wigner's motivation for utilizing banded random matrices was a physical interest in the so-called {\it strength function}. The relation to physical aspects of the strength function was for the first time introduced in application to molecular potential energy curves by Rice \cite{rice33}; the useful formulation can be found in the book \cite{BM}. The strength function $F_{k}(E)$ refers to the fragmentation of a basis state $|k\rangle$ over exact eigenstates
$|\alpha\rangle$ of the full Hamiltonian (\ref{1}),
\begin{equation}
H|\alpha\rangle=E^{\alpha}|\alpha\rangle.                     \label{7}
\end{equation}
We introduce the expansion of the eigenstates in the form
\begin{equation}
|\alpha\rangle=\sum_{k} C^{\alpha}_{k} |k\rangle, \quad |k\rangle=\sum_{\alpha}C_{k}^{\alpha\ast}
|\alpha\rangle;                                                                  \label{8}
\end{equation}
here and below, the notations use the low indices for the basis states, and upper indices for the exact eigenstates. The expansion of $|\alpha\rangle$ depends on the choice of the basis and in our problem the natural choice is the basis of the eigenstates of $H_{0}$: if $H_{0}$ corresponds to the mean field, the first expansion (\ref{8}) shows the degree of mixing of ``simple" mean-field states in the actual wave function of an eigenstate.

Wigner defined the {\it strength function} (SF) as the weighted level density [see below Eqs. (\ref{14}) and (\ref{15})],
\begin{equation}
F_k(E) = \sum_{\alpha} |C^{\alpha}_k|^2 \delta (E-E^{\alpha}).           \label{9}
\end{equation}
The strength function is normalized,
\begin{equation}
\int dE\,F_{k}(E)=\sum_{\alpha} |C^{\alpha}_k|^2 =1,               \label{10}
\end{equation}
and its {\sl centroid}, $\overline{E}_{k}$,  coincides with the {\sl diagonal} matrix element of the full Hamiltonian in the basis state $|k\rangle$,
\begin{equation}
\overline{E}_{k}\equiv \int dE\,F_{k}(E)E=\sum_{\alpha}E^{\alpha}|C^{\alpha}_k|^2=\langle k|H|k\rangle. \label{11}
\end{equation}
The inclusion of diagonal matrix elements of $V$ in $H_{0}$ eliminates the degeneracy often encountered in the mean-field picture of non-interacting particles.

The second moment of the SF,
\begin{equation}
\sigma_{k}^{2}=\int dE\,F_{k}(E)(E-\overline{E}_{k})^{2},                       \label{12}
\end{equation}
characterizes the dispersion of the fragmentation. As it is easy to see, the dispersion can be found as a sum of all squared {\sl off-diagonal} matrix elements along the $k$-th line of the original Hamiltonian matrix,
\begin{equation}
\sigma_{k}^{2}=(H^{2})_{kk}-(H_{kk})^{2}=\sum_{k'\neq k}|H_{kk'}|^{2}.       \label{13}
\end{equation}
It is practically important that both quantities, centroid and dispersion, can be found from the matrix of $H$ without actual diagonalization. The higher moments of the strength function can be also defined in a similar way.

With the aid of the {\sl density of states} normalized to the total dimension ${\cal N}$ of the Hilbert space,
\begin{equation}
\rho(E)=\sum_{\alpha}\delta(E-E^{\alpha}), \quad \int dE\rho(E)={\cal N},                       \label{14}
\end{equation}
the SF (\ref{9}) can be written as
\begin{equation}
F_{k}(E)=\rho(E)\langle|C^{\alpha}_{k}|^{2}\rangle_{E^{\alpha}=E}               \label{15}
\end{equation}
where the averaging $\langle ... \rangle $ is performed over a number of states with energy
close to $E$. This expression justifies the term frequently used, especially in condensed matter physics, {\sl local density of states} (LDOS), for which the lattice sites are taken as
an unperturbed basis thus representing the density of electrons at a specific  site.

In theoretical analysis, quite often it is more convenient to treat the average (\ref{15}) as
the {\sl ensemble average}, assuming that it gives the same result. In a numerical approach, it is
also possible to use the so-called ``moving window average" when the averaging is performed for a specific eigenstate with energy $E$  smoothing the function $|C^{\alpha}_k|^2$ by changing the index $k$. Such a procedure is justified if the global structure of eigenstates remains the same under a small shift of the eigenenergy $E$. While at high level density the components of individual states can strongly fluctuate, the SF assumes a smooth envelope in the energy representation for a fixed $k$ (or, which is the same, along the scale of the unperturbed energy $\epsilon_{k}$ of $H_0$ associated with the basis state $|k\rangle$), see, for example, Ref.~\cite{Frazier1996}). In other words, the strength function is a smoothed projection of a basis state $|k\rangle$ onto the exact states $|\alpha\rangle$, {\it given in the energy representation}.  If normalized to the mean energy level spacing, the strength function $F_k(E)$ characterizes an effective number $N_{{\rm pc};k}$ of principal components of stationary states $|\alpha\rangle$ which are present in the basis state $|k\rangle$.

Alternatively, one can speak of a smooth projection of an exact state $|\alpha\rangle$ onto the
basis states $|k\rangle$ expressed in the energy representation, due to the one-to-one correspondence between the index $k$ and the corresponding energy $\epsilon_k$ of an eigenstate of $H_0$. Then we come to the envelope of an exact eigenstate of $H$ in the basis of $H_0$, termed
{\it shape of the eigenstate}, that will be discussed later in connection to the chaotic structure of eigenstates. The weight matrix of components,
\begin{equation}
w_k^{\alpha}\equiv \left| C_k^{\alpha}\right| ^2=\left| C_k(E^{\alpha})\right| ^2,      \label{16}
\end{equation}
contains important information about the structure of {\it both}, eigenstates and strength functions.

It is important to note that the strength function and the shape of eigenstates have a well defined classical limit in the case of fully chaotic classical motion. If the system under consideration has a clear classical analog, the unperturbed energy $E_0$ is not constant along a classical chaotic trajectory of the total Hamiltonian $H=E$. Instead, it ergodically fills the volume created by projecting the surface corresponding to $H_0$ onto that of $H$. The form of the distribution of the energies $E_0$ is the classical counterpart of the quantum strength function. Conversely, if one keeps the unperturbed energy $E_0$ fixed, the bundle of trajectories of the total Hamiltonian $H$ that reaches the surface $H_0=E_0$ has a distribution of total energy $E$. In the quantum case, this distribution is nothing but a smooth envelope of an eigenstate of $H$ given in the energy representation. The quantum-classical correspondence for the strength functions, as well as for the envelopes of eigenstates, has been thoroughly studied in Refs.~\cite{WIC98,BGI98,MIU00,LMI00,BISS00,LMI01,I01,LMI02,BCIC02,BFH03,BFH03} for various models of interacting particles.

From the viewpoint of time-dependent evolution, the strength function shows the spread of excitation initially concentrated in a specific basis state $|k\rangle$ over other basis states due to the residual interaction $V$. If the interaction does not modify the boundary conditions, the unitarity of the transformation implies that the coefficients $(C^{\alpha}_{k})^{\ast}$ describe the spreading of the simple state $|k\rangle$ over stationary states $|\alpha\rangle$. For time-reversal invariant systems, these amplitudes can be taken as real. The knowledge of the matrix $C_k^\alpha$ for a given basis $|k\rangle$ and of the energy spectrum $E^\alpha$ gives complete information about the system.

\begin{figure}[htb]
\centering
\includegraphics[width=0.40\textwidth]{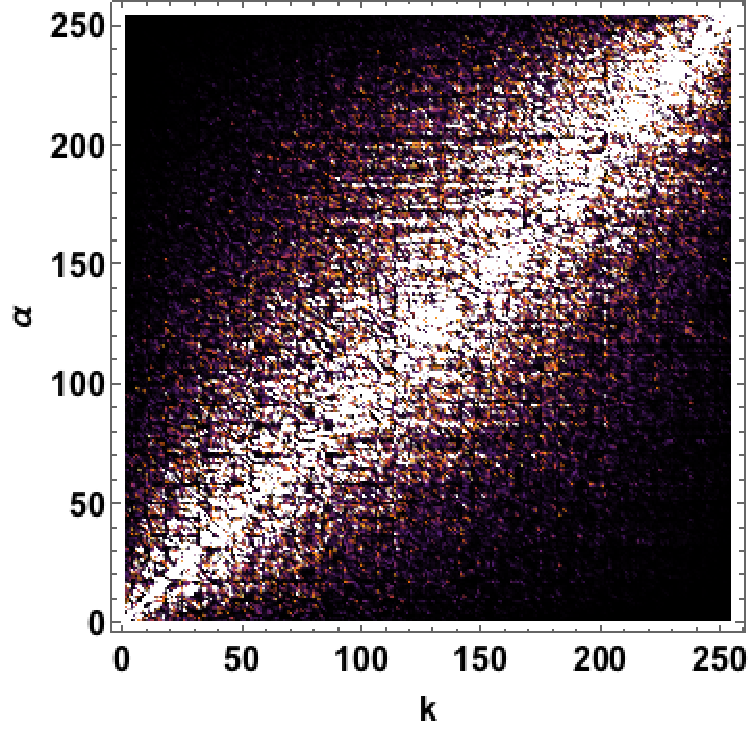}
\caption{Matrix $w_k^{\alpha}$ of squared components of eigenstates for Model 1 of interacting
spins 1/2 with $\mu =0.5$. Enhanced lightness corresponds to large values, black points
 stand for vanishing or very small values (after \cite{Santos2012PRE}).}
\label{w-n}
\end{figure}

An example of the structure of the matrix $w_k^{\alpha}$ is shown in Fig.~\ref{w-n}
for the model of four interacting spins-1/2 on a twelve-site one-dimensional lattice. One can see that both the strength functions (different $\alpha$ for a fixed $k$) and the eigenstates
(different $k$ for a fixed $\alpha$) are, globally, quite similar, occupying only a fraction of
the basis $|\alpha\rangle$ and $|k\rangle$, respectively. The restricted spreading is due to the
finite range of interaction between the particles reflected by the band-like structure of the
total Hamiltonian $H$ in the chosen mean-field basis. The knowledge of the matrix of components $w_k^{\alpha}$ allows one to predict the conditions for the onset of chaos and thermalization,
as will be discussed later.

When the number of significant components $C_{k}^\alpha$ in Eq.~(\ref{8}) is large,
the averaging and the summation can be replaced by an integration. Strictly speaking, this is correct when these components fluctuate around their mean values and can be considered as pseudo-random quantities. As discussed below, the condition of a large number of pseudo-random components in exact eigenstates can be used as a criterion of {\sl chaos} in quantum systems. In the case of a completely random perturbation $V$, the procedure of introducing a smooth energy dependence for the SF is well supported provided the number $N_{{\rm pc};k}$ of principal components $C_{k}^\alpha$ is large enough.

\begin{figure}[ht]
\centering
\subfigure{\includegraphics[scale=0.45]{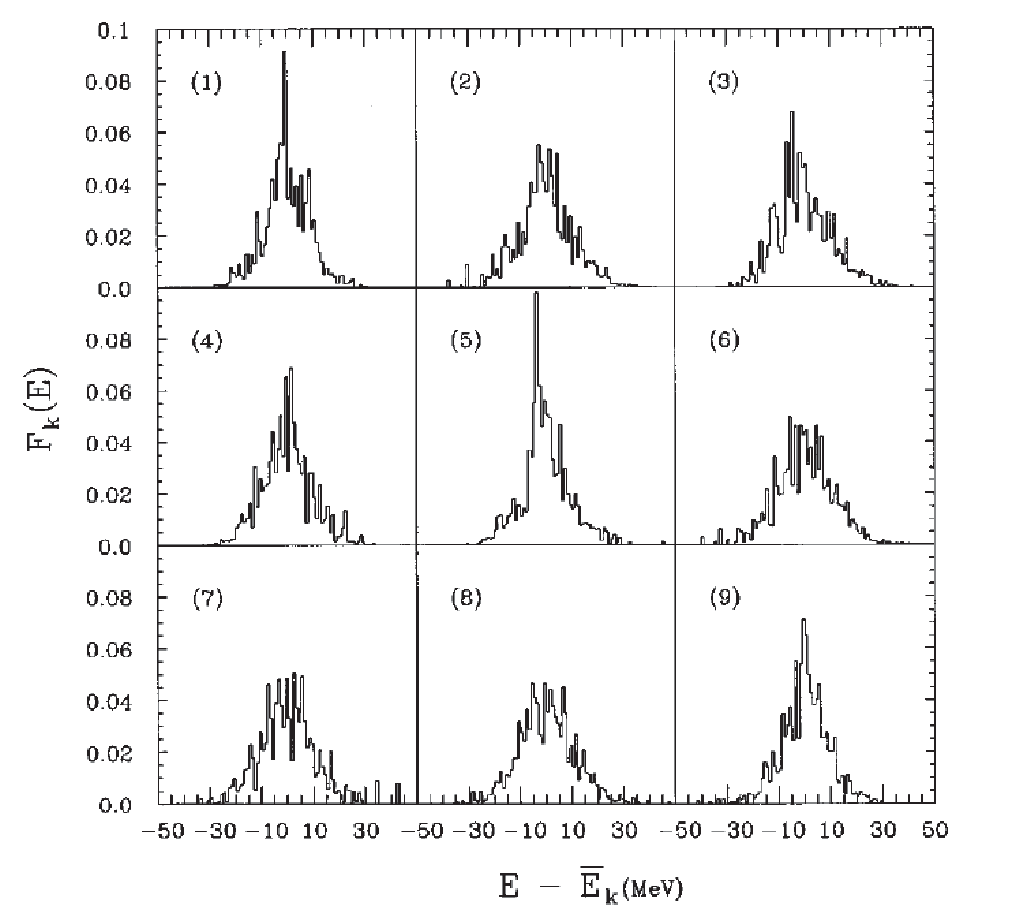}}
\subfigure{\includegraphics[scale=0.45]{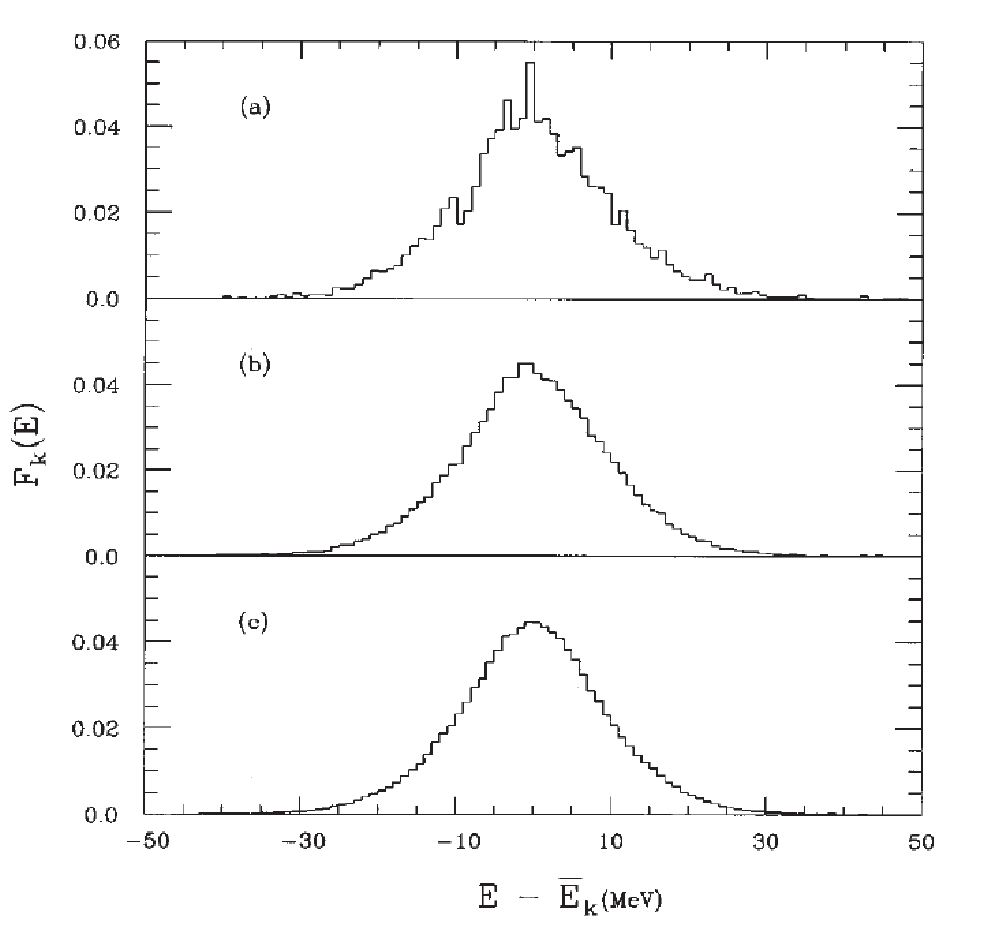}}
\caption{Left: the strength functions for individual $0^+ 0$ basis states $|k\rangle$ in the middle of the spectrum (histograms), versus the energy distance $E-\overline{E_k}$ from the centroid of the unperturbed state $|k\rangle$. Right: the strength function averaged over 10, 100 and 400 $J^{\Pi}T=0^+ 0$ basis states in the middle of the spectrum, panels (a)-(c), respectively. The bin size is 1 MeV (after \cite{Frazier1996}).}
\label{few-ldos-zele}
\end{figure}

We can illustrate this approach with examples from the nuclear shell model \cite{Frazier1996}. The
exact diagonalization of the semi-phenomenological Hamiltonian matrix describing the low-energy
spectrum of $^{28}$Si in the orbital space truncated to the $sd$-shell allows one to find all
energies and wave functions with conserved quantum numbers (in this example $J^{\pi}T=0^{+}0$) in the mean-field basis of non-interacting particles. The space dimension in this example is equal to ${\cal N}=839$ which is sufficiently large to extract statistical properties. The left part of Fig.~\ref{few-ldos-zele} shows nine individual strength functions $F_{k}(E-\overline{E}_{k})$ in the middle of the spectrum. On the right part of the same figure one can see that strong fluctuations are rapidly smeared by averaging over 10, 100, or 400 states. As a result, we come to the generic SF as a bell-shape function around the centroid $\overline{E}_{k}$.

The analytical evaluation of the strength function was found to be extremely difficult even for the relatively simple Wigner BRM model (\ref{2}) (for details and discussion, see Ref.~\cite{FCIC96}). Only in some limiting cases it was possible to derive an explicit expression. In order to understand how the typical shape of the strength function depends on the control parameters of the Hamiltonian $H$, we start with what is often considered as
a {\sl standard model} \cite{BM,Frazier1996}.

\subsection{Standard model of strength functions}

In this approach we single out a specific basis state $|k\rangle$ and assume that the rest of the
Hamiltonian matrix is pre-diagonalized so that we have there an intermediate basis $|\nu\rangle, \;\nu=1,...,{\cal N}-1$, with diagonal matrix elements as ``unperturbed" energies $e_{\nu}$, whereas the original state $|k\rangle$ with unperturbed energy $\overline{E}_{k}$ is coupled to the states $|\nu\rangle$ through the matrix elements
\begin{equation}
h_{k\nu}=\sum_{k'\neq k}H_{kk'}\langle k'|\nu\rangle.                      \label{17}
\end{equation}
This construction is playing here the role of a {\sl doorway} in the sense that any external perturbation acting on the state $|k\rangle$ will percolate to other states only through those coupling matrix elements. The idea of doorway going back to Bohr's compound nucleus is fruitful in many situations where the action of an external agent on a complex system can be considered as a multi-step process started at a specific (non-stationary) state; a more general consideration can be found in Ref.~\cite{AZdoorway}. In our case, all neighboring states are of a comparable degree
of complexity. It seems natural to begin with the assumption that the omission of one ``generic" state does not strongly distort the statistical properties of the dense spectrum, so that the density of states $|\nu\rangle$ is close to the genuine density of eigenstates of the total $H$.

The matrix with diagonal entries $e_{\nu},\overline{E}_{k}$ and off-diagonal elements $h_{k\nu}=h_{\nu k}^{\ast}$ can be diagonalized analytically in order to get the stationary states $|\alpha\rangle$ whose energies $E^{\alpha}$ are the roots of the dispersion equation
\begin{equation}
X_{k}(E^{\alpha})\equiv E^{\alpha}-\overline{E}_{k}-\sum_{\nu}\,\frac{|h_{k\nu}|^{2}}{E^{\alpha}-e_{\nu}}=0.
\label{18}
\end{equation}
The new roots $E^{\alpha}$, which do not depend on the choice of the excluded state $|k\rangle$, are located in between the values of $e_{\nu}$, except for a possible special {\sl ``collective"} state that can emerge at the edge of the spectrum. For a given energy $E^{\alpha}$, the squared amplitudes of the wave functions $|\alpha\rangle$ are given by the corresponding residues of the exact Green function $G(E)=(E-H)^{-1}$ in the poles (\ref{18}),
\begin{equation}
w^{\alpha}_{k}=(C^{\alpha}_{k})^{2}=\left(\,\frac{dX_{k}(E)}{dE}\,\right)^{-1}_{E=E^{\alpha}}=
\left[1+\sum_{\nu}\,
\frac{|h_{k\nu}|^{2}}{(E^{\alpha}-e_{\nu})^{2}}\,\right]^{-1}.   \label{19}
\end{equation}
The further analysis requires additional knowledge.

The standard model of strength functions is based on several critical assumptions: (i) the
background spectrum $e_{\nu}$ covers a large interval of energies that, for a state $|k\rangle$
far from the edges, can be considered infinite; (ii) this spectrum is rigid with relatively weak
fluctuations around the mean value of the level spacing, $D$, as is the case for canonical
random matrices; (iii) the coupling strengths $|h_{k\nu}|^{2}$ are uncorrelated with energies
$e_{\nu}$ and fluctuate around their typical value $\langle h^{2}\rangle$; and (iv) the spectrum is dense so that $D^{2}/\langle h^{2}\rangle<1$ which means that the residual interactions are sufficiently strong. We can note that in realistic nuclear calculations the {\sl uniformity} of the eigenstates is indeed observed: the dispersion (\ref{12}) is practically the same for all many-body states originating from the same occupation of single-particle orbitals, both in standard shell-model examples \cite{ZelevinskyRep1996} and in the modern multi-configurational approach \cite{pillet12}.

Using these assumptions, one can approximately calculate the sum in Eq.~(\ref{18}) and come
to the SF of {\sl Breit-Wigner shape},
\begin{equation}
F_{k}(E)=\,\frac{1}{2\pi}\,\frac{\Gamma_{k}}{(E-\overline{E}_{k})^{2}+\Gamma^{2}_{k}/4},    \label{20}
\end{equation}
where the {\sl spreading width} $\Gamma_{k}$ (sometimes denoted as $\Gamma^{\downarrow}$ in distinction to $\Gamma^{\uparrow}$ that describes the width with respect to the decay into continuum) does not depend, in this approximation, on the choice of the typical state $|k\rangle$ being given by the {\sl Fermi golden rule},
\begin{equation}
\Gamma=2\pi\,\frac{\langle h^{2}\rangle}{D}.                                  \label{21}
\end{equation}
In spite of the form typical for perturbation theory, the Fermi golden rule result has broader applicability. The original Wigner model (\ref{2}) assumed an equidistant spectrum of the background states which is not necessary. The conditions for the validity of the golden rule can be formulated \cite{FCIC96} as the double inequality $D\ll\Gamma\ll\Delta_{V}$, where $\Delta_{V}$ is the bandwidth of the matrix  $V$ in the energy scale, $\Delta_{V}\sim bD\equiv b/\rho_{0}$, characterizing the effective energy width of the interaction $V$. This means that the interaction is sufficiently, but not excessively, strong.

\begin{figure}[ht]
\centering
\subfigure{\includegraphics[scale=0.45]{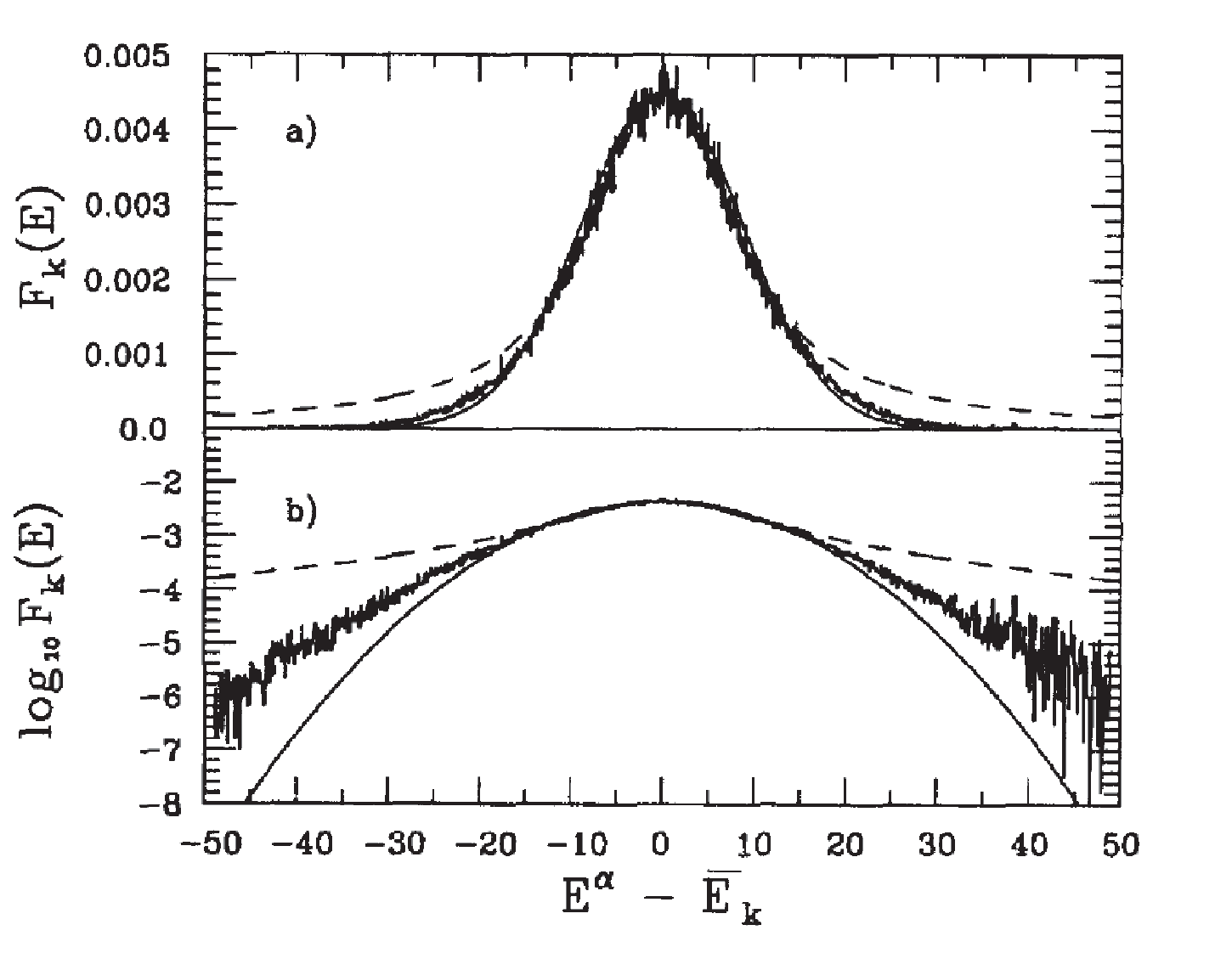}}
\caption{The Breit-Wigner fit (solid curves) and Gaussian fit (dashed lines) to the central part of the 400-state strength function (histograms), panel (a), and the same fits on the logarithmic scale, panel (b). The bin size is 100 keV (after \cite{ZelevinskyRep1996}).}
\label{ldos-zele}
\end{figure}

The result (\ref{20}) corresponds to a pure exponential decay in time, $\propto e^{-\Gamma t}$,
of the survival probability  of the initial quasistationary state $|k\rangle$ with the mean
lifetime $\tau\sim\hbar/\Gamma$. Similarly to the radioactive decay \cite{peres80}, this cannot
be exact \cite{peshkin14}, which would lead to the divergent energy dispersion (\ref{12}) (effectively, an infinite energy range of interaction). However, the golden rule might give a satisfactory approximation to the central part of the SF. Fig.~\ref{ldos-zele} shows that, in a realistic shell-model example, such an average description turns out to be reasonably good away from the remote wings.

In the case of a full random matrix standing in place of $V$, the perturbation mimics an extremely large bandwidth, $b \rightarrow \infty$. Therefore, in this case only the left part of the  inequality, $D\ll\Gamma$, comes into play, and the Breit-Wigner form of the SF emerges for any strong perturbation. In realistic models with one-body chaos, such as quantum billiards, the dependence of the width of the SF on the strength of interaction with rigid walls may be tricky \cite{Frahm97,WAV10}.

For a more generic model (\ref{2}) with a finite perturbation range (corresponding, for example, to the inter-particle interaction), the form of the SF changes with increasing $V$ and eventually approaches the standard semicircle with the radius $R=2v\sqrt{2b}$
\cite{wigner1951,wigner1955,wigner1957,wigner1958,lauritzen95,FCIC96} determined by the specific
interaction $V$ and the width $b$ of the band. As a result, one can conclude that, generically, the limiting form of the SF (when increasing the perturbation strength) is given by the {\it density of states} defined by the perturbation $V$ only. This fact was discussed when analyzing the SF for the TBRI matrices, see details in Ref.~\cite{FI97b}.

\begin{figure}[htb]
\vskip 0.4 cm
\centering
\includegraphics[width=0.5\textwidth]{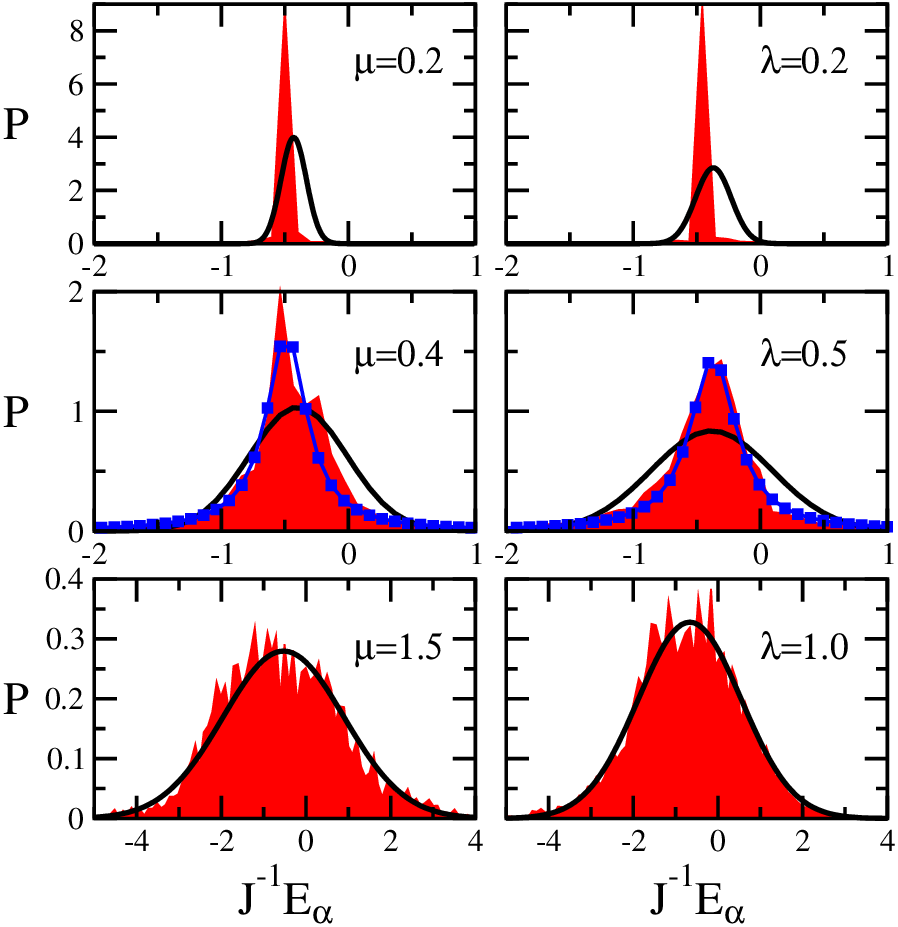}
\caption{(Color online.)  Strength functions for Model 1 (left) and Model 2 (right) for basis states away from the edges of the spectrum of the total Hamiltonian $H$ (as in Fig.~22). Numerical distribution: red shaded area; fitted Breit-Wigner shape in the middle panels: squares. Full curves are Gaussians with the variance $\sigma_k^2$ obtained from the Hamiltonian matrix through Eq.~(3.7);  $L=18$ and ${\cal S}^z_{tot}=-3$ (after \cite{Torres2014AIP}).}
\label{fig:SF}
\end{figure}

It was demonstrated that the shape of the SF in the model (\ref{3}) undergoes the transition from the Breit-Wigner to the Gaussian. Such a transition is typical for closed systems of interacting particles. For the models 1 and 2 of interacting spins, Eqs.~(\ref{5}) and (\ref{6}), this transition is perfectly manifested by the numerical data presented in Fig.~\ref{fig:SF}. Moreover, as shown in Ref.~\cite{IC06}, the change of the SF to the Gaussian corresponds to the predictions for Wigner BRM. Namely, the transition between the two different shapes, Breit-Wigner and Gaussian, occurs at  $\Gamma \sim \sigma_k$, where $\Gamma$ is the Breit-Wigner width (\ref{20}) and $\sigma_k^2$ is the variance of the strength function defined by Eq.~(\ref{13}). Note that $\Gamma$ is proportional to the square, $v^2$, of the perturbation strength, unlike $\sigma_k$ which is proportional to $v$. As will be argued below, such a change of the shape of the SF is directly related to the onset of strong chaos revealed both by the chaotic structure of eigenstates and by the statistical relaxation of an initial excitation to a steady-state distribution. Analytical attempts to derive the generic form of SF covering the whole transition from the BW to Gaussian, can be found in Refs.~\cite{FI00,Kota2001,AGK04,Chavda2004,KotaBook}.

%

The ``standard" model approximately works also for collective states, as giant resonances in nuclei which indeed play a role of doorway states in the processes of excitation by an external field of
corresponding symmetry, or for any simple mode coupled with the background of complicated states through weakly fluctuating matrix elements. It can be generalized for the practically important case when the states under consideration do in fact belong to the continuum \cite{SZ97}; then the interrelation between the characteristic spreading time and decay time of the background states, or between the widths $\Gamma^{\uparrow}$ and $\Gamma^{\downarrow}$, plays the main role. This leads to interesting phenomena observed in the experiment with an increase of energy: restoration of purity of the collective state and, on the other hand, loss of collectivity when the excitation
proceeds through the background states which decay before feeding the collective state. Here the main role is played by the competition between the characteristic time of fragmentation of a collective resonance into background states of a complicated nature and the continuum decay of those states. This physics is outside of the scope of this review.

\subsection{Beyond the standard model}

There are various effects which restrict the applicability of the golden rule (\ref{21}). In the limit of very strong coupling, a {\sl doubling phase transition} is expected that can be seen already from the equations of the preceding subsection. The Breit-Wigner shape corresponds to a  spreading width much smaller than the region $a\sim {\cal N}D$ covered by the background states. In the opposite limit, $\Gamma\gg a$, the secular equation (\ref{18}) predicts \cite{Frazier1996} that the strength is accumulated in two peaks on both sides of the unperturbed centroid,
\begin{equation}
E^{\alpha}\approx \overline{E}_{k} \pm \left[\sum_{\nu}h_{k\nu}^{2}\right]^{1/2}.    \label{22}
\end{equation}
One can note that this limit determines the extremal spreading width $\Gamma\approx 2\sigma$ in terms of the dispersion (\ref{13}) of the original doorway state. This effect, also known in quantum optics, was observed in the Mott metal-insulator phase transition \cite{zhang93} as
transfer of the quasiparticle strength to the so-called Hubbard bands. In nuclear physics it
appears, in the strong coupling limit, in the spreading widths of giant resonances \cite{lewenkopf94} and of highly-excited single-particle states interacting with many phonon
modes \cite{stoyanov04}.

The physical mechanism of the doubling transition can be understood by analogy with the formation
of collective modes usually described by the random phase approximation. The background states
interact among themselves through the selected state $|k\rangle$. If this interaction is sufficiently strong a collective state is formed as a coherent combination of the background states. Having the same quantum numbers, the original state and the newborn collective mode repel each other and form the two peaks predicted in (\ref{22}) which concentrate the significant fraction of the total strength [in the limit (\ref{22}) the whole strength is evenly divided between the two peaks]. If the coherent interactions responsible for the excitation of collective modes are absent and the system is close to chaoticity, the corresponding limit of the SF will be a semicircle with radius $R\approx\sigma_k$, where $\sigma_k^2$ is again the average second moment (\ref{12}) and the effective spreading width is $\Gamma=2\sqrt{3}\,\sigma_k$ \cite{ZelevinskyRep1996,lauritzen95}.

As it usually happens, realistic atoms and nuclei are typically between the two limits, and the shape of the SF evolves from the Breit-Wigner behavior at the center to the Gaussian behavior with faster decreasing wings. In many practical examples, the spreading width is not proportional to the square of the coupling matrix elements as in Eq.~(\ref{21}) but it is rather a quantity of the order of $\sigma$, i.e. depends {\sl linearly} on the magnitude of the interaction. Being a product (\ref{15}) of the level density and the average weight of simple components, the SF in the Gaussian regime is rather stable when the interaction is broadening the distribution keeping more or
less intact the average shape. In application to banded random matrices, various limits of the spreading width were studied in Refs.~\cite{FLW91,LF93}.

The actual manifestations of Breit-Wigner and Gaussian strength functions are seen in the spreading widths of multiple giant resonances (overtones) observed in  nuclei \cite{bertulani94,lewenkopf94}. Such collective excitations can be treated as two- and three-phonon vibrational levels. In the harmonic approximation reasonable for the qualitative description, the spreading width $\Gamma_{2}$ of the double excitation is given by a convolution of two single-phonon strength functions with widths $\Gamma_{1}$. In the Breit-Wigner limit this would lead to $\Gamma_{2}\approx 2\Gamma_{1}$, while for the Gaussian shape $\Gamma_{2}\approx \sqrt{2}\,\Gamma_{1}$, in better agreement with experiment. In such considerations it is implicitly assumed that the phonon spreading width does not significantly change with excitation energy: the density of background states grows with the excitation energy while the coupling matrix elements are quenched. For the giant resonances, as well as for isobaric analog resonances \cite{harney86}, this is a reasonable assumption based on the so-called {\sl $N$-scaling} \cite{blin73,FS82}.

Consider a transition between a ``simple" state $|k\rangle$ and a ``chaotic" state $|\alpha\rangle$ by
a ``simple" operator $\hat{Q}$, let say of one- or two-body nature,
\begin{equation}
\langle k|\hat{Q}|\alpha\rangle=\sum_{k'}Q_{kk'}C^{\alpha}_{k'}.                     \label{23}
\end{equation}
Simple operators can change orbits of only one or two particles and, due to {\sl selection rules},
the number of non-zero matrix elements $Q_{kk'}$ is small, let say $\nu_{Q}$. For typical mixture amplitudes $C^{\alpha}_{k}$ of the order $1/\sqrt{N}$, where $N$ is the number of the principal components of the state $|\alpha\rangle$, and for typical single-particle matrix elements $q$, we come to the estimate of suppression,
\begin{equation}
\langle k|\hat{Q}|\alpha\rangle \approx\,\frac{q\nu_{Q}}{\sqrt{N}}.                   \label{24}
\end{equation}
Similarly, we can roughly estimate the mixing between two chaotic states,
\begin{equation}
\langle\beta|\hat{Q}|\alpha\rangle=\sum_{kk'}C^{\beta\ast}_{k}Q_{kk'}C^{\alpha}_{k'}.       \label{25}
\end{equation}
In the same spirit, taking into account the random character of the mixing coefficients and assuming the same degree of complexity $N$ of the states $|\alpha\rangle$ and $|\beta\rangle$ we find
\begin{equation}
\langle \beta|\hat{Q}|\alpha\rangle\approx \sqrt{N} \,\frac{1}{\sqrt{N}}\,q\nu_{Q}\,\frac{1}{\sqrt{N}}
= \,\frac{q\nu_{Q}}{\sqrt{N}},                                                           \label{26}
\end{equation}
that is {\sl the same suppression} as in the previous case (\ref{24}). This way of arguing is based mainly on the assumption of the {\sl uniform degree of complexity} of mixed chaotic states, in the same energy region they ``{\sl look the same"} from the viewpoint of any ``{\sl simple}" operator \cite{percival73}. This is, as we will stress later, the root to understanding the thermalization in an isolated system.

The existence of $N$-scaling leads to important trends of {\sl chaotic enhancement} of perturbations. One of the spectacular effects is related to weak interactions which are known to violate parity conservation. Large manifestations of parity violation were observed in scattering of longitudinally polarized slow neutrons off various nuclei. Usual estimates of this effect as
coming from the mixing of single-particle levels of opposite parity predict the relative difference of total cross sections for left and right neutron helicities on the level of $10^{-7}\div 10^{-8}$. The actual effect is in many cases much greater, up to 10\%, see review articles
\cite{bowman93,frankle93,FGrib95,mitchell99}. As mentioned earlier, the narrow neutron resonances at very low energy correspond to long-lived compound states of a neutron captured by the target nucleus. The long lifetime of such a resonance allows the intrinsic equilibration (its relation to thermalization will be discussed later). The presence of many, $N\gg 1$, incoherent simple components in the wave function of a compound nucleus is a signature of quantum chaos. In heavy nuclei at neutron threshold energy, $N$ reaches $10^{6}$. According to Eq.~(\ref{26}), the matrix elements of the weak interaction mix the complicated states of opposite parity, in this case $s$- and $p$- neutron resonances, are suppressed by a factor $1/\sqrt{N}$. However, in this energy region the combinatorial level density is increased proportionally to the same factor $N$. Then the golden rule estimate (\ref{21}) shows the {\sl enhancement} of parity non-conservation by $\sqrt{N}$, of the order $10^{3}$. An additional factor of the same order comes from the kinematic gain in the neutron width for mixed resonances, $\Gamma^{(n)}_{s}/\Gamma^{(n)}_{p}\sim 10^{3}$.

Another related effect is the enhancement of the asymmetry of fission products with respect to
the neutron spin direction in fission of heavy nuclei induced by slow polarized neutrons; here there is no kinematic factor but the chaotic mechanism again produces the enhancement of the order
$\sqrt{N}\sim 10^{3}-10^{4}$. As follows from the experiment, the asymmetry is the same in numerous decay channels which differ in the mass distribution of fragments and kinetic energy distribution
\cite{kotzle00}. This independence confirms that the enhancement is generated in the compound
nucleus, prior to the choice of a certain decay channel by a fissioning nucleus. For more details concerning the violation of fundamental symmetries and chaotic enhancement we refer the reader to review articles \cite{FGrib95,mitchell99,AZ08}. There are also numerous examples of chaotic enhancement in tunneling or mixing of very different configurations in many-body systems
that occur at close energies, for example different shapes \cite{shapeenh} or long-lived isomers
\cite{AZ14}.

One can notice that in the regime of $N$-scaling the numerator and the denominator of the golden
rule (\ref{21}) have approximately the same $1/N$ dependence. This is equivalent to the {\sl saturation} of the spreading width when the system reaches the chaotic limit. {\sl Isobaric analog resonances} (IAR) in nuclei are the excited states, frequently in the continuum, which belong to the  same isospin multiplet as the ground state of the neighboring nucleus \cite{auerbachIAR}. A vast set of data accumulated on the IAR \cite{harney86} shows that their typical spreading width is indeed of the same order of magnitude in different nuclei, for very different spins, isospins and excitation energy.

Finally, the remote wings of the SF are probably neither Breit-Wigner nor Gaussian. Vast numerical material stored in the experience with large-scale shell-model diagonalization demonstrated \cite{Frazier1996,ZelevinskyRep1996} that the characteristic behavior at the tails in such cases is exponential. The exponential regime starts at a distance larger than $2\sigma$ from the centroid. This fact, analyzed and mathematically justified in \cite{HVZ99}, was laid in the foundation of the practical method of {\sl exponential convergence} \cite{HBZ02,HBZ03,HKZ03}: a regular progressive truncation of large Hamiltonian matrices ordered in a certain way is performed starting with few steps allowing full numerical diagonalization; after that the exponential regime sets in and the exact energy levels of low-lying states, first of all of the ground state, can be obtained by straightforward exponential continuation in a function of the dimension. The steady pressure of many small admixtures of highly excited chaotic states gradually included by the increase of the dimension  brings the low-lying states down. Based on ideas of quantum chaos, this approach works without exceptions. The exact knowledge of the ground state energy is necessary for astrophysical reactions, where the level density is a vital ingredient of all calculations. Recently, on this basis, a successful algorithm for calculating spin- and parity-dependent level density was constructed \cite{SHZ11,SHZ12}.

\section{Chaotic Eigenstates}

\subsection{Qualitative discussion and a little of history}

The problem of quantum chaos was initially referred to one-body quantum systems, fully deterministic but with strong chaos in the classical limit. As was discovered numerically, the properties of such quantum models as the kicked rotor \cite{CCFI79,CIS81,Izrailev90} and fully chaotic
billiards \cite{MK79,casati80,Brody81,Bohigas84} strongly depend on whether  the motion is regular or chaotic in the corresponding classical counterparts. It was understood that, unlike classical chaos that is due to the local instability of motion, in quantum chaotic systems the properties of spectra and eigenfunctions have to be compared with those described by full random matrices. It was argued  \cite{BT77,CCG85} that for integrable systems the nearest level spacing distribution $P(s)$ is generically quite close to the Poisson distribution emerging as a result of absence of correlations between eigenvalues (see also discussion and references in Ref.~ \cite{Izrailev90}).

For a long time the numerical check whether the form of $P(s)$ is close to the Wigner-Dyson distribution has served as the main tool for the characterization of quantum chaos. On the other hand, it was also numerically observed \cite{MK79} that the eigenfunctions of the stadium billiard have a quite complicated structure in the position representation. These results have led to the conjecture that the eigenstates of chaotic billiards may be compared to plane waves with random amplitudes \cite{B77}. Later, the complex structures of eigenstates were confirmed for many autonomous systems, as well as for time-dependent systems with external periodic perturbations (for references, see, for example, \cite{Izrailev90}).

Initially, the study of quantum chaos was restricted to models of a single particle interacting with external fields. As a result of extensive studies, currently the theory of {\it one-body chaos} is developed in great details (see for example, \cite{G90,HaakeBook,StockmannBook,ReichlBook} and references therein). On the other hand, many problems of {\it many-body chaos} occurring in quantum systems of interacting particles are not resolved yet. Recently, the burst of interest in many-body chaos has been triggered by the remarkable progress in experimental studies of trapped systems of bosons and fermions \cite{MetcalfBook} and large-scale exact diagonalization of Hamiltonian matrices for systems of interacting particles.

It is understood now that the level spacing distribution $P(s)$, although serving as a common test for distinguishing between integrable and non-integrable models, is not effective in application to many-body chaos. First, this quantity that requires precise knowledge of relatively long consecutive series of energy levels of fixed symmetry often is far from being experimentally accessible. Second,  the level spacing statistics is typically a weak characteristic of quantum chaos that appears already in the first stages of the process of chaotization. Third, the presence or absence of the Wigner-Dyson level spacing distribution
cannot be a necessary condition for classical chaos: for instance, a transition from the Poisson to Wigner-Dyson distribution was found in Ref.~\cite{Fausto} in the energy spectrum of the Bunimovich billiard, which is known to be fully chaotic. Finally, in many realistic systems the behavior of various observables is not directly related to the spectral statistics and continues to evolve with the strength of the perturbation after the function $P(s)$ has been stabilized.

In this situation the knowledge of the structure of the eigenstates turns out to be decisive in understanding regular or chaotic properties of realistic systems. In what follows, we define {\it quantum chaos} in terms of the (chaotic) structure of eigenstates, rather than in terms of the level statistics. This idea of shifting the definition of chaos onto {\it individual} eigenstates, rather than just on the spectral statistics, has been exploited in Ref.~\cite{SG84}. Specifically, it was suggested to define quantum chaos occurring in an individual eigenstate by the vanishing correlation function between the eigenstate components in an appropriate basis.

\begin{figure}[ht]
\centering
\subfigure{\includegraphics[scale=0.6]{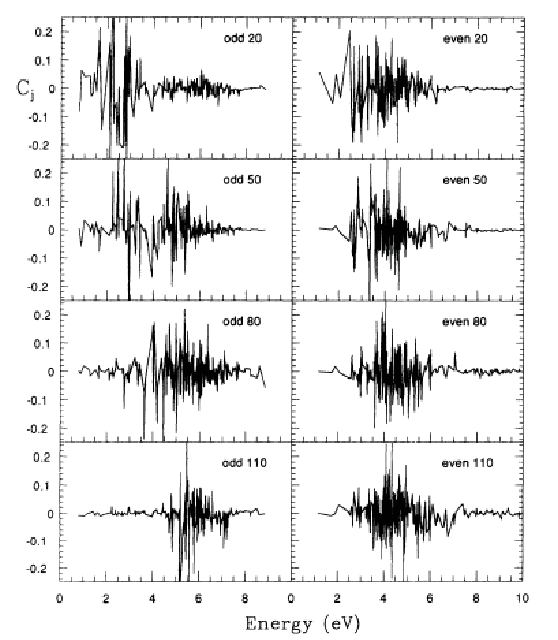}}
\subfigure{\includegraphics[scale=0.8]{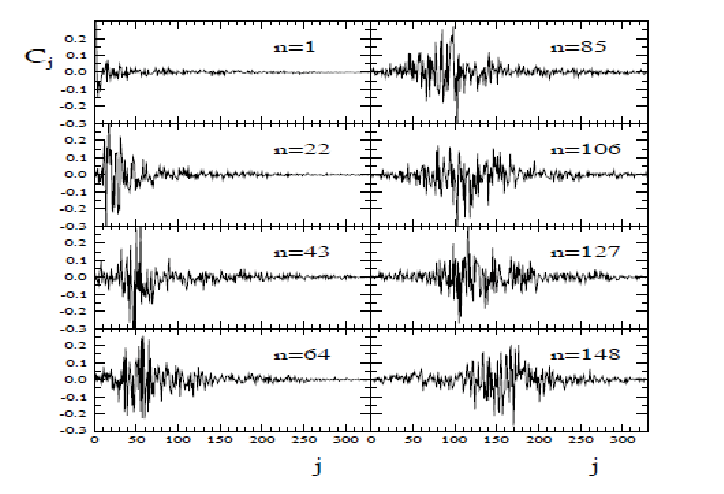}}
\caption{LEFT: The odd, $J^{\pi}=4^{-}$, and even, $J^{\pi}=4^{+}$, excited eigenstates of the cerium atom. Shown are the components $C_j$ of eigenstates in terms of basis (mean field) states, given in the energy representation due to one-to-one relation between $|j \rangle$ and $E_j$. RIGHT: Individual eigenstates $C_j$ in the basis of $H_0$ of the model (\ref{3}) with random two-body interaction for 4 particles and 11 single-particles levels with the perturbation strength $V_0=0.12$ and $d_0=1.0$. Here $n$ stands for the level number with $n=1$ as the ground state (after \cite{FGGK94}).}
\label{EF-examples}
\end{figure}

In Ref.~\cite{c85} the emergence of chaotic eigenstates was found when analyzing experimental data for the rare-earth cerium atom. It was shown that excited eigenstates of four valence electrons with the total angular momentum and parity $J^\Pi = 1^{+}$ are random superpositions of a number of basic states. Although this number was found to be relatively small as compared with chaotic eigenstates in heavy nuclei \cite{BM}, one can speak about chaotic atomic states. Later on, intensive analytical and numerical studies \cite{FGGK94} have confirmed the onset of chaos in both eigenstates and spectrum of the cerium atom. With the use of the relativistic configuration-interaction method it was shown that the structure of eigenstates of odd and even levels of this atom with angular momentum $J=4$ above 1 eV excitation energy becomes similar to that of compound states in heavy nuclei. It was found that the atomic stationary states are random superpositions of about $N_{{\rm p.c.}} \sim 100$ components built of the $4f,\,6s,\,5d$, and $6p$ single-electron orbitals. Thus, even four interacting electrons in the mean field of an inert core create chaotic eigenstates, see few examples in Fig.~\ref{EF-examples}. In this figure the examples of excited eigenstates from the TBRI matrices for four fermions and eleven single-particle orbitals are shown.  One can see that the eigenstates of the random matrix model (\ref{3}) are qualitatively the same as those of the real atom with no random parameters. These data demonstrate that the TBRI model can effectively describe generic properties of realistic physical systems which are completely deterministic.

Here and below we discuss the structure of the {\it many-body} states presented, as in Eq.~(\ref{8}),  either in the chosen basis of the mean-field $H_0$ or in the energy representation corresponding to this field. Being considered as {\it chaotic}, these eigenstates do not occupy the whole many-body basis of $H_0$ because of the finite range of the interaction. This is in contrast with one-body  chaos where chaotic single-particle eigenstates typically cover the whole basis of $H_0$. Such a situation occurs, for example, for chaotic billiards where the interaction with the boundaries couples all basis states and, therefore, it can be considered as infinitely strong. Thus, the assumption of fully chaotic and extended eigenstates as random superpositions of plane waves \cite{B77}, often used as a justification of quantum chaos, may be not valid for isolated systems of interacting particles. As shown below, instead, one has to consider the emergence of chaotic eigenstates in the {\it energy shell} defined by the projection of the ``unperturbed" Hamiltonian $H_0$ onto the total Hamiltonian $H$.

Let us consider, in a more general context, stationary quantum states of an isolated chaotic system in the region of sufficiently high excitation energy and level density. As we argued above, such states in a narrow energy window and with the same values of exact constants of motion ``look the same" \cite{percival73}. The corresponding wave functions are exceedingly complicated in any ``simple" basis, for example in the mean field representation that is natural for a many-body system separating in the best way regular dynamical features from incoherent collision-like interactions responsible for the onset of chaos. Any measurement of a macroscopic variable has a typical duration $\Delta t$ and covers an energy interval $\Delta E\sim\hbar/\Delta t$. In this interval we have a large number of orthogonal stationary states. The point is that due to their complete mixing (in an appropriate basis of the macroscopic devices) and random phases of the components of the wave functions, the results of such a measurement in practice do not depend on the instantaneous, in reality random, phase relationships between these numerous components.
Otherwise, the whole idea of thermodynamical equilibrium would not work.

The description in terms of chaotic wave functions in fact agrees with the argumentation used in the foundations of equilibrium statistical mechanics. Similar properties of equilibrated systems are stressed in  {\sl Statistical Physics} by Landau and Lifshitz \cite{LL58}: ``{\sl It may again be mentioned that, according to the fundamental principles of statistical physics, the result of the averaging is independent of whether it is done mechanically over the exact wave function of the stationary state of the system or statistically by means of the Gibbs distribution. The only difference is that in the former case the result is expressed in terms of the energy of the body, and in the latter case as a function of its temperature}". Close ideas were put in the foundation of statistical mechanics by Krylov \cite{krylov72} who used the term {\sl mixing} for the process of equilibration. Each generic wave function of a complicated system has essentially the same macroscopic properties as a thermal ensemble for the same values of global conserved quantities. As we understand now, the mechanism responsible for the mixing is provided by many-body quantum chaos. Even the pioneering idea by Boltzmann was based on the ``molecular chaos" that is required to justify the kinetic approach to the equilibrium. Similar ideas form the basis for the practical truncation of the formally exact infinite sequence of many-body correlational functions (Bogoliubov-Born-Green-Kirkwood-Yvon hierarchy \cite{BBGKY}). In a modern language, one needs to have phase incoherence of the components of the initial state. The decoherence can come from the surrounding or, even in a closed system, from the presence of many degrees of freedom coupled through their interaction. The interaction will create such complicated superpositions of simple states that the advent of quantum chaos is unavoidable.

From this viewpoint, the concept \cite{bohr36} of the compound nucleus has the same spirit; the famous photograph of Bohr's wooden model with an incoming ``neutron" shows that the equilibration comes from strong intrinsic interactions. If the lifetime of an open system is of the order of or greater than the Weisskopf time $\hbar/D$ (with $D$ as the mean level spacing) of the typical periodic motion inside the system, the intrinsic degrees of freedom in a nearly closed system mix, creating compound states which practically serve as representatives of quantum chaos similarly to the predictions of the canonical Gaussian ensemble. Early studies of many-body quantum chaos mainly used the so-called ``{\sl nuclear data ensemble}" of neutron resonance data \cite{haq,bohigas88}. Later a broad material from large-scale atomic and nuclear many-body calculations \cite{FGGK94,Horoi1995,ZelevinskyRep1996,WMRMP} was added to the discussion that gave an impulse to return \cite{ZelevinskyRep1996,ann} to the problem of interrelation between quantum chaos and thermal equilibrium.

Coming on a new level to the above statement of Landau and Lifshitz, we can claim that properties
of individual chaotic eigenstates are essentially equivalent to those of the equilibrium thermodynamic ensemble \cite{Peres1984,Deutsch1991,Srednicki1994,Horoi1995,FI97b,ZelevinskyRep1996,ann,Rigol2008}.
In the recent literature, this idea sometimes is called the {\sl eigenstate thermalization hypothesis}. Below we shall discuss some of the problems related to the time-dependent equilibration process \cite{Polkovnikov2011RMP}; here, we have to stress that the actual problem is to understand the conditions under which the statistical approach works (i.e. the conditions for the onset of chaotic states), rather than the statement of Landau and Lifshitz itself.

A prototypical system of interacting constituents is the gas of hard spheres, where the classical
chaoticity was rigorously established \cite{sinai70}. The exact results were obtained since the interaction was actually reduced to the excluded volume of the spheres. The quantum analog of the hard sphere gas was considered in \cite{Srednicki1994} where it was shown that a chaotic initial state leads to the statistical equilibrium corresponding to the type of statistics (bosons, fermions or distinguishable particles). Similar results, not limited by the gas of hard spheres,
were derived earlier by Van Hove \cite{vanHove1959a,vanHove1959b,vanHove1959c}. In general, the main properties of the equilibrated system do not depend on the details of the initial state; as in statistical thermodynamics, they are defined by the constants of motion while the time required for equilibration is determined by the strength of the interactions.

Below we present the results of the detailed analysis of exactly diagonalizable matrix models which describe realistic systems of interacting quantum particles. The essential difference with the standard thermodynamics formulated in the textbooks for large systems ({\sl thermodynamic limit}) is that here we consider {\sl finite} systems without an external heat bath. The equilibration occurs as a result of internal interactions leading to many-body quantum chaos. Therefore, one can
introduce (not uniquely) analogs of usual thermodynamic characteristics $-$ entropy, temperature, etc. $-$ which however may depend on the type of the measuring device  ({\sl thermometer}).

\subsection{Information entropy}

The simplest characteristic of a chaotic state $|\alpha\rangle$ related to the statistical
interpretation is {\sl the information (Shannon) entropy} $S^{\alpha}$ defined in terms of the components $C^{\alpha}_{k}$, see Eq.~ (\ref{8}), of this state with respect to a certain basis $|k\rangle$,
\begin{equation}
I^{\alpha}=- \sum_{k}w^{\alpha}_{k}\ln(w^{\alpha}_{k}),
\quad w^{\alpha}_{k}\equiv |C^{\alpha}_{k}|^{2}.
                                                                                    \label{27}
\end{equation}
This quantity can be found for any individual stationary state \cite{Horoi1995}; it can change from zero, when the selected representation basis $|k\rangle$ coincides with that of stationary states of the Hamiltonian and $w^{\alpha}_{k}=\delta^{\alpha}_{k}$, to the maximum value of $\ln {\cal N}$ for a ``microcanonical" state when all basis states are represented with equal probability,
$w_{k}^{\alpha}=1/{\cal N}$, where ${\cal N}$ is the total dimension of the Hilbert space.

This basis dependence can cast doubt on the significance of the information entropy as a physical
indicator of the complexity of the state. On the other hand, here we express the interrelation
between two bases and this aspect carries an important physical knowledge. In practice, selecting the basis $|k\rangle$ in the many-body problem as that of the mean field in a system like a complex atom, nucleus, molecule or nanostructure, we separate the regular dynamics from the chaotic features and the information entropy is an additional useful tool for quantifying the {\sl relative} complexity \cite{Horoi1995}. In condensed matter applications, especially in the presence of disorder, the appropriate representation basis can be that of lattice sites when the information entropy describes the degree of localization of stationary electron states. Alternatively, the Bloch wave basis can be more convenient, in particular for problems of electron-phonon interactions or quantum signal transmission. As a characteristic of information, the same definition (\ref{27}) can be applied to any, not necessary stationary, quantum state $|\alpha\rangle$. The process of quantum evolution of a generic state will show irreversible equilibration with growth of entropy. An originally excited state $|k\rangle$ has zero entropy in the original basis, but the interactions will lead to the fragmentation of this state corresponding to a simple analog of the $H$-theorem or the second law of thermodynamics.

In the Gaussian random matrix ensembles all eigenstates have a similar degree of complexity. In the Gaussian orthogonal ensemble (GOE) of large dimension, ${\cal N}\gg 1$, the real components $C^{\alpha}$ of a generic eigenstate have a Gaussian distribution,
\begin{equation}
P^{\alpha}(C^{\alpha})=\sqrt{\,\frac{{\cal N}}{2\pi}\,}\exp\left(-\,\frac{{\cal N}}{2}(C^{\alpha})^{2}\right), \label{28}
\end{equation}
with the variance $\overline{(C^{\alpha})^{2}}=1/{\cal N}$, compare Eq. (\ref{24}). The information entropy of such a state is \cite{Izrailev90}
\begin{equation}
\overline{S}=\ln(0.482 {\cal N})+{\cal O}(1/{\cal N}).                                        \label{29}
\end{equation}
For a realistic system, this value serves as an upper limit of complexity.

The spectral properties typically reveal a {\sl secular behavior} as a function of the excitation energy so that it makes sense to introduce a {\sl local} characteristic of complexity for the states $|\alpha\rangle$ in a given spectral interval. Such an interval is characterized by the local number of significant components, $N^{\alpha}$, that determines, according to Eq.~(\ref{29}), the local information entropy $\overline{S^{\alpha}}$ for the chaotic eigenstates. Instead of $S^{\alpha}$ one can use the corresponding {\sl localization length} in the Hilbert space,
\begin{equation}
S^{\alpha}_{I}=\exp(S^{\alpha}),                                        \label{30}
\end{equation}
that defines the local number of principal components as $l^{\alpha}_{S}=0.482 N^{\alpha}$.

\begin{figure}[ht]
\centering
\includegraphics[width=0.6\textwidth]{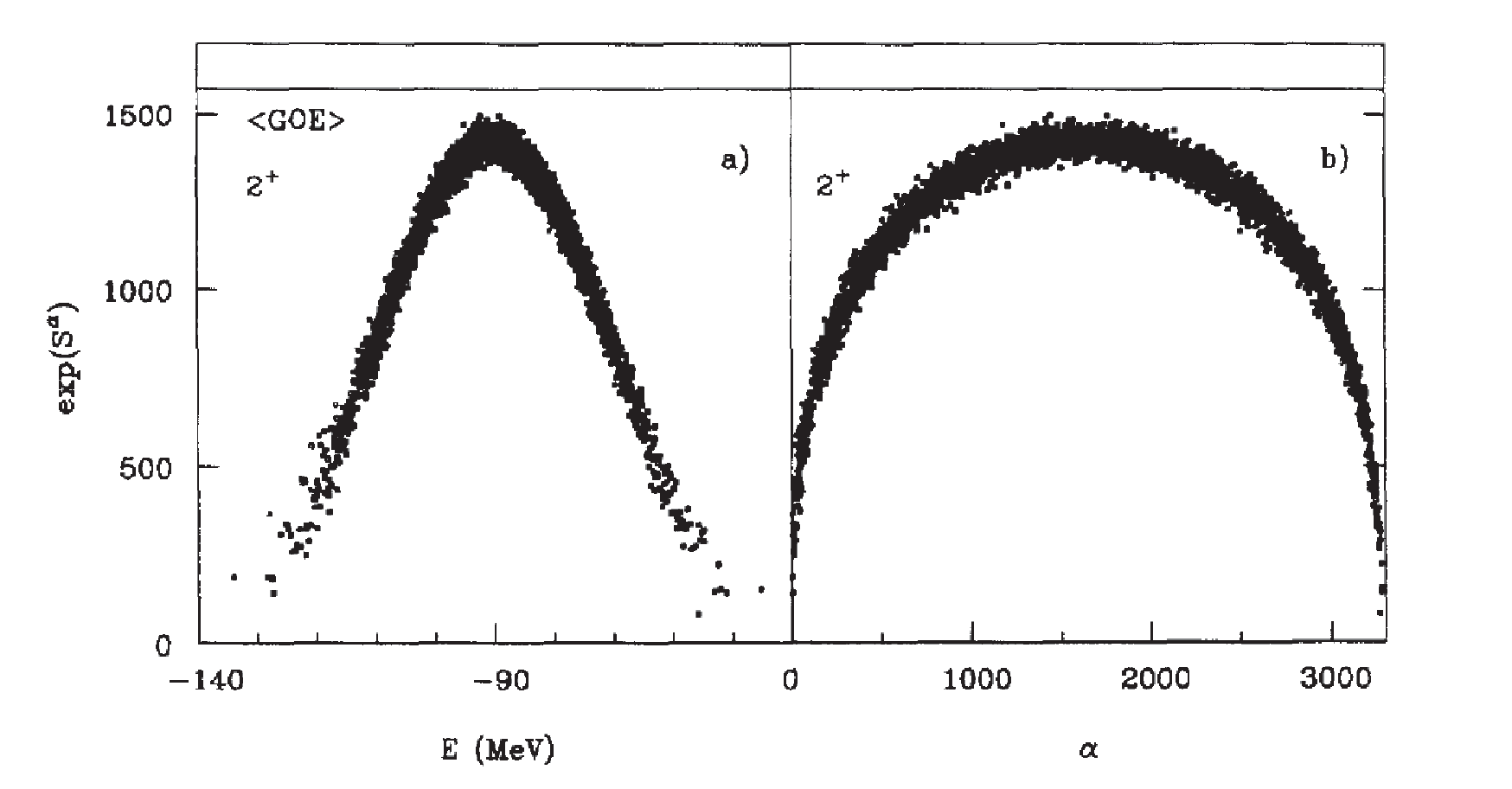}
\caption{Localization length, Eq. (\ref{30}), of all 3276 $J^{\pi}T=2^{+}0$ states in the $sd$-shell model \cite{ZelevinskyRep1996} for $^{28}$Si. Left panel uses the actual scale of excitation energy; in the right part the states are uniformly listed in order of increasing energy. The horizontal line shows the GOE limit, Eq. (\ref{29}), equal to 1578 (after \cite{ZelevinskyRep1996}). }
\label{LocL-zele}
\end{figure}

Fig.~\ref{LocL-zele} shows the localization length (\ref{30}) for all ${\cal N}=3276$ states with the same
quantum numbers of spin, parity and isospin, $J^{\Pi}T=2^{+}0$, found in the exact shell-model diagonalization for the $^{28}$Si nucleus. The model contains an inert core of $^{16}$O and twelve valence nucleons moving in the $sd$-shell in the mean field of the core and interacting through the full Hamiltonian consisting of all 63 independent two-body matrix elements allowed by the conservation laws of angular momentum and isospin in this orbital space. The strengths of the matrix elements were earlier fit by shell-model practitioners using the data for low-lying states individually resolved in the experiment; some of the matrix elements cannot be defined from the data so that their values were assigned more or less arbitrarily within reasonable limits. The results for all states are presented in two equivalent scales, as a function of excitation energy and in the unfolded form for levels uniformly ordered by increase of energy. The shape of the distribution, being similar to that of the level density \cite{ZelevinskyRep1996}, is typical for a system with finite Hilbert space (of course, for realistic nuclear physics, only the part of the left branch can be juxtaposed to data because of the space truncation). In the mean field basis, the stationary states rapidly become more and more complicated (delocalized in the Hilbert space) as the excitation energy grows; the maximum of the distribution is already close to the GOE value (\ref{29}) for $N^{\alpha}={\cal N}$. For our purpose here, the most important feature is the fact that the information entropy or the localization length are smooth functions of the excitation energy, with rather small fluctuations. In this sense the characteristics of individual states become {\sl thermodynamic} quantities.

\begin{figure}[ht]
\centering
\includegraphics[width=0.6\textwidth]{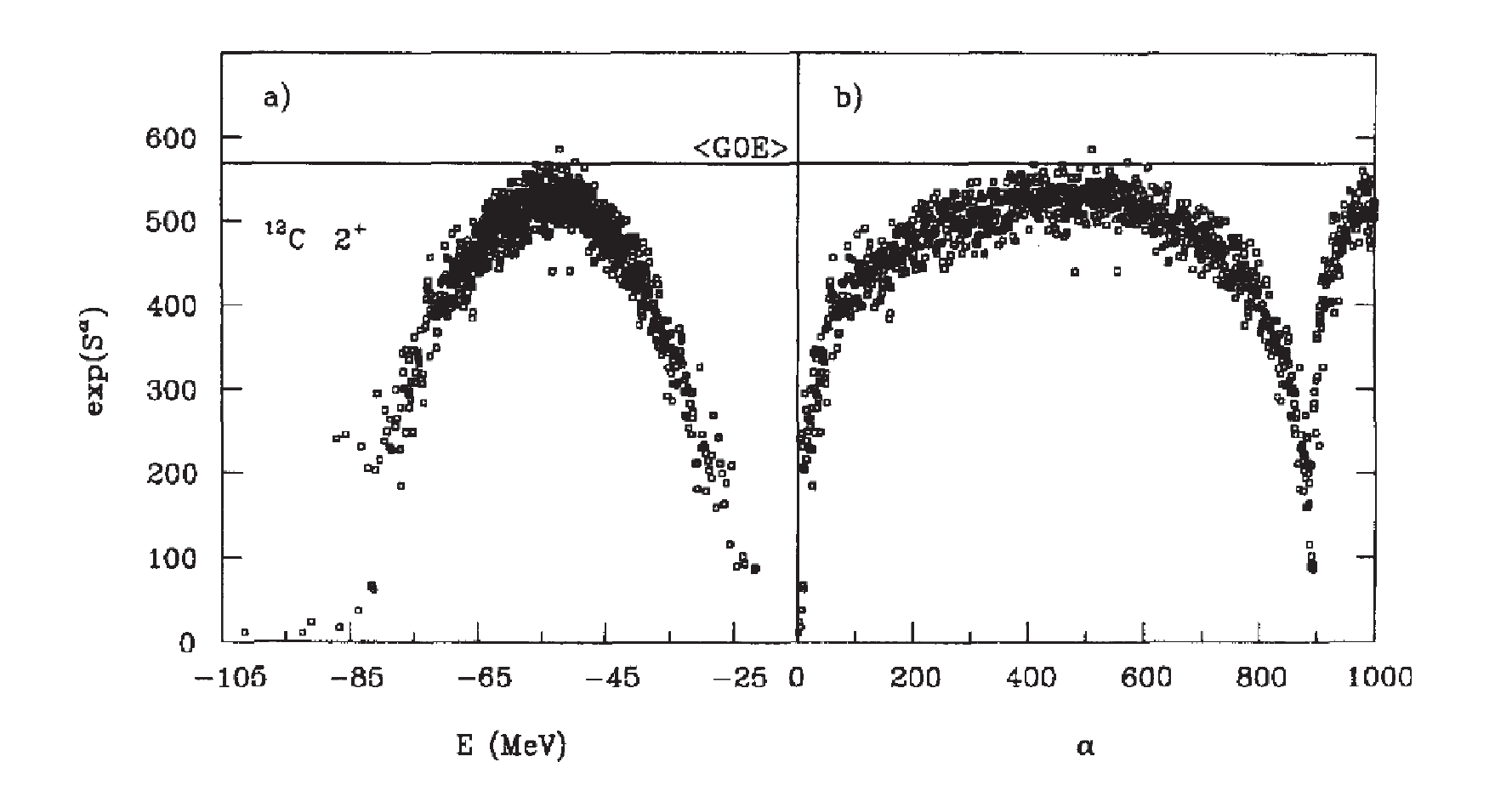}
\caption{Localization length, eq. (\ref{30}), of 1183 $J^{\pi}T=2^{+}0$ states in the four major oscillator shell model \cite{ZelevinskyRep1996} for $^{12}$C. The right branch corresponds to the spurious states with excitation of center-of-mass motion separated by the energy shift (after \cite{ZelevinskyRep1996}).}
\label{LocL-zele-1}
\end{figure}

The direct physical meaning of information entropy is emphasized by a consideration of Fig.~\ref{LocL-zele-1} where this quantity is calculated for a different type of the shell model. The carbon isotope $^{12}$C is analyzed in a bigger space of four major shells. With enriched interaction, the information entropy of the states in the centroid of the distribution already reaches the GOE limit. However, the shell model with cross-shell transitions includes unphysical excitations of the center-of-mass. These spurious states have to be eliminated, in particular for determining the physical level density used in the theory of nuclear reactions, especially in astrophysics. There are special methods of statistical spectroscopy for doing this \cite{HZ07,SHZ11,SHZ12} based on the correct counting of possible center-of-mass excitations
for a given scheme of active orbitals. In the shell model, alternatively, the separation of spurious states might be done by adding to the actual Hamiltonian special terms of global kinetic energy with a large positive coefficient; then spurious states are artificially pushed up in energy from real excited states. Fig.~\ref{LocL-zele-1} in the ordered energy scale clearly shows these states as a separate branch with a similar behavior of the information length. This means that the information entropy (or the length) is smart enough to discriminate the states of comparable complexity but different physical nature.

\subsection{Moments of the distribution of components}

A role similar to the information entropy or information length can be played by other structural moments of stationary wave functions,
\begin{equation}
M_{n}^{\alpha}=\sum_{k}(w^{\alpha}_{k})^{n}; \quad M_{1}^{\alpha}=1.                            \label{31}
\end{equation}
For a local Gaussian distribution of the components this leads to
\begin{equation}
M_{n}^{\alpha}=\,\frac{(2n-1)!!}{(N^{\alpha})^{n-1}}.                                        \label{32}
\end{equation}
The quantity $M^{\alpha}_{2}$ determines the so-called {\sl inverse participation ratio}. This allows for a complementary definition of the local {\sl number of principal components} through the average square of the probability, $N_{{\rm pc}}^{\alpha}=(M^{\alpha}_{2})^{-1}=N^{\alpha}/3$.

From those definitions we see that the information entropy and the number of principal components depend on complementary characteristics of the structure of the eigenstates. Namely, the information entropy is more sensitive to small components of the wave function, while the inverse participation ratio emphasizes large components. However, if we can speak of a universal distribution of the components, those different characteristics have to be interrelated. For the ``microcanonical" wave functions, both definitions give the same value, $l_{S}=1/M_{2}={\cal N}$. In the case of the Gaussian distribution a universal value of the ratio emerges,
\begin{equation}
\frac{l_{S}^{\alpha}}{N_{{\rm pc}}^{\alpha}}=1.44.                                            \label{33}
\end{equation}
As a matter of fact,
in many atomic and nuclear examples of shell-model calculations, the ratio is close to this number for the states far from the edges of the spectrum.

The moments (\ref{32}), similarly to the information entropy, are not basis-invariant (except for the normalization $M_{1}$). We can repeat the arguments given above that the representation dependence can provide an additional information concerning the nature of stationary states and the dynamics in the system; examples can be found in Ref.~\cite{ZelevinskyRep1996}. It was also shown there that it is possible to study the behavior of complexity as a function of the strength of residual interaction with respect to the characteristic scales for non-interacting particles in the mean field. In degenerate models, for example if the mean field is taken as that of an isotropic harmonic oscillator, the residual interaction is effectively very strong, so that the majority of eigenstates even at relatively low excitation energy reveal the complexity on the GOE limit. A too strong interaction eliminates the process of chaotization bringing immediately the eigenstates into the chaotic regime. One can say that under such conditions the ``thermometer" using such measures of complexity is not working properly $-$ the system is ``too hot" to start with, and the chaotic dynamics destroys the remnants of regular motion.

The measures of complexity given in this and in the previous subsections reflect only the absolute values of the components of eigenfunctions. As in statistical equilibrium, the detailed information concerning the relative phases of the components is lost. A collective non-chaotic state, like that of a nuclear giant resonance, in principle can appear with a large localization length being a {\sl coherent} combination of many basis states, for example of simple particle-hole excitations in a system of interacting fermions. However, such states (i) have a very low statistical weight and (ii) as a rule are not stationary $-$ as waves in continuous media they are damped into genuinely stationary states of  complicated nature as we mentioned earlier in relation to the growth of entropy. The practical manifestation of such states is usually a broad bump in the energy dependence of the system response to some excitation process (see Section III on strength functions). This response has usually a large spreading width $\Gamma^{\downarrow}$ that, in the case of nuclear giant resonances, is much bigger than the decay width into continuum
$\Gamma^{\uparrow}$. The presence of such collective waves is similar to the phenomenon of {\sl scars} in simple models of chaotic mechanics \cite{kaplan99}. A possible tool for discriminating collective and chaotic states through the phases of their components is given by the phase correlator to be briefly discussed below.

\subsection{Invariant correlational entropy}

Although we have argued that the basis dependence of the usual measures of complexity could be useful for gaining supplementary physical information, one should be careful in the interpretation of results obtained with the aid of such measures. A simple example mentioned in \cite{ZelevinskyRep1996} shows possible dangers. In a routine tight-binding model of a particle in a periodic ${\cal N}$-well potential, the number of principal components found in the lattice site basis is $N_{\rm pc}=(2/3)({\cal N}+1)$ for all standing Bloch waves. In the absence of any chaotic interaction, this exceeds the GOE limit but has nothing to do with chaos or thermalization being generated just by the choice of the representation. Another crucial problem is related to the possibility of {\sl phase transitions}, such as magnetic spin alignment, restructuring of a crystal, or nuclear shape deformation at a certain excitation energy (temperature). Then the mean field basis has to be correspondingly changed. It would be useful to have a measure of complexity {\sl invariant with respect to the choice of basis} and still reflecting possible phase transformations. In this case the {\sl phase relationships} between the components can be taken into account as well.

One possible construction \cite{SBZ98} using the {\sl density matrix} of individual eigenstates can be introduced in the following way. Let the states $|k\rangle$ form an arbitrary complete orthonormal basis of many-body states. This determines the set of amplitudes $C^{\alpha}_{k}$ for stationary states $|\alpha\rangle$. We assume that the Hamiltonian of the system, and therefore eigenstates and their components in any basis, depend on a random parameter (or parameters). The role of such parameters can be played by the coupling constants of the same Hamiltonian varying in small intervals. For a given distribution of these parameters ({\sl noise} applied to the system), the averaging (shown by the overline) defines the density matrix found for a given state $|\alpha\rangle$ that is assumed to adiabatically follow the change of the parameters,
\begin{equation}
{\cal D}_{kk'}^{\alpha}=\overline{C^{\alpha}_{k}C^{\alpha\ast}_{k'}},                 \label{34}
\end{equation}
normalized as Tr$({\cal D}^{\alpha})=1$. The expectation value of a physical observable ${\cal O}$ over the ensemble characterized by the density matrix (\ref{34}) is given by Tr$({\cal OD}^{\alpha})$. Let us stress again that we still consider here a single eigenstate $|\alpha\rangle$. If all states under consideration belong to the same global class with fixed constants of motion, the corresponding energy terms evolving under a continuous change of the noise parameters do not cross, so the symbol $\alpha$ refers to the evolution of a given energy level.

This construction naturally leads to the entropy of a given state,
\begin{equation}
S^{\alpha}=-\,{\rm Tr}({\cal D}^{\alpha}\ln{\cal D}^{\alpha}).                    \label{35}
\end{equation}
All such expectation values are expressed by traces and are
therefore basis independent. Moreover, the result depends on the phase relationships and correlations between the components of the wave function, which justifies the term {\sl invariant correlational entropy}.

If the neighboring states are strongly mixed and ``look the same", the above averaging can be performed over these neighboring states of approximately the same energy, instead of noise parameters. This leads to the thermodynamic description characterized by the corresponding entropy and temperature. Further, if the mixing is chaotic and the components of the wave functions are uncorrelated, the averaging preserves only the {\sl diagonal} components, $k=k'$. Then we come to the information entropy and the corresponding thermal interpretation. In the canonical Gaussian random matrix ensemble all eigenstates are similar, while in realistic systems one has to choose only the  completely mixed neighboring states. The absence of correlations in general is a property of a consistent choice of the mean field (the basis) and the residual interaction (mixing agent) as we discussed earlier. From this viewpoint the equivalence of descriptions shown in the preceding subsections is quite natural. Similar arguments were given in a different context in Ref. ~\cite{elze94}.

Now let us consider the case when we indeed average a given energy level over the noise parameters. In particular, we can take some parameters of an actual Hamiltonian as fluctuating in certain, usually small, limits. The strength of a specific term in the interaction Hamiltonian can be responsible for a phase transition. This happens, for example, for the pairing strength. In macroscopic superconductors, the Cooper instability of a normal Fermi gas with respect to the particle attraction near the Fermi surface \cite{cooper56} occurs at an arbitrary weak strength. In a finite system, such as an atomic nucleus or atoms in a trap, the discreteness of the single-particle spectrum near the Fermi surface requires that the pairing attraction be stronger than some critical value \cite{belyaevKgl}. At weaker attraction, the standard BCS theory being asymptotically exact for macroscopic superconductors gives zero correlation energy when applied, for example, to semi-magic nuclei like $^{48}$Ca; the presence of significant correlations can be revealed by the exact diagonalization of the pairing Hamiltonian \cite{VBZ01}.

\begin{figure}[ht]
\centering
\includegraphics[width=0.7\textwidth]{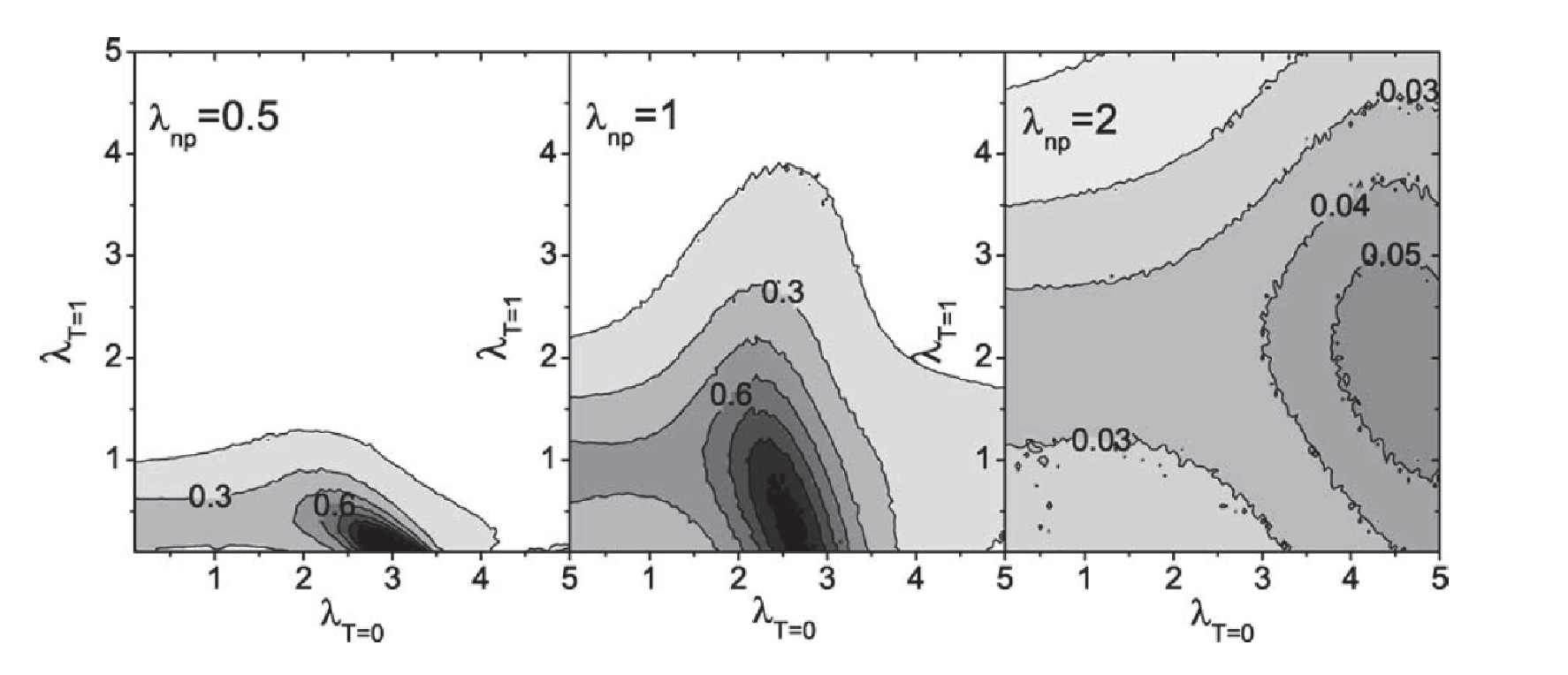}
\caption{(Color online.) The invariant entropy of the ground state of $^{24}$Mg computed on the plane $\lambda_{T=1} -  \lambda_{T=0}$. The three panels show the evolution of the phase diagram depending on the overall scale of non-pairing matrix elements $\lambda_{\rm np}$. The averaging interval of $\delta=0.05$ was used, and the entropy is scaled (divided) by $0.015$ (after \cite{VZ03}).}
\label{phase-diagram}
\end{figure}

Let the parameter in the interaction Hamiltonian responsible for the phase transformation fluctuate around its critical value. Due to the phase transition, the wave functions and observables will be noticeably fluctuating between the two phases. This should lead to the maximum of the corresponding entropy as a function of the noise strength. In this way, the invariant entropy serves as an {\sl indicator for the phase transition regions} in the phase diagram of a finite system.

As an example, in Fig.~\ref{phase-diagram} we show the phase diagram for the nucleus $^{24}$Mg where two types of pairing, isovector ($T=1$ of the $pp,nn$, and $np$ pairs) and isoscalar ($T=0$, quasideuteron $np$ pairs), can compete. The variable parameters here are the interaction strengths of these two types of pairing, $\lambda_{T=1}$ for the main isovector pairing known to prevail in ground states of nuclei, and $\lambda_{T=0}$ for the hypothetical neutron-proton $T=0$ pairing. Three panels correspond to different values $\lambda_{{\rm np}}$ of remaining, non-pairing, parts of the two-body interaction. Each panel shows a contour map of the invariant correlational entropy calculated for the ground state wave function that is subject to random changes of $\lambda _{T}$ within small intervals used for averaging the density matrix. In the actual ground state of $^{24}$Mg, the standard isovector pairing dominates while the phase transition to the isoscalar phase (the black spot corresponding on the first and second diagrams to the maximum of correlational entropy) could be possible if the corresponding interaction strength $\lambda_{1}$ were stronger by approximately a factor of 3. This estimate agrees with another one derived from a different perspective \cite{bertsch10}.

For a realistic strength of pairing in the usual isovector channel, the majority of states do not feel the pairing phase transition that influences only the few low-lying states in the class with $J^{\Pi}T=0^{+}0$ quantum numbers. This is seen as well in the Fermi-liquid analysis: the presence of pairing is noticeable only at the low edge of the energy spectrum while at higher energy the normal Fermi-liquid description is sufficient. The role of non-pairing interaction components is important in mesoscopic systems as seen from the map at strong $\lambda_{{\rm np}}$, third panel of Fig. 10; such incoherent interactions are capable of removing or weakening correlations and enhance chaotic trends of the dynamics. These parts of interaction are playing a key role in smoothing the total density of states \cite{SHZ11,SHZ12} while in the mean-field approach this level density shows unphysically strong oscillations.

We can mention here that the pairing interaction by itself is incapable of chaotizing the system. In a large system it is essentially a part of the mean field introduced by the Hartree-Fock-Bogoliubov method when new quasi-particles are based on the existence of the macroscopic condensate. The situation is slightly different for small systems with pairing (nuclei \cite{VZB02}, atoms in traps \cite{armstrong12}) where the regular properties of the ground and few low-lying states coexist with some chaotic features of excited states. Having only pairing interaction we cannot bring the system to full chaos; increasing the pairing strength we just return to different but still regular dynamics. Only intermediate values of the pairing strength introduce the elements of a non-fully developed chaos and the corresponding trend to thermalization.

In such situations, another related quantity can be useful in the analysis, namely the {\sl phase correlator} of a given eigenstate \cite{armstrong12},
\begin{equation}
{\cal P}^{\alpha}= \frac{1}{{\cal N}}\,\sum_{kk'}C^{\alpha}_{k}C^{\alpha}_{k'},                   \label{36}
\end{equation}
that represents the average of all matrix elements of the density matrix ${\cal D}^{\alpha}$ for a given eigenstate $|\alpha\rangle$. If the wave function components are not correlated, their signs could be more or less random, and the main contribution to this quantity originates from the diagonal terms giving ${\cal P}^{\alpha}\sim 1/{\cal N}$. Collective states are usually characterized by strong correlations between the signs of components (we can repeat here that the discussion of collectivity as well as chaoticity only makes sense with respect to a certain basis). Such correlations are recorded by the phase correlator, giving values larger than $1/{\cal N}$. For a unique super-correlated ``microcanonical" state, where all the amplitudes are equal to $1/\sqrt{{\cal N}}$ and of the same sign, the extreme limit of the phase correlator is ${\cal P}= 1$. This allows one to distinguish between chaotic and collective states both being delocalized in the selected basis.

\subsection{Eigenstates versus energy shell}

As we already noted, in isolated systems of interacting particles, chaotic eigenstates, as a rule, occupy only a part of the available (unperturbed) basis. Therefore, the definition of such states involves two requirements, a large number $N_{{\rm pc}}$ of ``principal components" (practically, $\sqrt {N_{{\rm pc}}} \gtrsim 10$, see Ref.~\cite{FI97b}), and the absence of correlations between these components. The first condition can be mapped onto a large value of the localization length defined via either the Shannon entropy or the inverse participation ratio. As for the second condition, to justify the absence of strong correlations in the structure of eigenstates is not a simple task. Instead, a semi-analytical approach has been suggested \cite{FI97a,BGI98,WIC98,BISS00,LMI00,I01,LMI01,BFH03,Santos2012PRL,Santos2012PRE} that allows one to predict the conditions under which chaotic eigenstates emerge depending on the model parameters. This approach is based on the concept of {\it energy shell} with respect to which the structure of eigenstates should be compared.

The properties of eigenstates in connection to the energy shell have been firstly
analyzed in Refs.~\cite{Casati1993,Casati1996} with the use of the Wigner band random matrix model (\ref{2}). The approach developed there is given in general form and can be effectively applied to realistic physical models. Based on numerical data obtained in \cite{FLW91}, it was understood that the global structure of eigenstates can be described by a relatively simple theory with the only scaling parameter $\beta_{{\rm loc}}$ \cite{Casati1993},
\begin{equation}
\beta_{{\rm loc}}=l/l_{{\rm max}}, \qquad l_{{\rm max}}=a \rho_0 \Delta_E, \qquad \Delta_E = 2 R=4v \sqrt {2b},
                                                                                             \label{37}                                                                          
\end{equation}
which is the ratio of actual localization length $l$ to its maximal value, $l_{{\rm max}}$. Here $R$ is the radius of the semicircle determined by the density of states of the interaction $V$, $b$ is the width of the band, and $a$ is a constant of order unity which depends on how the localization length is measured. It is important that the localization length cannot be larger than $l_{{\rm max}}$, and the latter is determined by the number $1/\rho_0$ of basis states occupying the energy shell
$\Delta_{E}$. Thus, the parameter $\beta$ is restricted by the value $\beta=1$; it is increasing with an increase of interaction $V$. For isolated systems with a large number of particles, the density of states defined by $V$ has typically a Gaussian shape (see examples in Refs.
\cite{FI97a,Santos2012PRL,Santos2012PRE}). Therefore, the width $\Delta_E$ can be also associated with the square root of the variance $\sigma^{2}$ of the density of states
defined by the interaction term in Eqs.~(\ref{2}) and (\ref{3}), see also Eq.~(\ref{13}).

As discussed above, the localization length $l^{\alpha}$ of an individual state $|\alpha \rangle$ can be defined through its Shannon entropy $S^{\alpha}$ or the inverse participation ratio $M^{\alpha}_2$. Depending on this choice, the factor $a$ in Eq.~(\ref{37}) is different, however it remains of the order unity \cite{Casati1993,Casati1996}. In order to reduce fluctuations, in Ref.~\cite{Casati1993} the localization length $l$ that enters Eq.~(\ref{37}) was obtained by averaging the Shannon entropy $S^{\alpha}$ or $(M^{\alpha}_{2})^{-1}$ over all eigenvectors from an ensemble of Wigner BRM (\ref{2}) with fixed parameters ${\cal N},\rho_0,b,V$. The detailed analysis \cite{Casati1996} of the structure of eigenstates in the Wigner BRM ensemble has revealed a transition from localized to delocalized states with the increasing scaling parameter $\beta_{{\rm loc}}(\lambda)$ where $\lambda=b^{3/2}/(\rho v)$. It should be stressed that here the notion of localization/delocalization refers to the ``energy shell" which means a corresponding segment of the whole unperturbed basis. This can serve as the core of the many-body localization occurring in both disordered \cite{A90,alt97,FI97b} and regular isolated systems of interacting particles \cite{Santos2012PRL,Santos2012PRE}. Due to the finite range of inter-particle interaction, the part of the unperturbed many-body basis occupied by the eigenstates of the total Hamiltonian can be much smaller than the basis dimension (sometimes, infinite). On the other hand, the number of components in chaotic eigenstates can be very large, thus, allowing one to treat them as pseudo-random ones. In what follows, we assume that the localization length is smaller than the available set of basis states defined by the width of the energy shell. Correspondingly, by delocalized states we mean chaotic eigenstates that densely fill the whole energy shell provided by the strength and type of the inter-particle interaction.

In fact the energy shell determines the shape and the width of the strength function when it reaches its limit on increasing the interaction $V$. Indeed, for relatively small (however, non-perturbative) values of $V$, the SF is known to have the Breit-Wigner shape with the half-width $\Gamma$. The latter parameter is proportional to $v^2$ where $v$ is the mean value of the matrix elements of $V$ directly participating in the creation of the SF. On the other hand, the width $\Delta_E$ of the energy shell is given by $\sigma$ which is the square of the variance of the interaction $V$, therefore, $\Gamma \ll \Delta_E$ for small $v$. With an increase of $v$, the shape of the SF changes from  Breit-Wigner to Gaussian. In this case $\Gamma$ tends to $\sigma$, and the latter  fully defines the shape of the SF. This crossover is clearly demonstrated by the data in Fig. 4, where for the Models 1 and 2 the strength functions are shown as a function of the interaction strengths, in comparison with the shape of energy shells (solid curves). As we show below, this crossover corresponds to the onset of delocalization of chaotic eigenstates with respect to the energy shell.

\begin{figure}[ht]
\centering
\subfigure{\includegraphics[angle=-90,scale=0.35]{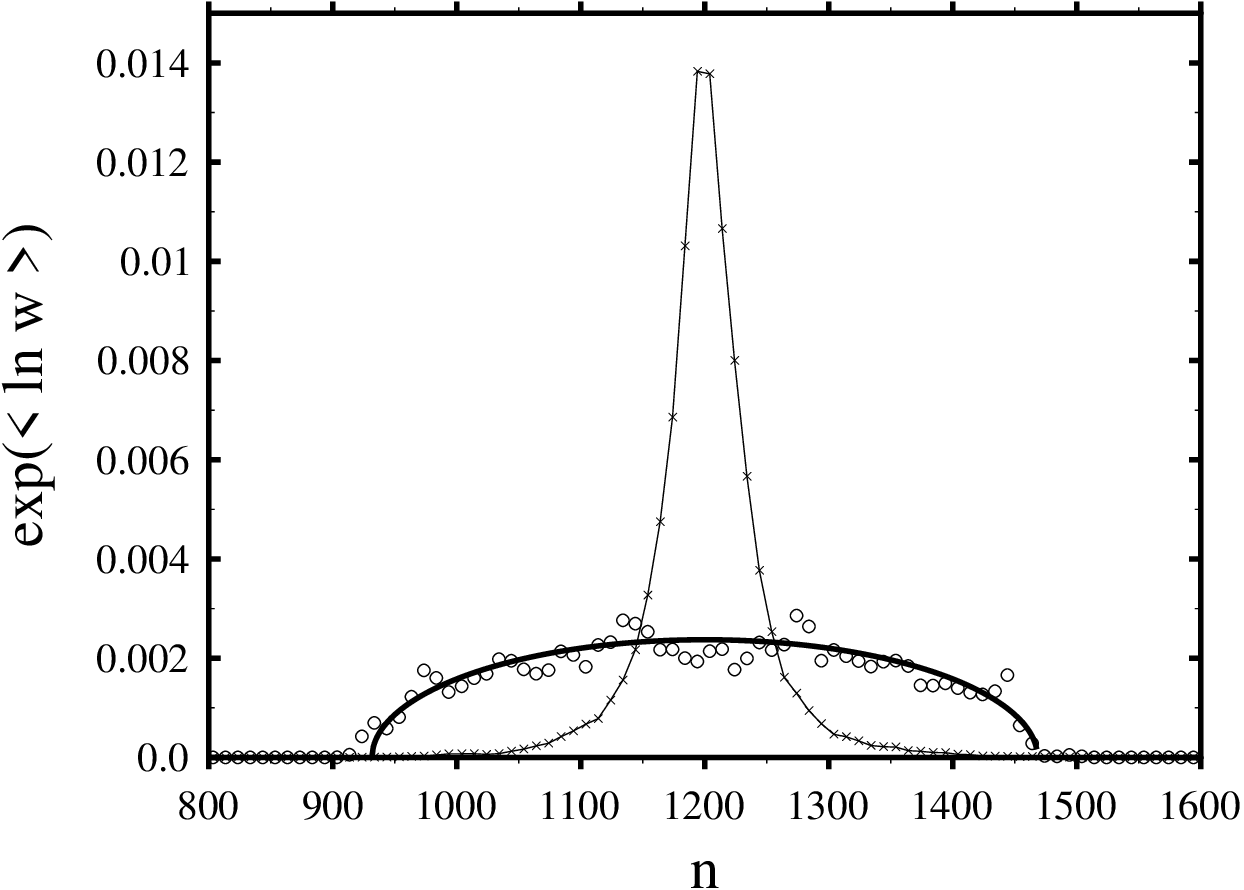}}
\subfigure{\includegraphics[angle=-90,scale=0.35]{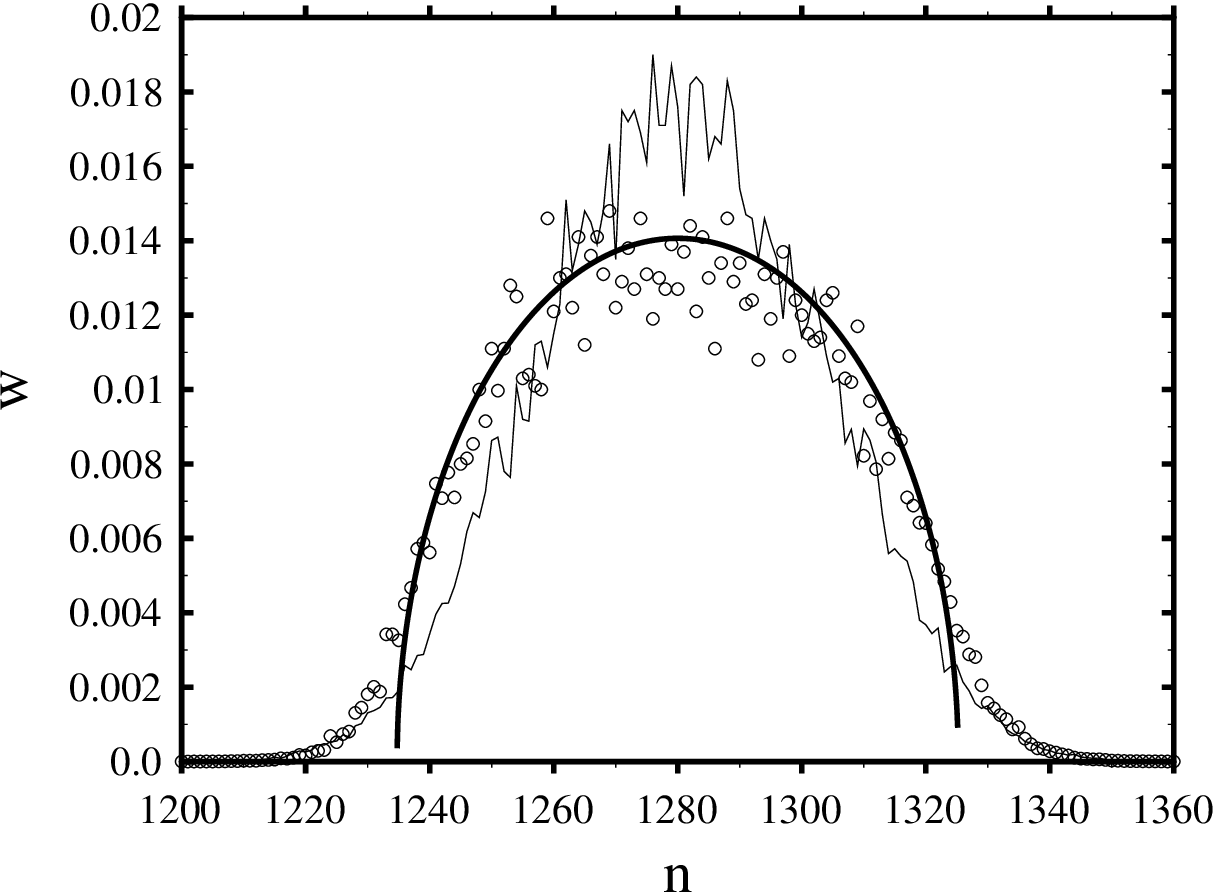}}
\caption{Structure of localized (left panel) and ergodic delocalized (right panel) eigenfunctions for the Wigner band random matrices. Here $N=2560, v=0.1, b=16, \rho_0=40$ on left panel, and $N=2400, v=0.1, b=10, \rho_0=300$ on right panel. Thick curves stand for the semicircle, and thin curves are obtained by the average of 300 eigenfunctions with respect to their centers. Circles denote the average of the same eigenfunctions with respect to the centroids of their energy shells. For the ergodic eigenfunctions all distributions look the same, apart from  fluctuations. For the localized eigenfunctions with $\beta_{{\rm loc}}=0.24$, the average with respect to their centroids $n_c$ shows the localization inside the energy shell, while the other average remains close to the semicircle with $\beta_{{\rm loc}}=0.99$ (after \cite{Casati1996}). }
\label{boris}
\end{figure}

The illustration of two different kinds of eigenstates is given in Fig.~\ref{boris}. The energy shell is shown by the thick smooth curve, in accordance to the semicircle shape of the density of states covered by the perturbation $V$, see eq.~(\ref{2}). Each circle in both panels refers to the average of $w^{\alpha}_n=\left(C^{\alpha}_n\right)^2$ taken over a number of eigenstates $\alpha$. Thus, the distribution of points gives the average shape of eigenstates in the original basis. When the value of the scaling parameter $\beta_{{\rm loc}}$ is small (left panel), the number of principal components $N^{\alpha}_{{\rm pc}}$ is smaller than the total number of basis levels inside the energy shell. Since in this case the position of the centroid of an individual eigenstate is strongly fluctuating inside the energy shell, the average was performed after shifting the eigenstates to their individual centroids. In this way, the full curve on the left panel manifests the localized structure of eigenstates. Contrary to that, the circles are obtained by the average {\it without} such a shift. The comparison of the distributions with and without shifts allows one to detect whether the eigenstates are localized or extended in the energy shell.

The second situation shown on the right panel of Fig. \ref{boris} refers to the eigenstates that fill the whole energy shell; this filling is practically ergodic with no big difference between the two types of averaging. A small deviation of the distributions (of both types of averaging) from the semicircle can be attributed to the insufficiently large value of the scaling parameter. These data clearly demonstrate the transition from localized to delocalized states with respect to the energy shell. In Ref. \cite{Casati1993} it was shown that when the scaling parameter $\beta_{{\rm loc}}$ approaches 1, the statistics of level spacings tends to the Wigner-Dyson distribution revealing that the filling of the energy shell by chaotic eigenstates corresponds to the onset of quantum chaos.

The relevance of the eigenvalue statistics to the chaotic structure of eigenstates has led in Ref. \cite{Casati1993} to an unexpected result: a one-to-one correspondence was numerically found between the rescaled localization length $\beta_{{\rm loc}}$ and the repulsion parameter $\beta$ in the parametrization of the nearest level spacing distribution $P(s)$. The latter was obtained by the fit of $P(s)$ to the Brody distribution \cite{B73} that approximately interpolates between the Poisson and Wigner-Dyson statistics of level spacings. The relation between $\beta_{{\rm loc}}$ and $\beta$ was found to be linear, $\beta_{loc}=\beta$ with a  high accuracy (the deviation is less than $1\%$). Being qualitatively natural since both quantities reflect the process of chaotization, this intriguing result still has no quantitative explanation. Indeed, unlike the level statistics which is basis-independent, the localization length is defined with respect to a certain basis. A similar result has been recently obtained for the one-dimensional tight-binding Anderson model \cite{SIZC12}, and the same relation between $\beta_{{\rm loc}}$ and $\beta$ was found over a large range of the model parameters (see also \cite{Fo13,Fo13a,Robnik13}).

\begin{figure}[ht]
\centering
\includegraphics[width=0.6\textwidth]{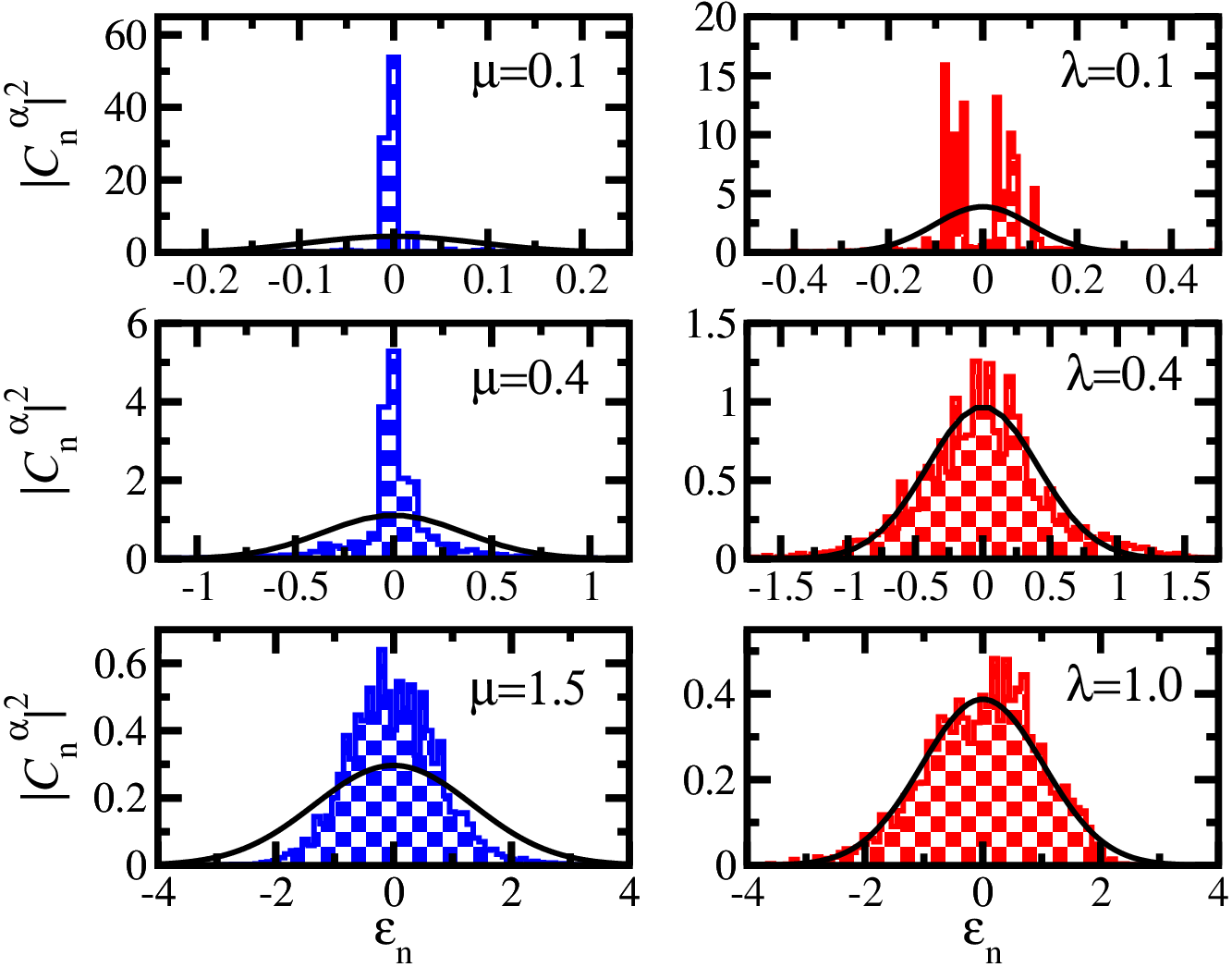}
\caption{(Color online.)  Structure of eigenstates for Model 1 (\ref{5}) (left) and Model 2 (\ref{6}) (right) in the energy representation, obtained by averaging over five even states in the middle of the spectrum. Histograms correspond to the average of $w^{\alpha}_n=|C^{\alpha}_n|^2$ after shifting the centroid to the center of the energy shell. Solid curves correspond to the Gaussian form of the energy shells. Here $\varepsilon_n$ stands for the unperturbed energy (after \cite{Santos2012PRE}).}
\label{fig:EF}
\end{figure}

The onset of chaos due to delocalization of eigenstates in the energy shell is a generic scenario for many-body isolated systems, both random and deterministic. As an illustration, we refer to the results \cite{Santos2012PRL,Santos2012PRE} concerning the study of two models (\ref{5}) and (\ref{6}) of interacting spin-$1/2$ particles. The data in Fig.~\ref{fig:EF} demonstrate the spread of eigenstates on
increasing the perturbation, $\mu$ and $\lambda$. For the non-integrable Model 2, the critical strength of perturbation resulting in a complete filling of the energy shell is $\lambda_{cr} \approx 0.4$. For the integrable Model 1, at $\mu >\mu_{cr} \approx 0.5$, the eigenstates may be treated as delocalized and chaotic-like, however, they do not fill completely the energy shell. This lack of ergodicity can be attributed to the integrability of this model. Even for the integrable model the individual eigenstates look quite chaotic, see details in \cite{Santos2012PRE}. As we discuss below, quench dynamics for {\it both} models turn out to be quite similar, and the statistical description of the dynamics works well for both models.
It should be stressed that for the non-integrable Model 2 the same critical value $\lambda_{cr} \approx 0.5$ emerges when exploring the level spacing distribution $P(s)$. A clear transition from the Poisson to Wigner-Dyson form of $P(s)$ was detected numerically in Refs. \cite{Santos2012PRL,Santos2012PRE} for Model 2, while for Model 1 the distribution remains Poisson for any $\mu$. These data indicate that although the spectrum statistics helps to distinguish between integrable and non-integrable models, the global properties of eigenstates with respect to the energy shell are quite similar.

As shown in Refs.~\cite{FI97b,Santos2012PRL,Santos2012PRE} the simplest analytical estimate for the emergence of chaotic eigenstates filling the energy shell can be obtained by exploring the structure of the Hamiltonian matrix without  diagonalization. To do this, one has to compare the off-diagonal matrix element $V_{nm}$ that couples the selected basis states $|n\rangle$ and $|m\rangle$ to the energy difference $d_f$ between the levels corresponding to these states. This spacing $d_f$ turns out to be much larger than the mean level spacing $D$ between {\it all} states, since the two-body nature of the interaction imposes some restriction to the coupling between many-body states. This fact is reflected by a high fraction of zero matrix
elements of $V_{nm}$, even in the TBRI model (1) in which all {\it two-body} matrix elements
are non-zero (see Fig.~\ref{matrix99} and details in Refs.~\cite{FI97b}). For this reason, one has to exclude those states that are coupled with vanishing {\it many-body} matrix elements \cite{A90}. Numerical data obtained for the TBRI model as well as for the physical Models 1 and 2 (see Eqs.(\ref{5}) and (\ref{6})), indeed, give the correct values for the interaction strengths above which the eigenstates of the total Hamiltonian look random and extended in the energy shell. For example, the above approach gives the same critical values $\lambda_{cr} \approx 0.4$ and $\mu_{cr} \approx 0.5$ as numerical data \cite{Santos2012PRL,Santos2012PRE}.



\section{Thermalization}

\subsection{Emergence of thermalization in isolated systems}

As discussed above, in many physical situations the eigenstates of an isolated system of interacting particles can be treated as chaotic superpositions of their components in an appropriate many-particle basis. This fact has been used in Ref.~\cite{FGGK94,Flambaum1996a,Flambaum1996b,FI97a,FI97b} for developing the statistical approach to the description of various observables, based on two key ingredients, the notion of the strength function and the shape of eigenstates in the basis of non-interacting particles (or quasiparticles). In this way a natural question arises about the possibility of thermalization in isolated systems in spite of the absence of a heat bath. In the canonical description of conventional statistical mechanics, thermalization is directly related to the temperature defined by the heat bath; any definition of temperature (for example, through the kinetic energy of an individual particle, statistical canonical distribution, or via the density of states) gives the same result. Contrary to that, in isolated systems of a finite number of particles, these temperatures can be different, and the difference increases with the decrease of the particle number. This is also known for classical systems; the detailed study of different temperatures has been done for interacting classical spins moving on a ring \cite{Borgonovi2000}. With an increase of the particle number $N$, all definitions of temperature tend to the unique value, corresponding to the standard result of the thermodynamic limit, $N \rightarrow \infty$. In this limit (both in classical and quantum mechanics), the statistical behavior of systems emerges irrespectively of whether the system under consideration is integrable or non-integrable, (see, for example, Refs.\cite{B45,B81,AGL75,G75,C86}).

Thus, the definition of ``thermalization" in application to isolated mesoscopic systems is obscure; for this reason we prefer to speak of thermalization in a broader context, namely, as
the existence of statistical relaxation to a steady state distribution. As rigorously shown in Refs.~\cite{B45,Bogolubov70}, in the thermodynamic limit $N\to\infty$ even for a completely integrable system an infinitely small subsystem (probe) exhibits statistical relaxation, e.g. the emergence of  the Gibbs distribution.
 With the motion not even ergodic, one can wonder how statistical relaxation can occur since
this requires mixing. In the system considered by Bogoliubov this is explained
by a perturbation spectrum of the probe oscillator that becomes {\it continuous} in the limit $N\to \infty$, a condition necessary for mixing (see discussion in \cite{C86}). For finite $N \gg 1$ the spectrum is discrete which implies a quasi-periodic behavior for any observable. However, the characteristic time for revivals is typically so large that on a finite time scale (which, however, could be larger than the lifetime of the Universe) a perfect relaxation occurs. In this sense, quantum chaos can be considered as ``temporary chaos", however, indistinguishable from that occurring in classical mechanics on a finite time scale.

A new situation emerges for isolated systems consisting of a finite number of particles. Although in this case the energy spectrum is discrete, the behavior of a system can reveal strong statistical properties, and the adequate theory is based on the notion of quantum chaos. First of all, such a theory has to give an answer to the basic question: {\it when} (or under which conditions) the statistical description is possible and practically useful for a quantum isolated system. To answer this question one has to know the mechanism responsible for the onset of statistical behavior. In classical mechanics this problem was solved by the concept of local instability of motion leading to chaotic behavior in spite of completely deterministic equations of motion. In quantum mechanics the Schr\"odinger equation is linear which means absence of local instability for the time-dependent wave function. For this reason, the chaos that emerges in quantum systems was termed ``linear chaos" by Chirikov \cite{c97}, in order to stress the principal difference from deterministic chaos occurring in non-linear classical systems.

The situation is somewhat easier when a quantum system has a well defined classical limit and the corresponding classical motion is strongly chaotic. Such a situation was originally referred to as ``quantum chaos", the term that is nowadays used in other applications such as optics, acoustics, etc. In contrast to classical mechanics, the description of a quantum system is based on the energy spectrum of stationary eigenstates, so that the emergence of quantum chaos in a closed system has to be quantified in corresponding terms. This allows one to speak of quantum chaos in a more general context, namely, including either systems without a classical limit or disordered systems, in contrast with  deterministic ones.

As argued in Refs.\cite{ZelevinskyRep1996,ann,FI97b}, for an isolated system of a finite number of particles (that can be quite small), the mechanism of thermalization is due to the interaction between particles. When the interaction strength exceeds some critical value, or the level density becomes sufficiently high, the many-body wave functions become extremely complicated (``chaotic") and this leads to thermalization. In a broad sense, this is understood as emergence of relaxation to a steady-state distribution allowing for a statistical description. In fact, at such complexity of stationary eigenfunctions expressed in the appropriate basis, the statistical description seems to be the only one reasonable. A direct link between chaotic eigenstates and the conventional statistical distributions (Boltzman, Fermi-Dirac and Bose-Einstein ones) was analytically established in Ref.~\cite{Srednicki1994} for the billiard models.
Concerning  the conditions for the onset of many-body thermalization due to the inter-particle interaction, the basic ideas and their implications were reported in Refs.~\cite{FI97b,I01} demonstrating that the role of a heat bath is played by a sufficiently strong interaction between particles.

\subsection{Fermi-liquid description}

According to Landau, a non-superfluid system of interacting fermions can be considered as a result of an adiabatic transformation of the perfect Fermi gas when we switch on the residual interaction and gradually convert this gas into a Fermi liquid. If so, the energy spectrum of the system can be classified in the gas-like way although the quantitative characteristics, such as effective mass, susceptibilities etc. can change. We can expect that the actual excited states of the system, including very complicated ones, can be reasonably described with the language of a Fermi gas of dressed quasiparticles distributed over single-particle levels $|\lambda)$ with effective mean-field energies $\epsilon_{\lambda}$ according to the mean {\sl occupation numbers} $n_{\lambda}$. For our purposes, this brings in the idea of a {\sl single-particle thermometer} that measures the effective single-particle temperature $T^{\alpha}_{{\rm s-p}}$ of individual eigenstates $|\alpha\rangle$ in terms of the grand-canonical Fermi distribution,
\begin{equation}
f^{\alpha}_{\lambda}=\,\frac{1}{e^{(\epsilon_{\lambda}-\mu^{\alpha})/T^{\alpha}_{{\rm s-p}}}+1}.   \label{38}
\end{equation}
For a given many-body state $|\alpha\rangle$ and all single-particle levels $\lambda$, the distribution function (\ref{38}) should work with the same values of effective temperature $T^{\alpha}$ and chemical potential $\mu^{\alpha}$ determined, up to fluctuations, by the position $E^{\alpha}$ of the state under consideration in the energy spectrum.

\begin{figure}[ht]
\centering
\includegraphics[width=0.65\textwidth]{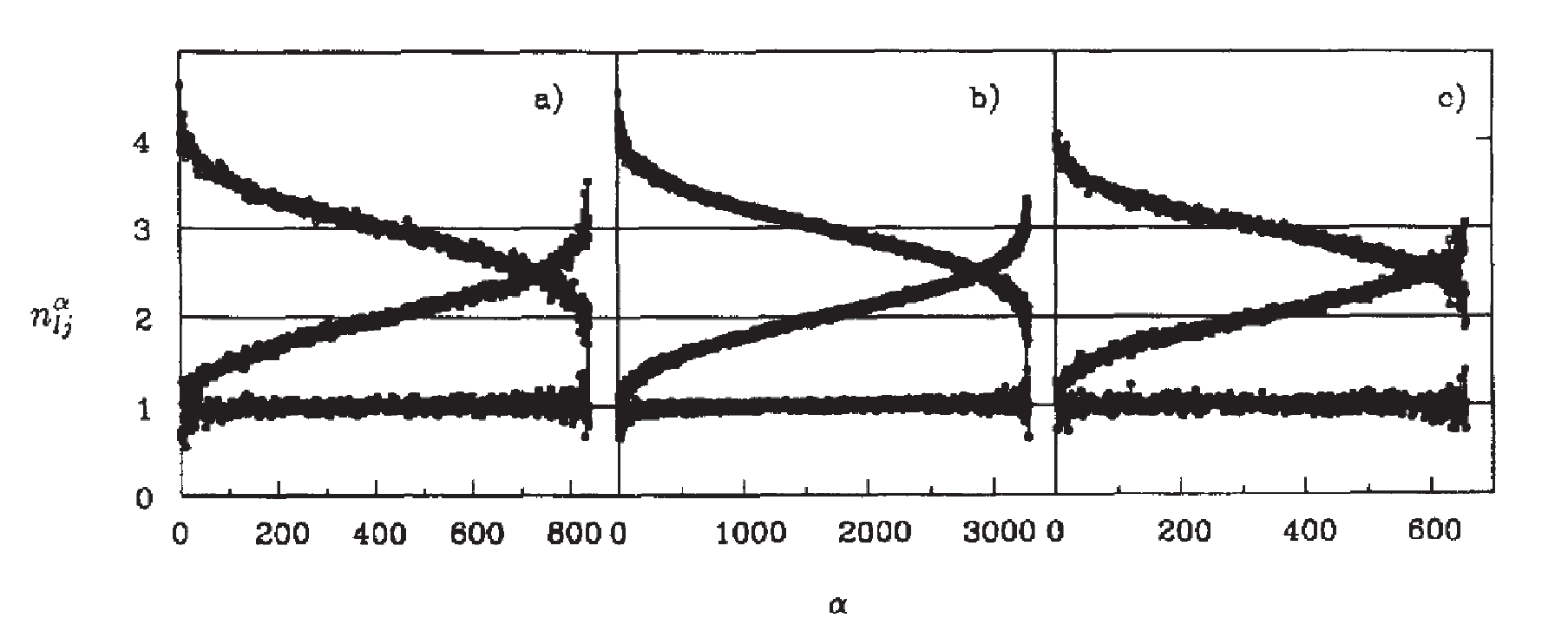}
\caption{Single-particle occupancies found for exact stationary wave functions in the $sd$-shell model calculations for $^{28}$Si. Three panels refer to different classes of states with quantum numbers $J^{\Pi}T$ equal to $0^{+}0$, panel {\sl a)}, $2^{+}0$, panel {\sl b)}, and $9^{+}0$, panel {\sl c)}. Three thick lines (in fact, points for all individual states) for each panel correspond to the occupancies of $s_{1/2},\;d_{3/2}$ and $d_{5/2}$ orbitals, from bottom to top on the left-hand side of each panel (after
\cite{ZelevinskyRep1996}).}
\label{occupation}
\end{figure}

Atomic and nuclear calculations show that the Fermi-liquid model indeed describes well the single-particle occupation numbers for exceedingly complicated stationary many-body eigenfunctions. The chemical potential varies very little along the spectrum while the effective temperature again manifests the average behavior that we have already seen in the description utilizing the detailed information on the wave functions. In the example of the nuclear spherical shell model, when single-particle quantum numbers are orbital, $\ell$, and total, $j$, angular momentum, Fig.~\ref{occupation} demonstrates that the occupancies of each orbital, $n^{\alpha}_{\ell j}=f^{\alpha}_{\ell j}(2j+1)$, with the exception of a small number of states at the edges, smoothly evolve along the energy spectrum. At the lowest excitation energy, the pairing interaction would require a BCS-like parametrization of the occupation numbers; the nucleus close to its ground state is rather a small drop of superfluid matter. This will not play a considerable role at energies above the pair breaking threshold (energy gap). In the center of the spectrum, all occupancies $f^{\alpha}_{\ell j}$ become equal to 1/2 as expected for infinite temperature in a system like $^{28}$Si with the half-occupied orbital space, $sd$-shell in this case. Different classes ($J^{\Pi}T$) of global quantum numbers for a system as a whole behave nearly identically although, because of a different total number of many-body states $|\alpha\rangle$ in each class, the fluctuations are slightly different. This is another example of the  correlations between separate subclasses of the Hilbert space induced by the interaction (earlier we mentioned analogous effects of a random interaction). A similar calculation for $^{24}$Mg (1/3 of orbital space filled) confirms the same conclusions. In the original Landau's theory of a normal Fermi-liquid, only the weakly excited states were classified in analogy to a Fermi-gas. Here we see that in a closed mesoscopic system, the self-consistent interaction between constituents creates complicated states where the single-particle occupancies evolve as in a Fermi-gas with the effective temperature smoothly changing along the entire spectrum of excited states.

\begin{figure}[ht]
\centering
\includegraphics[width=0.45\textwidth]{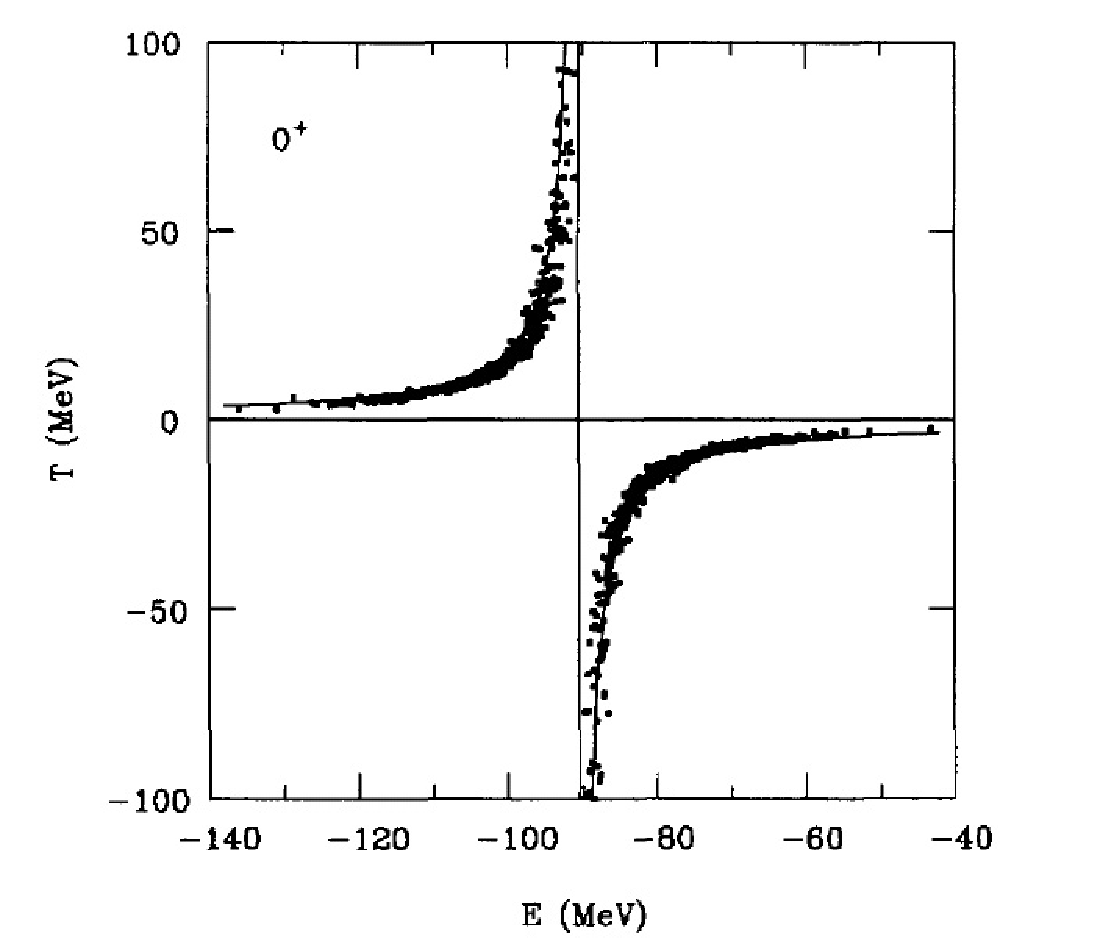}
\caption{Comparison of effective temperatures for the set of many-body states $J^{\Pi}T=0^{+}0$ found in the $sd$-shell model calculations for $^{28}$Si. Solid curves: thermodynamic temperatures found from the level density according to Eq. (\ref{39}). Dots: effective temperatures $T^{\alpha}_{{\rm s-p}}$ extracted from the single-particle occupancies, Eq. (\ref{38}), for all individual states (after \cite{ZelevinskyRep1996}).}
\label{temper-zele}
\end{figure}

It is instructive to compare the single-particle temperature $T^{\alpha}_{{\rm s-p}}$ of individual many-body states with the conventional {\it thermodynamic temperature} which can be defined via the thermodynamic entropy $S(E)$ of the microcanonical ensemble found as a function of energy averaged over a small energy window rather than for individual eigenstates. In statistical mechanics the entropy is determined ($k_{B}=1$) as logarithm of the statistical weight, $S(E)=\ln \Omega (E)$, and the corresponding temperature $T$ of the ensemble is
\begin{equation}
T=\left(\frac{\partial S}{\partial E}\right)^{-1}.                 \label{39}
\end{equation}
In this language, for an isolated system with a finite number of particles and single-particle states, the growing level density (the left half of the spectrum) describes the temperature raising from zero to infinity, while the right half corresponds to negative temperature for the states with the inverse population of mean-field orbitals. Assuming that the level density $\rho(E)$ in a finite space can be modeled by a Gaussian \cite{French1970,mon75,Brody81,ZelevinskyRep1996,HKZ03} with the centroid at $E=E_{0}$ and width $\sigma_{E}$, we obtain
\begin{equation}
S(E)=-\ln\rho(E) \quad \leadsto \quad T(E)=\,\frac{\sigma_{E}^{2}}{E_{0}-E}.      \label{40}
\end{equation}

The data in Fig.~\ref{temper-zele} clearly show that the thermodynamic temperatures essentially coincide, both in the regions of positive and negative temperatures, with the single-particle temperatures of {\sl individual} many-body states determined from the actual single-particle occupancies of each orbital. A view through a magnifying glass would show deviations at the edges of the spectrum for a small number of eigenstates. One reason is the presence of pairing correlations; the level density may carry the remnants of pairing effects resulting in phase transitions. However, the correct microscopic calculations of the level density \cite{SHZ11,SHZ12} show that, except for the very edges of the spectrum, the incoherent collision-like parts of the interaction smooth out in a small system the pairing modulation of the level density; as discussed earlier, phase transitions in such systems are also significantly washed out.

The analysis \cite{ZelevinskyRep1996} reveals that the different measures of complexity of individual eigenstates are practically equivalent to each other and to the thermodynamic consideration based on the level density under the important condition: an {\sl appropriate thermometer}. This is the point where the choice of the representation for the measures using the components of the wave function in a certain basis becomes important. As we have already argued, in a many-body system of interacting particles, the mean field representation separates in the best way regular and chaotic features of dynamics. The solution of the many-body problem should be self-consistent. Taking arbitrarily weak or arbitrarily strong residual interactions, we would introduce the discrepancy between different parts of the description. The information entropy of the majority of states would become either uniformly low or very high (everywhere close to the GOE limit) without demonstrating the spectral evolution. The single-particle thermometer still works at weak interaction but becomes useless in the limit of too strong residual forces when all available orbitals along the spectrum are occupied uniformly as for  infinite temperature.

\subsection{Microcanonical versus canonical distribution}

 An approach to thermalization in isolated systems of interacting particles based on the ground of chaotic eigenstates was suggested in Refs. \cite{ZelevinskyRep1996,Flambaum1996b,FI97b}. The main object is the set of single-particle occupation numbers $n_s$ defined by the standard expression,
\begin{equation}
n^\alpha_s=\left\langle \alpha\right| \hat n_s\left| \alpha\right\rangle
=\sum_k\left| C_k^{\alpha}\right| ^2\left\langle k\right| \hat n_s\left|
k\right\rangle                                                           \label{41}
\end{equation}
with $\hat n_s=a_s^{\dagger }a_s$ the occupation number operator for the mean-field orbital $s$ that has (for fermions) eigenvalues (occupation numbers) equal to $1\,$or $0$. We remind that here $C_k^{\alpha}$ are the components of a specific many-body state $\left| \alpha \right\rangle$ (exact eigenstate of the full Hamiltonian $H$) presented in the basis of $H_0$, see Eq.~(\ref{8}). The actual occupancy $n_s^{\alpha}=n_s(E^{\alpha})$ is the sum of the probabilities of filling the basis states used as the elements constructing the exact state $|\alpha\rangle$. The typical regular evolution of the occupancies along the spectrum of stationary states in a closed many-body system of interacting fermions is illustrated by Fig.~13.

The key point of the approach is that for a chaotic eigenstate $|\alpha\rangle$ the occupation numbers $n_s^{\alpha}$ are given by {\it smooth} functions of the total energy $E=E^{\alpha}$ of the system, due to a large number of {\it strongly fluctuating} components $C_k^{\alpha}$ contributing to the stationary wave function $|\alpha\rangle$. In this situation one can safely perform an averaging {\it inside} these eigenstates (for example, with a moving window), or, equivalently, over neighboring states with the same values of the global constants of motion within a small energy interval centered at $E^\alpha$. Equivalence of different kinds of averaging is in the spirit of conventional statistical mechanics for systems in contact with a thermostat. As we have already argued, the role of the thermostat is played by the interaction producing a complex structure of eigenstates. Such an average is similar to the microcanonical one since it is done at fixed total energy $E$. Thus, we have
\begin{equation}
n_s(E)=\sum\limits_k\overline{\left| C_k^{\alpha}\right| ^2}
\left\langle k\right| \hat n_s\left| k\right\rangle
=\sum\limits_k F^{\alpha}(E)\left\langle k\right| \hat
n_s\left| k\right\rangle,                        \label{42}
\end{equation}
where the $F-$function (also called shape of eigenstates),
\begin{equation}
F^{\alpha}(E) \equiv \langle \left| C_k^{\alpha}\right| ^2 \rangle _{E_k=E},   \label{43}
\end{equation}
is introduced similarly to the strength function (\ref{15}). As discussed earlier, the $F-$function and the strength function $F_k(E)$ originate from the same matrix of components $w^{\alpha}_k=|C^{\alpha}_k|^2$ obtained by the diagonalization of the total Hamiltonian $H$ in the $k$-basis: the strength function is the average projection of a {\it basis} state onto the exact ones, and the $F-$function is the projection of an {\it exact} state onto the basis states. Both these projections are considered in the energy representation rather than
in the basis representation, an important point  for the analysis.

Apart from the condition of a large number of components in an eigenstate, one has to be sure that these components are effectively uncorrelated [compare the discussion of the correlator (\ref{36})]. Otherwise, the averaging is not justified, and the distribution of $n_s$ will depend on tiny details of eigenstates instead of being a smooth function of the total energy (and possibly of other exact constants of motion). From Eq.~(\ref{42}) one can also see that the distribution $n_s$ depends on two ingredients:  the $F$-function and the
mean single-particle occupancies $ \left\langle n\right| \hat n_s\left| n\right\rangle $
which  absorb the requirements of quantum statistics  (a similar approach works for bosons as well).
A special remark should be made  for  systems with a well defined classical limit.
In such a case, the $F-$function can be found from the classical Hamiltonian, and the problem of the shape of this function, which is central for the following analysis, is considerably simplified (see \cite{I01} and references therein).
It is instructive to compare the microcanonical distribution (\ref{42}) of occupation numbers $n_s(E)$ with the standard {\it canonical distribution},
\begin{equation}
n_s(T)=\frac{\sum_{\alpha}n_s^{\alpha}\,\exp (-E^{\alpha}/T)}{
\sum\limits_\alpha\exp (-E^{\alpha}/T)}.                   \label{44}
\end{equation}
Here $T$ is the temperature of a heat bath and the sum runs over exact eigenstates. The important difference between the  distribution (\ref{42}) and the canonical distribution (\ref{44})
is that in Eq.~(\ref{42}) the occupation numbers are calculated for the fixed total energy $E$ of a system and not for the fixed temperature $T$ in Eq.~(\ref{44}). Their interrelation comes through the identification of the running energy $E$ with the canonical mean energy at the temperature $T$,
\begin{equation}
E=\left\langle E\right\rangle _T=\frac{\sum\limits_{\alpha}E^{\alpha}\,
\exp (-E^{\alpha}/T)}{\sum\limits_{\alpha}\exp (-E^{\alpha}/T)}.     \label{45}
\end{equation}
The arguments are essentially similar to those traditionally used in statistical mechanics to demonstrate the equivalence of  canonical and microcanonical ensembles for large systems. Here we replace the summation over $\alpha$ by the integration over the density of exact
energy levels, $\rho (E^{\alpha})$,
\begin{equation}
\sum_{\alpha}n_s^{\alpha}\exp \left( -E^{\alpha}/T\right) \approx \int
n_s^{\alpha}\rho (E^{\alpha})\exp \left( -E^{\alpha}/T\right) dE^{\alpha}   \label{46}
\end{equation}
$$
\approx \sum_kn_s^{(k)}\int F^{\alpha}(E_k)\rho (E^{\alpha})\exp \left(
-E^{\alpha}/T\right) dE^{\alpha}=\sum\limits_kn_s^{(k)}\,F(T,E_k),
$$
where $ n_s^{k}=\left\langle k\right| \hat n_s\left| k\right\rangle $.
Here the function $F(T,E_k)$ is the canonical average of $F^{\alpha}$ ,
\begin{equation}
F(T,E_k)=\int F^{\alpha }\Phi _T(E^{(\alpha)})\,dE^{\alpha}   \label{47}
\end{equation}
with another canonically (thermally) averaged function,
\begin{equation}
\Phi _T(E)=\rho (E)\exp \left( -E/T\right).                    \label{48}
\end{equation}
In standard statistical mechanics, the product of an exponentially growing level density
and an exponentially decreasing thermal excitation factor produces the effective microcanonical
window $\Delta_{T}$. As a result, one can transform the canonical distribution (\ref{44}) into the form similar to the $n_s-$ distribution (\ref{44}),
\begin{equation}
n_s(T)=\frac{\sum\limits_kn_s^{(k)}\,F(T,E_k)}{\sum\limits_k\,F\,(T,E_k)}.    \label{49}
\end{equation}
This distribution can be used, for example, in the calculation of occupation numbers and other
mean values for a quantum dot in thermal equilibrium with environment.

In a large system, the position $E_m$ of the maximum of the distribution (\ref{49}) is defined by
\begin{equation}
\frac{d\ln \rho (E)}{dE}=\frac{1}{T},                              \label{50}
\end{equation}
and the width is given by
\begin{equation}
\Delta _T=\left| \frac{d^2\ln \,\rho (E)}{dE^2}\right| ^{-1/2}.      \label{51}
\end{equation}
For isolated systems of interacting particles the density of states is typically described
 by a Gaussian with  centroid $E_c$ and variance $\sigma ^2$, compare Eqs.~(\ref{39}) and
(\ref{40}). This leads to a Gaussian form of the thermal averaging function
$\Phi _T(E)$ with a shifted centroid,
\begin{equation}
E_m=E_c-\frac{\sigma ^2}T,                                    \label{52}
\end{equation}
and a  width $\Delta _T = \sigma$. For a large number of particles the narrow function $\Phi _T$ is close to a delta-function at $E=E_m$ and the $n_s-$ distribution is close to the canonical one, see Eq.~(\ref{47}). For isolated systems with a small number of particles, one should use the $n_s-$ distribution (\ref{43}), see details in \cite{Flambaum1996b,FI97b}.

\subsection{Onset of the Fermi-Dirac distribution}

Here we illustrate the emergence of Fermi-Dirac distribution with the use of the TBRI-model (\ref{3}). If the interaction $V$ is relatively weak, the many-body eigenstates cannot be
treated as chaotic. It follows that (i) the fluctuations of the components $C^\alpha_n$ in Eq.~ (\ref{41}) do not obey the standard central limit theorem, (ii) the averaging procedure is not supported and  (iii) the fluctuations of the occupations numbers $n_s$ as a function of the energy $E^\alpha$ are strong, so that they do not decrease with the number of principal components, $N_{{\rm pc}}$.

\begin{figure}[!htb]
\centering
\includegraphics[width=0.45\textwidth]{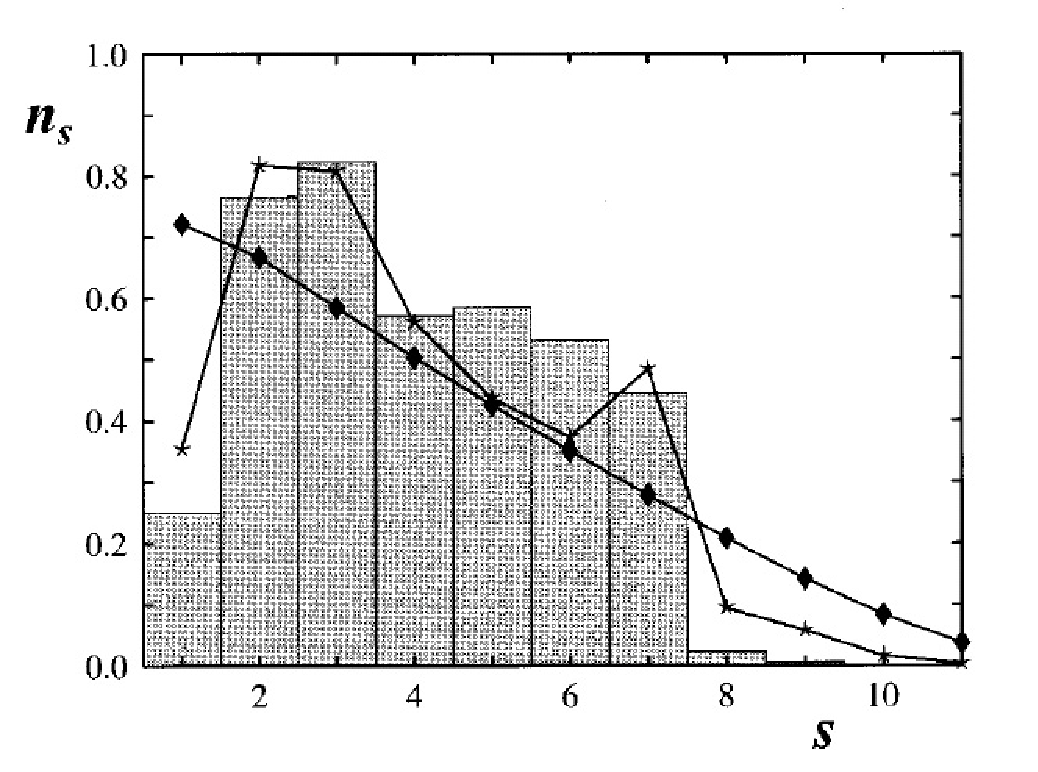}
\includegraphics[width=0.45\textwidth]{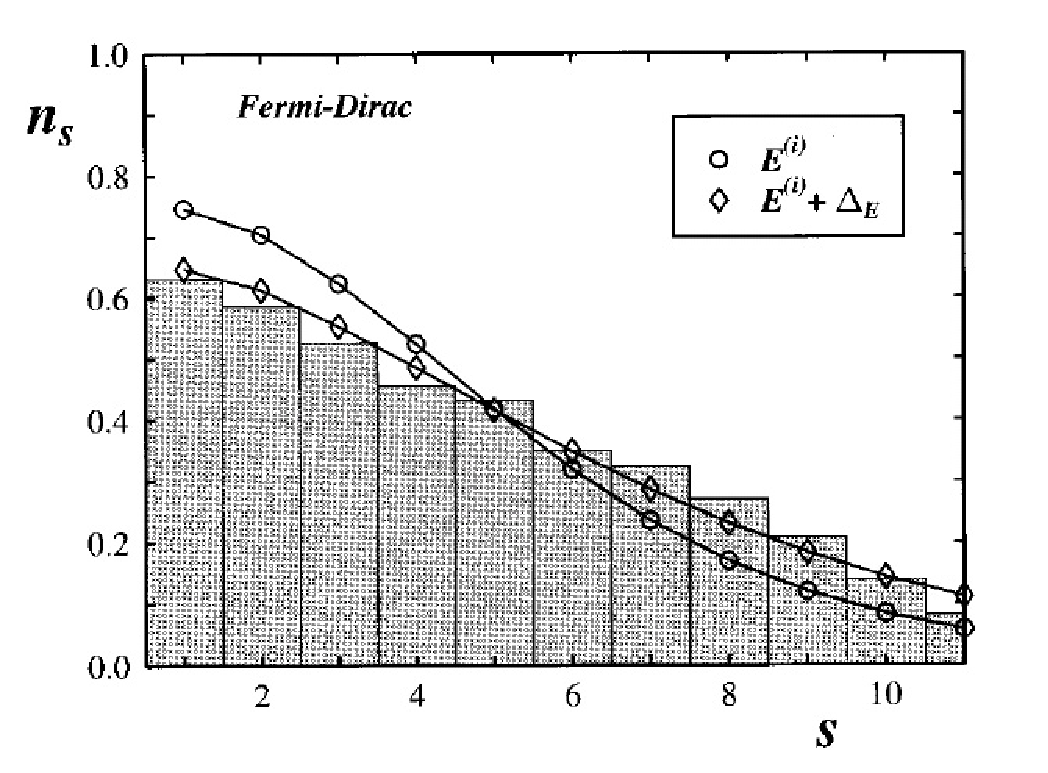}
\caption{The $n_s$-distribution for strongly interacting particles in the TBRI-model (\ref{3})
for 4 fermions on 11 orbitals and $\epsilon_{s}=s+1/s$ as single-particle energies. The histograms are obtained by averaging over a number of eigenstates with energies $E \approx 17.33 \gg E_g = 6.08$ which is much higher than the energy $E_g$ of the ground level. Left: weak interaction with $V=0.02$.  Full diamonds correspond to the standard Fermi-Dirac distribution and stars stand for $n_s$ computed by Eq.~(\ref{42}) with the $F-$function (\ref{43}) taken in the Breit-Wigner form. Right: strong interaction with $V=0.20$. Open diamonds correspond to the shifted energy in Eq.~(\ref{54}) with the total energy $E$ as the energy of eigenstates; the shift is defined by Eq.~(\ref{55}). Circles stay for the Fermi-Dirac distribution obtained from Eq.~(\ref{54}), however, without the shift $\delta_E$ (after \cite{FI97b}).}
\label{FI-fig3}
\end{figure}

This can be illustrated by the results of Refs.~\cite{Flambaum1996b,FI97a,FI97b} in the TBRI-model for just four fermions occupying eleven orbitals, see Fig.~\ref{FI-fig3}. For sufficiently weak interaction, the numerically obtained $n_s$-distribution is clearly different from the Fermi-Dirac distribution (full diamonds on left panel), being even not a monotonic function of the orbital energy $\epsilon_s$. In this case the averaging did not wash out the fluctuations in $n_s$ and the whole picture strongly fluctuates when changing the energy $E^\alpha$ of an eigenstate (see left panel in Fig.~15).

However, for strong interaction resulting in a large value $N_{pc}\gg 1$, the equilibrium
$n_s$-distribution emerges (see right panel in Fig.~15). In this case the fluctuations of the eigenstate components $C_k^{(\alpha)}$ have a Gaussian shape with respect to an envelope which is a smooth function of $E^{\alpha}$. This leads to small fluctuations of the occupations numbers $n_s$ in
accordance with the central limit theorem for the sum (\ref{41}), namely, with $\Delta n_s/n_s\sim N_{{\rm pc}}^{-1/2}\ll 1$ for $n_s\sim 1$. In this region the value of $N_{{\rm pc}}$ can be estimated as $N_{{\rm pc}}\sim \Gamma /D$ where $D$ is the mean level spacing for many-body energy levels. As a result, the $n_s-$distribution evolves smoothly when changing the energy of the system. Such a situation can be associated with the onset of thermal equilibrium and the emergence of the Fermi-Dirac distribution, see Fig.~\ref{FI-fig3}.

The data shown in Fig.~\ref{FI-fig3} can be compared with the canonical form of the Fermi-Dirac
distribution,
\begin{equation}
n_s=\frac{1}{1+\exp \left( \alpha +\beta \epsilon_s\right) }.     \label{53}
\end{equation}
In Eq.~(\ref{38}), the parameters equivalent to $\alpha$ and $\beta$ in (\ref{53}) were found for each individual many-body state $|\alpha\rangle$ of  a strongly interacting small nuclear system as local equivalents of the chemical potential and temperature which are smooth functions of energy. According to  Ref.~\cite{FI97b}, the parameters $\alpha$ and $\beta$ can be found by solving the following system,
\begin{equation}
\sum_sn_s=n,\,\,\,\,\,\,\,\,\,\,\,\,\,\,\sum_s\epsilon_s n_s=E+\delta_E,    \label{54}
\end{equation}
where the shift $\delta_E$ is given by by
\begin{equation}
\delta_E=\langle  E_k^{\alpha}\rangle- E^{\alpha} \simeq \frac{d\ln \rho_0}{dE} \sigma_k^2\simeq
\frac{\sigma_k^2}{\sigma _0^2}( E_c-E).                                             \label{55}
\end{equation}
Here $\sigma_k^2$ is defined by Eq.~(\ref{13}), $E_c$ is the centroid of the energy spectrum, and $\sigma_0^2$  is the variance of the unperturbed density of states for the Hamiltonian $H_0$. The parameter $\delta_E$ represents the difference between the energy of an individual state $E^{\alpha}$ and the mean energy of the system, taking into account that $\langle  E_k^{\alpha}\rangle- E^{\alpha} \neq 0$ away from the centroid $\langle  E_k^{\alpha}\rangle \approx \int F_k^{\alpha}(E_k,E^{\alpha})\rho (E^{\alpha})dE^{\alpha}$. The correction $\delta_E$ emerges since the interaction shifts asymmetrically the unperturbed levels that are lower and higher than energy $E_k$. In this way we take into account the modification of the energy spectrum due to the interaction between the particles; for details, see Refs.\cite{FI97b,FI97a}. The data in Fig.~\ref{FI-fig3} are compared with the results given in Eq.~(\ref{54}) with and without the energy shift. One can see that when the influence of the interaction is quite strong the energy shift should be taken into account in order to get the correct results.

An instructive example of the emergence of  Fermi-Dirac distribution in a realistic physical system,
namely, for multicharged gold ions Au$^{25+}$ near the ionization threshold, is given in
Ref.~\cite{GGF99}. With the use of an appropriate mean field, the authors were able to relate
excited eigenstates and spectra to the radiative capture of low-energy electrons. Moreover,
the enhancement of the recombination over the direct radiative recombination was found that explains experimental data. The situation is quite similar to the radiative neutron capture by complex nuclei in the $(n,\gamma)$-reaction, where the resonance mechanism emerging due to chaotic compound states is also much stronger than the direct capture \cite{FS84}.
\begin{figure}[!htb]
\centering
\includegraphics[width=0.6\textwidth]{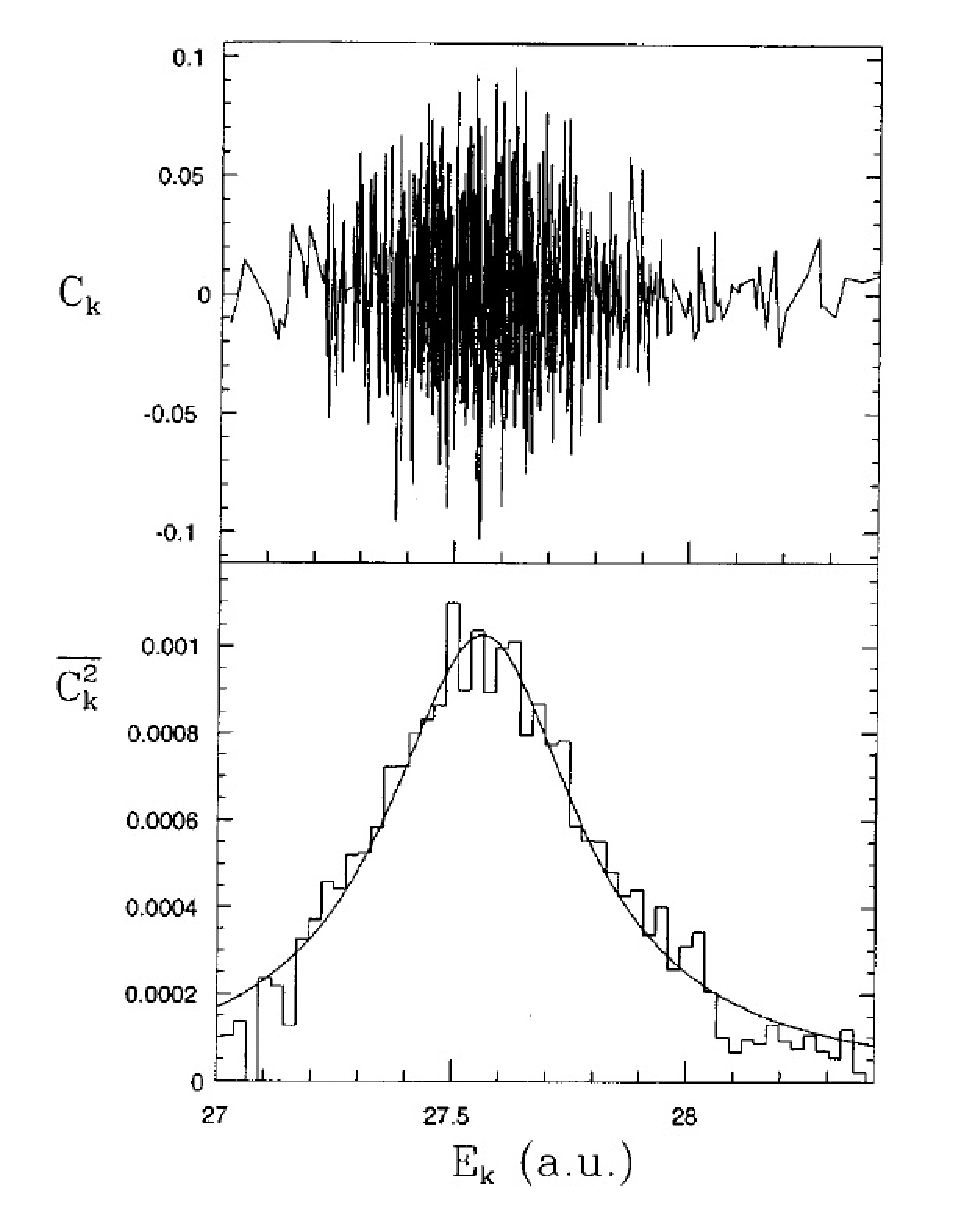}
\caption{Top: components of the 590th eigenstate $J^{\Pi} = 13/2$, and the least-square fit of $C_k^2(E)$ to the Breit-Wigner formula (bottom) (after \cite{GGF99}).}
\label{EFforGold}
\end{figure}
\begin{figure}[!ht]
\centering
\includegraphics[width=0.5\textwidth]{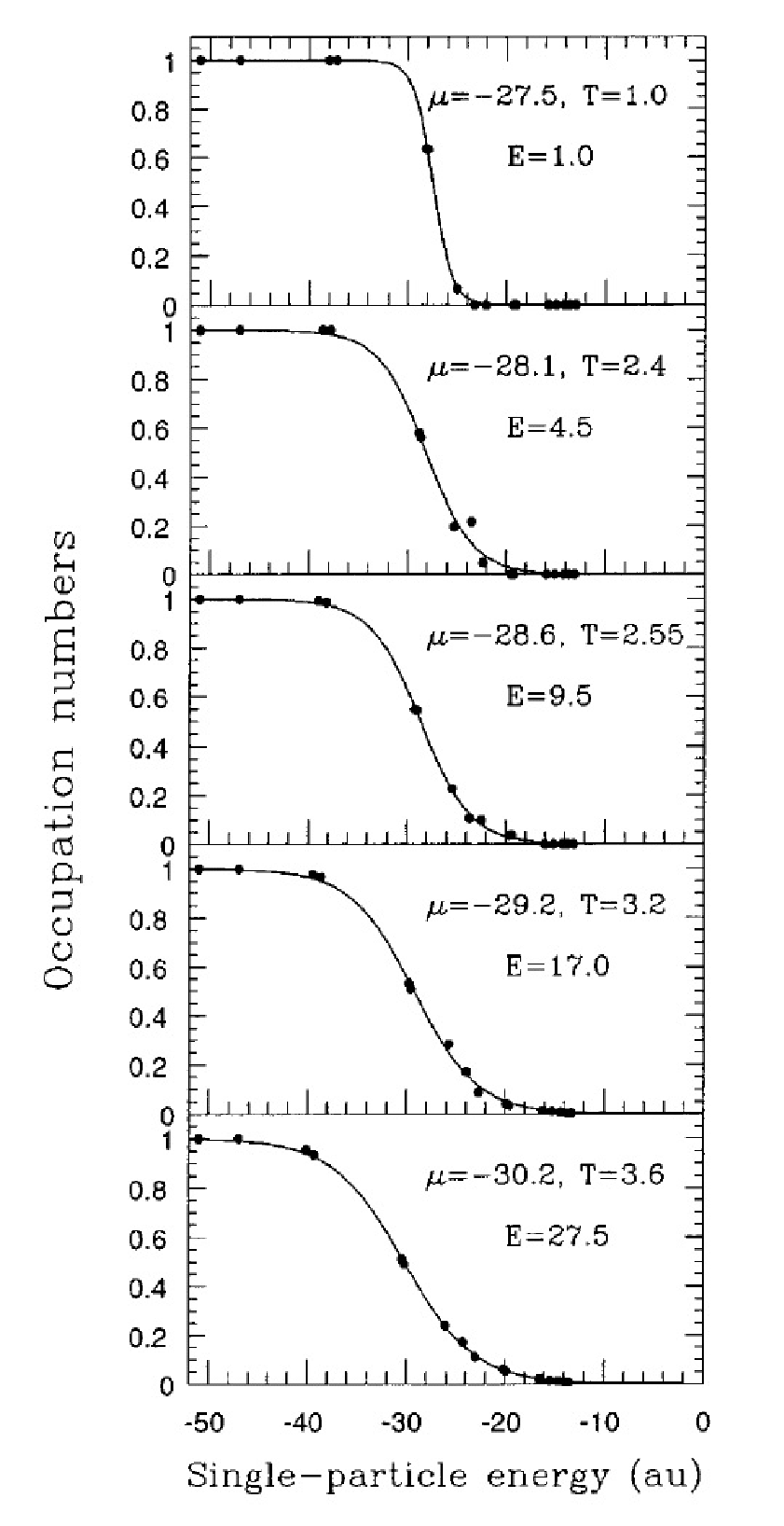}
\caption{Occupation numbers in Au$^{25+}$ calculated numerically according to Eq.(\ref{41}) at high excitation energies $E$. The temperature $T$ and the chemical potential $\mu$ are obtained with the best fits of the data to the Fermi-Dirac distribution (after \cite{GGF99}).}
\label{FDforGold}
\end{figure}

A typical example of the structure of a highly excited eigenstate of Au$^{25+}$ is presented in Fig.~\ref{EFforGold}. This eigenstate is obtained by the diagonalization of the Hamiltonian matrix of size $N=1254$ for $J^{\Pi}=(13/2)^{-}$. Its shape in the energy representation was found to have good correspondence with the Breit-Wigner formula. For such kind  of eigenstates, the statistical approach is expected to give reliable results for the $n_s$-distribution in the Fermi-Dirac form. This is confirmed in Fig.~\ref{FDforGold} where  few $n_s$-distributions obtained from {\it individual} eigenstates of $Au^{25+}$ are shown, as well as by similar results for the nuclear shell model, see Fig.~13. Note that, similarly to the earlier discussed nuclear shell model, the calculations are performed for fully deterministic systems with no random parameters in the Hamiltonian, so that the validity of the statistical description is based exclusively on the chaotic structure of the eigenstates.

\subsection{Different temperatures}

For a relatively small number of particles the temperature $T$ defined by the Fermi-Dirac distribution according to Eq.~(\ref{54}) can be different from both the {\it thermodynamic temperature}
$T_{{\rm th}}$, Eqs. (\ref{39}) and  (\ref{50}) and the {\it canonical temperature} $T_{{\rm can}}$ defined by the canonical distribution (\ref{44}). Different temperature scales and different appropriate thermometers already appeared in relation to Fig.~14.

\begin{figure}[!ht]
\centering
\includegraphics[width=0.6\textwidth]{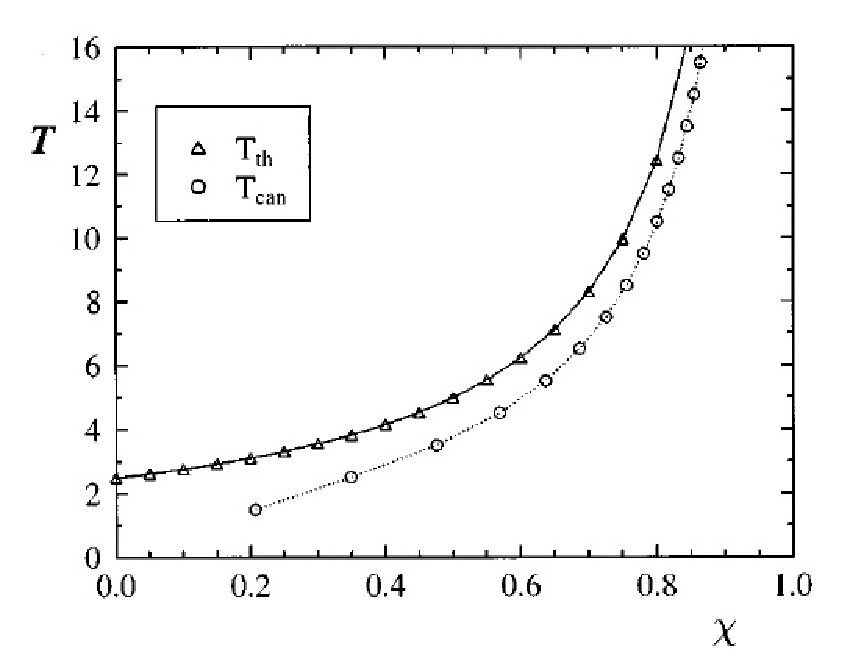}
\caption{Different temperatures versus the rescaled energy $\chi$ for the TBRI-model (\ref{3}) with 4 particles in 11 orbitals. Triangles stand for the thermodynamic temperature $T_{{\rm th}}$ and circles for canonical temperature $T_{{\rm can}}$. The data are obtained for $V=0.12$ and the mean spacing between single-particle levels $d_0=1$ (after \cite{FI97b}).}
\label{two-temp}
\end{figure}

Let us discuss the problem of different temperatures in more detail. Assuming a Gaussian  density of states $\rho(E)$, the thermodynamic temperature is given by,
$T_{{\rm th}}(E)=\sigma^2/(E_c-E)$ where $E_c$ and $\sigma$ are the centroid and the width of
the distribution $\rho(E)$. In Ref.~\cite{FI97b} it was found that the canonical temperature $T_{{\rm can}}$ can be expressed as
\begin{equation}
T_{{\rm can}}= \frac{\sigma^2}{E_c-E+\Delta},     \label{55}
\end{equation}
where the shift $\Delta$  is proportional to the width of the function $\Phi_T(E)$ defined
by Eq.~(\ref{48}). The difference between these two equations of state, $T_{{\rm th}}(E)$ and $T_{{\rm can}}(E)$, disappears for highly excited eigenstates, or in systems with a sufficiently large number $n$ of active particles. In finite systems, such as quantum dots, atoms, or nuclei, the number of active particles (particles in an open shell) is not large. For example, for the cerium atom we have $n=4$ \cite{FGGK94} and in the nuclear shell model for $^{28}$Si \cite{Horoi1995,
HZB95,ZelevinskyRep1996} $n=12$. Therefore, the correction to the thermodynamic temperature can be significant, especially at low energy. The energy dependence of the temperatures $T_{{\rm th}}$ and $T_{{\rm can}}$ for the TBRI-model with four interacting fermions occupying eleven orbitals is shown in Fig.~\ref{two-temp}. The comparison of the two temperatures reveals a noticeable difference for all rescaled energies $\chi=(E-E_{{\rm fermi}})/(E_c-E_{{\rm fermi}})$.

In the TBRI-model the inter-particle interaction is assumed to be completely random. As was argued, this model demonstrates generic properties that typically emerge in many-electron atoms, nuclei, and in other mesoscopic systems of interacting particles. In atomic and nuclear systems one can use realistic interactions as it was done, for example, in Figs.~11 and 12. Due to the complicated character of the interaction between particles at high level density, it is  plausible to expect that the genuine deterministic interaction can be substituted by random matrix elements. Now, it is quite instructive to see what is going on if the interaction is deterministic and the number of particles is very small. In this case the emergence of thermalization, especially, the notion of temperature, is far from being obvious.

 A detailed study \cite{Borgonovi1998}, however, shows that one can speak about thermalization and different temperatures even for a system of two interacting particles. The model here describes two coupled rotators with angular momenta vectors ${\bf L}$ and ${\bf M}$ and a simple Hamiltonian,
\begin{equation}
H=H_0+V=(L_z+M_z)+L_x M_x.       \label{56}
\end{equation}
A similar model is also considered for two interacting quasi-spins in the pairing problem of nuclear physics. The classical analog of this model thoroughly studied in Refs. \cite{FP83,FMP84,P84}
shows  strongly chaotic motion when the absolute values of both spins are large, $L \sim M\gg 1$.
 Therefore, one can expect that in this regime the quantum counterpart
 reveals the properties of quantum chaos.

\begin{figure}[!ht]
\centering
\includegraphics[width=0.5\textwidth]{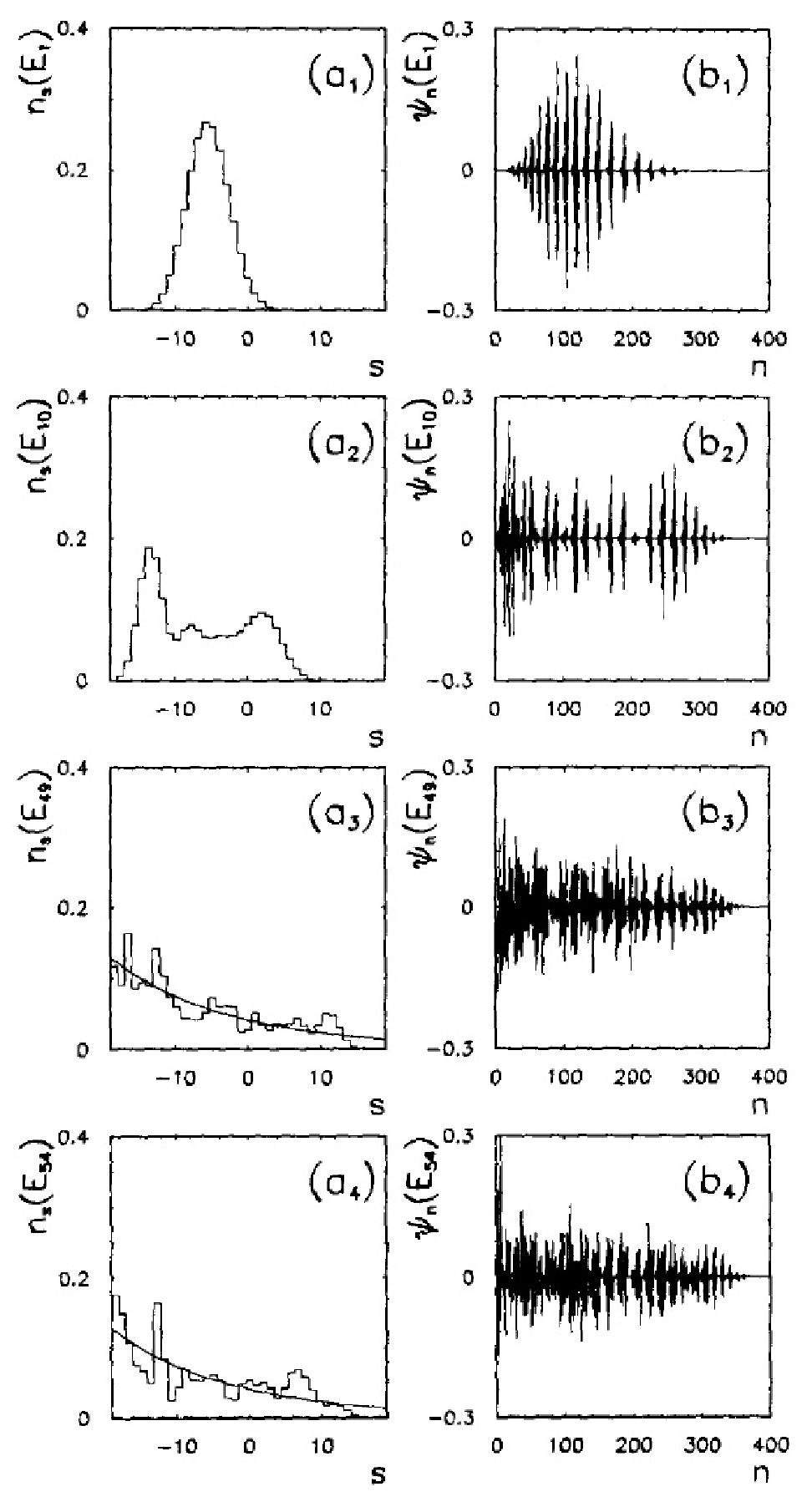}
\caption{Eigenfunctions (right column) and the corresponding occupation number distribution (left column) for $L=3.5, \, \ell=19$. The panels ($a_1, b_1$) are for the ground state, ($a_2, b_2$)
for the 10th state (classically integrable) , ($a_3, b_3$) are for the 49th state (with a chaotic phase space in the classical limit), and ($a_4, b_4$) correspond to the 54th state (with fully chaotic phase space).  Full curves on left panels are obtained by the best fit to the Bose-Einstein distribution  (after \cite{Borgonovi1998}).}
\label{EF-faus}
\end{figure}

The numerical study \cite{BGI98,Borgonovi1998} was restricted by the values $M=L$ and $1 < L < 10$. Typically, when the energy $|E|$ is close to the maximal value $E_{{\rm max}}=L^2+1$, the classical
 trajectories are regular, while for $E \simeq 0$ (at the center of the energy spectrum) the islands
of stability become very small and chaotic motion dominates. In the quantum problem the
angular momenta are quantized according to $L^2=M^2=\hbar ^2 \ell(\ell+1)$ where $\ell$ is
integer.
The Hamiltonian matrix is  more sparse than in the TBRI-model,
even if its  size can be large (${\cal N}=2\ell^2+2\ell+1$).
The properties of the eigenstates and the strength function were studied in great detail in Ref. \cite{BGI98}, showing the emergence of chaotic eigenstates in the region of classical chaos. Fig. \ref{EF-faus} shows few individual eigenstates and the distribution of occupation numbers obtained for these eigenstates. The Hamiltonian was presented in the symmetrized two-particle basis of non-interacting particles. As a result, the $n_s$-distribution has to be compared with the Bose-Einstein distribution.

As seen in Fig.~\ref{EF-faus}, the ground state eigenfunction, panel ($b_1$), even
if characterized  by a large number of components, has clear correlations which do not
allow for a statistical description of $n_s$. Indeed, the actual distribution of occupation numbers, panel ($a_1$) has a very specific form. The same happens for the  energy level $E_{10}$, see
 panels ($a_2$) and ($b_2$).
With an increase of the energy $E$, the structure of eigenfunctions becomes complicated, and the Bose-Einstein distribution emerges (and becomes close to the Boltzmann distribution due to the large values of quantum numbers). These results are important in view of two points. First, in correspondence with the basic principles of statistical physics, the conventional statistical properties of observables (in our case, the occupation number distribution) emerge on the ground of individual eigenstates. Second, even for a simple model of two particles with a deterministic interaction and well defined classical limit, the statistical description works quite well as expected from the quantum-classical correspondence.

\begin{figure}[!ht]
\centering
\includegraphics[width=0.5\textwidth]{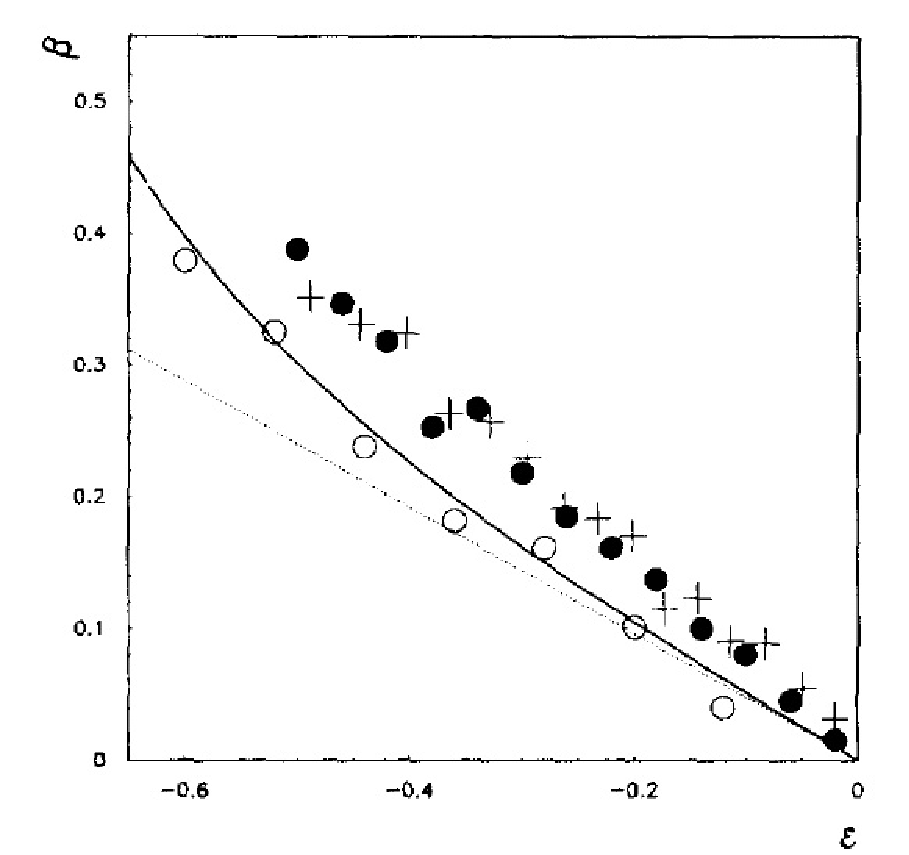}
\caption{Different inverse temperatures versus rescaled energy $\epsilon=E/E_{{\rm max}}$. Full curve:
$\beta_{{\rm can}}=1/T_{{\rm can}}$; dotted curve: $\beta_{{\rm th}}=1/T_{{\rm th}}$, full circles:
$\beta_{{\rm fit}}=1/T_{{\rm fit}}$, and crosses: $\beta_{{\rm BE}}=1/T_{{\rm BE}}$. Open circles stand for $\beta=1/T_{{\rm fit}}$ for random model, (after \cite{Borgonovi1998}).}
\label{temp-faus}
\end{figure}

Now, in view of our previous discussion of different temperatures, let us look at the data presented in Fig.~\ref{temp-faus}. One can see that the thermodynamic inverse temperature, $\beta_{{\rm th}}$ (full curve) computed by Eq.~(\ref{50}) with the use of the Gaussian density of states (that is well confirmed numerically), is very different from $\beta_{{\rm BE}}$ (crosses). The latter was found by solving two equations,
\begin{equation}
\sum_{s=-\ell}^{\ell} n_s^{BE}=2, \,\,\,\, \sum_{s=-\ell}^{\ell}\hbar s n_s^{BE}={\cal E},   \label{57}
\end{equation}
for temperature and chemical potential. Here the energy ${\cal E}=\sum_s \hbar s n_s$ is computed from the numerical values of the actual $n_s$. The $n_s-$distribution found in accordance with Eq.~(\ref{57}) fully takes into account the effect of the inter-particle interaction. The fact that the canonical temperature $T_{{\rm can}}$ does not correspond to the actual $n_s$-distribution is not surprising since for two particles one cannot expect the coincidence of microcanonical and canonical averages.

The temperature $T_{{\rm BE}}$ found from Eq.~(\ref{57}) turns out to be close to that defined from the one-parameter fit to the Bose-Einstein distribution with the only constraint of a finite number of particles,
$\sum_{-\ell}^{\ell} n_s^{{\rm BE}}=2$. The corresponding inverse temperature, $\beta_{{\rm fit}}$ (full circles), is close to $\beta_{{\rm BE}}=1/T_{{\rm BE}}$. This supports, on one hand, the significance of the fitting procedure with the Bose-Einstein distribution, and, on the other hand, the validity of this distribution for isolated systems via a proper renormalization of the energy ${\cal E}$, see previous Section.

An interesting observation in Ref. \cite{Borgonovi1998} refers to the ``randomized" model created by replacing non-zero off-diagonal matrix elements in the Hamiltonian by random variables with the same mean and variance as in the original deterministic model (\ref{56}). In this way we destroy the inherent correlations between matrix elements, however keeping the positions of the zero matrix elements that are due to the selection rules. For such a modified model, the inverse canonical temperature, $\beta_{{\rm can}}$ (open circles) did not practically change, while both, $\beta_{{\rm fit}}$ and $\beta_{{\rm BE}}\approx \beta_{{\rm fit}}$ considerably change and come closer to the canonical inverse temperature, $\beta_{{\rm can}}$. This indicates that the conventional canonical distribution of occupation numbers may appear even in isolated systems with a small number of particles, provided the dynamical correlations are weak due to a complicated interaction.

As for the standard thermodynamic inverse temperature defined by $\beta_T=d \ln \rho /dE$ with $\rho$ as the density of states of the total Hamiltonian, the dependence $\beta_{T}(E)$ (dotted line) is different from $\beta_{{\rm can}}$; however, it comes closer near the center of the energy spectrum. This result was attributed in Ref. \cite{Borgonovi1998} to the small number of particles. For the randomized model, $\beta_{{\rm th}}$ is close to $\beta_{{\rm fit}}(E)$ at high temperature where $\beta_{{\rm th}}$ and $\beta_{{\rm can}}$ tend to coincide; however, it deviates significantly at lower temperature. Note that near the edges of the spectrum the density of states is small, and the eigenstates are not chaotic any more.


\section{Statistical relaxation}

Above we have discussed the interplay of chaoticity and thermodynamics applied to the set of stationary states of a many-body Hamiltonian. We now turn our attention to the theory of
{\sl non-equilibrium} phenomena that was extensively developed during the last decades, being particularly influenced by the availability of new experimentally accessible systems, such as
optical lattices with a small number of atoms, and by the progress of quantum informatics. The notions of quantum fidelity \cite{Misz2009} quantum quench \cite{zyczkowski08,gleizer03,zurek85,kibble07} and quantum annealing \cite{santoro06,Das2008} require time-dependent descriptions, where echo-like returns \cite{Combescure2005,Gorin2006} and possible quantum phase transitions display interesting physics \cite{Polkovnikov2011RMP}. The method most closely related to our description operates with the so-called diagonal entropy that is a cousin of the information entropy. Many physical problems here are still not fully resolved. It was noticed long ago \cite{balian91} that, under an adiabatic evolution of the interacting systems, the characteristic time scales are much shorter than the Weisskopf time $\hbar/D$ corresponding to the inverse distance between the microscopic many-body energy levels and therefore to the characteristic periods of wave packets in this Hilbert space. Our previous considerations allow to hypothesize that the Weisskopf time describes the closest return of the wave packet, while it is not needed for statistical equilibrium: indeed, if all states in a given energy window ``look the same", the system does not need to probe every one of them.

Advances in the theory of non-equilibrium quantum dynamics have benefited from new experiments with highly controllable interacting quantum systems, where the evolution remains coherent
for a long time. They include solid-state~\cite{Cappellaro2007,Ramanathan2011,Kaur2013} and
liquid-state~\cite{Batalhao2014} NMR platforms, ultracold atoms, and molecules in optical
lattices~\cite{Bloch2008,Trotzky2008,Chen2011,Trotzky2012,Fukuhara2013,Yan2013,Hild2014}.
In the first case, the experiments are performed at room temperature and the dynamics is controlled with magnetic pulses. In the latter case, selected initial states can be prepared, and the strength of the interactions and the level of disorder can be engineered. In parallel to experiments, new numerical methods, such as those based on matrix product states and density-matrix
renormalization group (DMRG) ~\cite{Schollwock2005,Schollwock2011,Banuls2011} as well as
linked-cluster approaches~\cite{Rigol2014}, have made possible the simulations of the dynamics of large quantum systems and the analysis of infinite time averages in the thermodynamic limit.

Non-equilibrium quantum physics is a highly interdisciplinary subject. It reveals, for example, a symbiotic relationship with quantum information science. Some illustrations include the growth
of entanglement as a justifier for the limitation to short times of the time-dependent DMRG method~\cite{Eisert2010}; the manipulation of the dynamics of interacting quantum systems~\cite{Rego2009,Ramanathan2011,Khodjasteh2013} that emerged from the need to transfer and store information for quantum communication; and the further advances of quantum simulators
for adiabatic quantum computation and quantum annealing~\cite{santoro06,Das2008}. Non-equilibrium phenomena are also highly connected with the  characterization of quantum
transport behavior, whether it is ballistic or diffusive~\cite{Zotos1999,Karrasch2013}; with new physics emerging at the points of quantum phase transitions~\cite{Zurek1996,kibble07,Polkovnikov2011RMP}; and with the viability of thermalization, see Sec.~\ref{Sec:Equilibration}.

In this section, in order to study the statistical relaxation of observables, we concentrate on the
dynamics of isolated quantum systems prepared in an eigenstate of an initial (unperturbed) Hamiltonian $H_0$ and evolving by a full  Hamiltonian $H=H_0 +V$. This scenario is often referred to as {\sl quenched} dynamics; it corresponds to an (almost) instantaneous perturbation that converts $H_0$ into the final Hamiltonian $H$ during a time  interval much shorter than any characteristic time scale of the system. We will also discuss the case,  relevant for recent experimental studies, when the initial state is not an eigenstate of $H_0$. In our context,  this means that the initial state is not an eigenstate of the mean field basis, even if  analytical results are mainly available for this situation.

\subsection{Survival probability: theoretical background}
\label{SubSec:Return}

As a simple case of quench dynamics, we consider the situation where the initial state  is a basis state $|k_0\rangle$, i.e. an eigenstate of the unperturbed Hamiltonian $H_0$ in Eq.~(\ref{1}). The evolution of the state vector  is described by
\begin{equation}
|\Psi (t) \rangle  =\sum\limits_{\alpha}
C_{k_0}^\alpha|\alpha\rangle \exp(-iE^\alpha t),          \label{58}
\end{equation}
where the coefficients $C_{k_0}^\alpha = \langle \alpha |k_0\rangle$ are obtained from the expansion of the initial state over the stationary states $|\alpha \rangle$  of the total Hamiltonian $H=H_0+V$ and
$E^\alpha$ is the energy eigenvalue corresponding to the eigenstate $|\alpha\rangle$. The probability
\begin{equation}
w_{kk_{0}}=|A_{kk_{0}}|^2 =|\left\langle k|\Psi(t)\right\rangle|^2   \label{59}
\end{equation}
to find the system at time $t$ in the state $|k\rangle$ is determined by the amplitude
\begin{equation}
A_{kk_{0}}(t)= \left\langle k|\exp(-iHt)|k_{0}\right\rangle=
\sum\limits_\alpha C_k^{\alpha\ast}C^{\alpha}_{k_{0}}\exp(-i E^\alpha t).   \label{60}
\end{equation}

A quantity of particular interest is the {\it survival probability} $W_0(t)$ to find the system at time $t$ in the initial state $|k_0\rangle$,
\begin{equation}
W_0(t)\equiv w_{k_0 k_0}(t)=\left| A_{k_0 k_0}(t)\right| ^2,          \label{61}
\end{equation}
where we come to the Fourier image of the strength function,
\begin{equation}
A_{k_0 k_0}(t)=\sum\limits_\alpha|C_{k_0}^\alpha|^2\exp(-iE^\alpha t)
= \int F_{k_0}(E)\exp(-i Et)dE. \label{62}
\end{equation}
As seen earlier, for chaotic eigenstates the strength function $F_{k_0}(E)$ is a smooth function of energy which is the key point in the analytical evaluation of the time dependence of $W_0(t)$.

Numerical and analytical results for the survival probability in the TBRI model are considered
in Sec.~\ref{SubSec:Return} and for spin-1/2 models in Sec.~\ref{SubSec:ReturnSpin} . This quantity, also known as return probability and quantum fidelity, formally coincides with the Loschmidt echo (see reviews \cite{Gorin2006,Arseni2012})
when the two Hamiltonians involved in the echo are taken as $H_0$ and $H_0+V$.

According to perturbation theory, at short times the general result is valid,
\begin{equation}
W_0(t)\approx 1-\sigma_{k_0}^2t^2,           \label{64}
\end{equation}
where $\sigma_{k_0}^2$ is the energy variance of the initial state $|k_{0}\rangle$ projected onto the basis generated by the total Hamiltonian $H$, see Eq.~(\ref{13}). This variance defines the width of energy shell, and can be determined, prior to the diagonalization, by the sum of squares of off-diagonal matrix elements of the matrix of $H$ in the basis of $H_{0}$, namely, as $\sigma_{k_0}^2 = \sum_{k\neq k_0} V_{k_{0}k}^2.$ .
This expression is universal in the sense that it is exact for any kind of perturbation $V$ \cite{FI97b}. The initial time scale for the perturbative expression (\ref{64}) to be valid is very small, and the main interest is in the time-dependence of $W_0(t)$ beyond this time scale, which is entirely determined by the form of the strength function.
A similar initial stage is formally present in the usual description of irreversible radioactive decay but in such applications, as a rule, the initial moment of creation of an unstable state cannot be known precisely enough \cite{terentiev,peshkin14}.

As discussed above, in generic systems of interacting particles the strength function takes the Breit-Wigner form (\ref{20}), provided the interaction is relatively strong. The half-width $\Gamma_0$ of the BW is given by the Fermi golden rule and in our case reads as
\begin{equation}
\Gamma_0(E)\equiv \Gamma(E,k_0) \simeq 2\pi
\overline{\left| V_{k_0k}\right| ^2} \rho_f(E).                 \label{65}
\end{equation}
Here $\overline{\left| V_{k_0k}\right| ^2}$ is the mean square value of many-body matrix elements obtained by the average over $k$, and $\rho_f(E)$ is the density of the basis states directly
coupled to the state $|k_0\rangle$ by the interaction $V$. For TBRI, this density can be roughly evaluated as follows: the basis state $|k_0\rangle$ has $N_0$ non-zero matrix elements $\langle {k_0}'|V|k_0\rangle$ for ${k_0}' \ne k_0$. Let us define $k_m = {\rm Min}[{k_0}']$ and $k_M = {\rm Max}[{k_0}']$ taken over the $N_0$ values for given ${k_0}$. The values $k_m$ and $k_M$ determine the width of interaction in the chosen unperturbed many-body basis. One can associate with these two values the corresponding eigenvalues of the total Hamiltonian $E_{k_M}$ and $E_{k_m}$; in this way we have $\rho_f^{-1} \simeq (E_{k_M}-E_{k_m})/N_0$. It should be stressed that $\rho_f(E)$ is much smaller than the total density $\rho(E)$ of all many-body states due to many vanishing matrix elements $\langle {k_0}'|V|k_0\rangle$. As mentioned earlier, this  is attributed to the two-body nature of the interaction $V$.

Assuming the Breit-Wigner form of the strength function, the survival probability is described by the exponential time dependence,
\begin{equation}
W_0(t)\simeq \exp\left(-\Gamma_0 t\right),              \label{66}
\end{equation}
apart from a short-time regime $t < t_0$ when the quadratic decrease (\ref{64}) occurs. This situation is similar to what is known for radioactive decay into continuum, when the short initial stage is changed by the conventional exponential decay, see the recent short review article
\cite{peshkin14} and references therein.

For a long time it was assumed that the exponential decrease of the survival probability was the only regime which is physically relevant to the dynamics of systems with many interacting
particles. However, as discussed in Section III, now it is understood that in many situations the form of the strength function can be  close to a Gaussian. This is related to the finite width of the interaction in the energy space reflected by the band-like structure of $V$ in Eq.~(\ref{3}). For the TBRI model it was shown \cite{FI01a} that if $\Gamma_0<\Delta_E$, the form of the strength function, is, indeed,  Breit-Wigner. In the opposite limit, $\Gamma_0 \approx \Delta_E$, of a strong interaction, the leading dependence of $W_0(t)$ is  Gaussian,
\begin{equation}
W_0(t) \simeq \exp (-\sigma_{k_0}^2t^2).                  \label{67}
\end{equation}
This occurs on some scale $0 < t \leq t_c $~\cite{FI01a}, after which an exponential dependence may emerge, see below.

Different functions have been suggested to describe the shape of the strength function in the transition region between Breit-Wigner and Gaussian, for example  Student's $t$-distribution~\cite{Chavda2004,KotaBook}, interpolation of the two functions~\cite{FI00,FI01a}, and Voigt distribution~\cite{Torres2014PRAb}. Although the rigorous analytical description of this crossover is quite complicated, in order to evaluate $W_0(t)$, one can use the phenomenological expression suggested in Ref.~\cite{FI01a} for the strength function that depends on two key parameters, $\Gamma_0$ and $\sigma_{k_0}$, see also Ref.~\cite{IC06}. For the TBRI model with  $N_p$ Fermi  particles in  $N_s$ active orbitals, one has \cite{FI97b},
\begin{equation}
\sigma_{k_0}^2=\frac{v_0^2 }{12}N_p(N_p-1)(N_s-N_p)(N_s-N_p+3),        \label{68}
\end{equation}
where $[-v_0, v_0]$ is the range within which the two-body matrix elements are distributed randomly with a constant probability. The variance $\sigma_{k_0}^2$ turns out to be independent of the specific basis state $\left|k_0 \right \rangle$; as already mentioned, the same
property is practically well satisfied in various versions of the nuclear shell model \cite{ZelevinskyRep1996,pillet12},
where one cannot directly apply eq. (\ref{68}) because of the presence of exactly conserved constants of motion
dividing the Hilbert space into non-overlapping domains.

\begin{figure}[ht]
\centering
\subfigure{\includegraphics[scale=0.33]{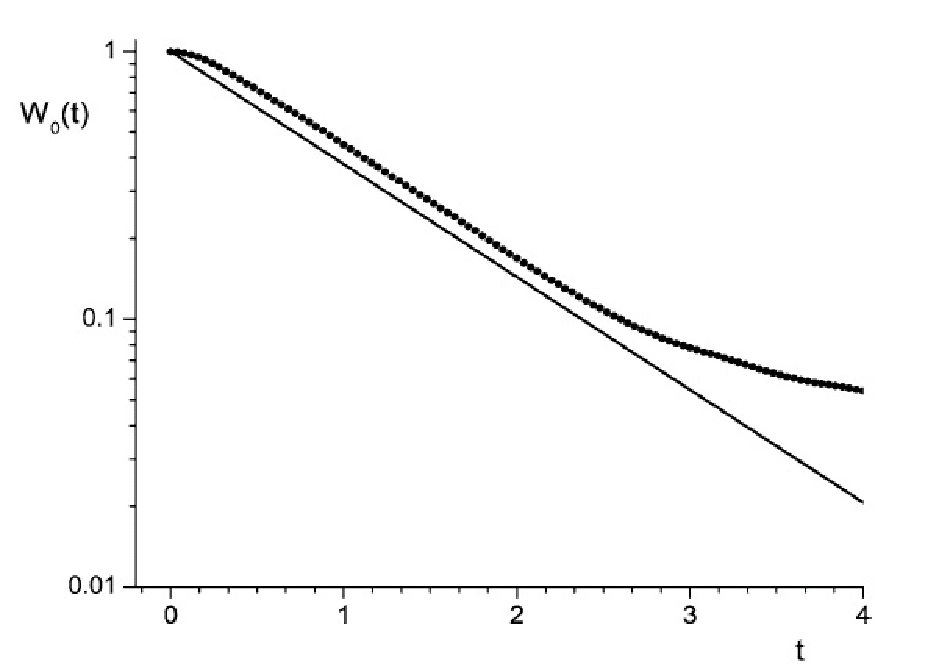}}
\subfigure{\includegraphics[scale=0.33]{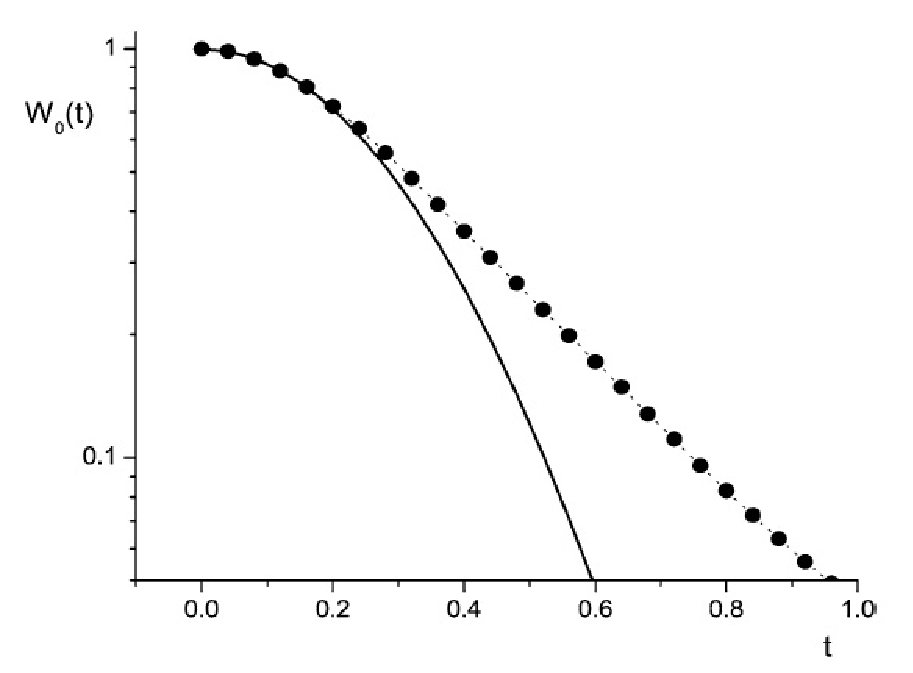}}
\subfigure{\includegraphics[scale=0.33]{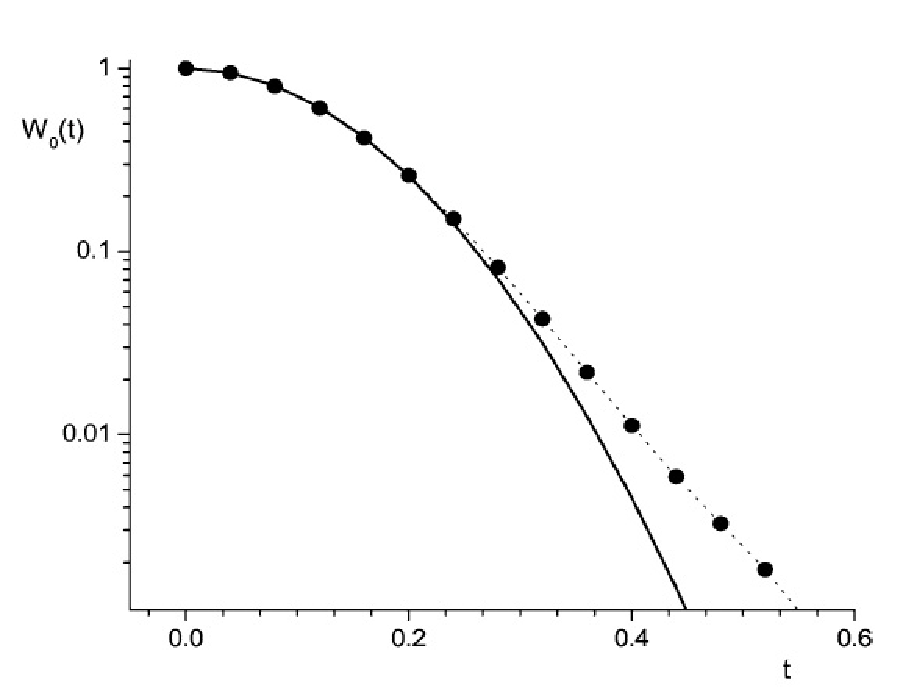}}
\caption{Survival probability $W_0(t)$ for TBRI model (\ref{3}) with 6 particles and 12 single-particle levels. Full circles correspond to numerical data. LEFT:  Weak non-perturbative
interaction, $v_0=0.12$, the strength function has the Breit-Wigner form. The straight line shifted for a better visualization is the linear fit. MIDDLE: Intermediate situation
with $v_0=0.25$ when both regimes, Gaussian and exponential, are characteristic of the dynamics. Solid curve is the theoretical expression (\ref{67}). RIGHT: Strong interaction, $v_0=0.5$, with the Gaussian shape of the strength function. Solid curve is the analytical dependence (\ref{67}) with $\sigma_{k_0}^2$ determined by Eq.~(\ref{68}) (after \cite{IC06}).} \label{W0}
\end{figure}

Thus, in the case of a not too strong perturbation, the decrease of the survival probability is exponential, and with an increase of the interaction $V$, one should expect a long time scale where the Gaussian form (\ref{67}) occurs for $t > t_c$. These predictions are confirmed by numerical data in Fig.~\ref{W0}. Here and below the time $t$ is measured in units $\hbar/d_0$ where $ d_0=\langle\epsilon _{s+1}-\epsilon _s\rangle$ is the mean spacing between single-particle levels $\epsilon_s$. The calculations were performed for $N_p=6$ fermions and $N_s=12$ single-particle states resulting in 924 many-body states. The fit to the exponential dependence (\ref{67}) determines $\Gamma_0 \approx 0.97$ that can be compared with the rough estimate (see details in Ref.~\cite{IC06}),
\begin{equation}
\Gamma_0 \approx
\frac{\pi \sigma_{ k_0}^2}{ d_0 (N_s-N_p)},                              \label{69}
\end{equation}
where $d_0$ is the mean level spacing between single-particle levels. With this estimate one obtains $\Gamma_0 \approx 1.03$ which is very close to the numerical data, see Fig.~\ref{W0}, left panel; here $\Gamma_0$ is measured in units $d_0/\hbar$.

In the intermediate regime between  Breit-Wigner and Gaussian, there are two time scales, see Fig.~\ref{W0}, middle panel. When $t \leq t_c\approx 0.3$, the decrease of $W_0(t)$ has the
Gaussian form (\ref{67}), whereas for $t > t_c$ the time dependence changes to exponential. The critical time $t_c$ that divides these two regimes can be estimated as
{\begin{equation}
t_c \approx \frac{\Gamma_0}{\sigma^2_{k_0}}.  \label{70}
\end{equation}
In fact, $t_c$ is the time to resolve the width, $\Delta_E \approx \sigma_{k_0}$, of the energy shell. If this width is very large, the exponential decrease starts on a short time scale. Such a situation is typical for one-body chaos, for example, for a particle in a billiard: since the interaction is due to the hard walls, the energy range of interaction is extremely large (actually, infinite), and the interaction couples all unperturbed states apart from the origin of the energy spectrum. That is why in such models the typical shape of the strength function is Breit-Wigner, independently of the interaction strength, see Ref.~\cite{WAV10}. Contrary, for isolated systems of interacting particles, such as the TBRI model, for relatively small values of $\Delta_E$, the Gaussian decrease of $W_0(t)$ starts from $t\simeq 0$ and lasts for a long time. According to this estimate, we have $t_c \approx 0.5$ which roughly corresponds to the data. The critical time $t_c$ is  not a well defined quantity and can be determined up to a numerical factor of the order unity. Remarkably, the time of the correspondence of the data to Eq.~(\ref{67}) turns out to be independent of the perturbation strength.

Finally, for a relatively strong interaction resulting in the Gaussian shape of the strength function, the numerical data reported in Fig.~\ref{W0} (right panel) manifest a long
Gaussian-like decrease of the survival probability up to very small values of $W_0$, in contrast to the intermediate regime. The deviation from the Gaussian dependence towards the exponential one (linear slope after $t\approx 0.4$) begins at small values of $W_0(t)$. Therefore, practically the decrease of the survival probability is described by the dependence (\ref{67}). The detailed analysis performed in Ref.~\cite{FI01a} has revealed an unexpected result. It was analytically shown that, having an exponential tail for the time after which the Gaussian decay ends, a {\it non-conventional decay} emerges asymptotically for a large time, given by $W_0(t) \approx C \exp (-\Gamma_0 t)$, where the constant $C$ can be very large, $C \approx \exp[\frac{1}{4} \{\Gamma_0 ^2/(\Delta_E)^2\}]$. Such a strong deviation from the conventional decay, $W_0(t) \approx \exp (-\Gamma_0 t)$ can be important in realistic physical systems. It should be noted that the exponential decay cannot last to arbitrarily long time because then the energy uncertainty of the initial state would be infinite. The long-time regime should be of power law or something like that being defined by the threshold, the lowest end of the energy spectrum
\cite{peshkin14}.

\subsection{Survival probability: spin-1/2 systems}
\label{SubSec:ReturnSpin}

Here we focus on realistic,  completely deterministic, spin-1/2 systems [see Eqs.~(\ref{5}) and (\ref{6})].  We show that the analytical approach developed for the TBRI model works also for deterministic models provided the eigenstates of the total Hamiltonian $H$ can be treated as chaotic. Since the survival probability is the Fourier transform of the strength function, the shape and filling of the latter regulate the decay of the first. This allows for the following conclusions~\cite{Torres2014PRA,Torres2014NJP,Torres2014PRE,Torres2014PRAb,Torres2014AIP,TorresKollmarARXIV,TorresLocalization}, which are illustrated below: (i)  the decay for integrable and chaotic Hamiltonians may be very similar; (ii) it may be exponential, Gaussian and even faster than Gaussian, the fastest behavior being limited by the energy-time uncertainty relation;  (iii) it slows down as the energy of the initial state moves away from the middle of the spectrum.

\subsubsection{Integrable vs Chaotic Regime}

The gradual broadening of the strength function as the perturbation increases for the integrable Model 1 and  for the chaotic Model 2 was shown in Fig.~6.
Here, that figure is translated to the non-equilibrium scenario. The initial states are the basis states of the unperturbed Hamiltonian $H_0$. They are evolved by $H$  being labeled by the energies
\begin{equation}
\tilde{E}_{k_0}=\langle k_0 |H |k_0 \rangle                      \label{71}
\end{equation}
chosen away from the edges of the spectrum of the total Hamiltonian $H$. Note that $\tilde{E}_{k_0}$ is equal to $E_{k_0} = \langle k_0 |H_0 |k_0 \rangle$ only at the center of the energy band (see discussion in Ref.\cite{FI97b}). The corresponding decay of the survival probability is illustrated in Fig.~\ref{fig:fidelitySpin}.
\begin{figure}[htb]
\centering
\includegraphics*[width=4.in]{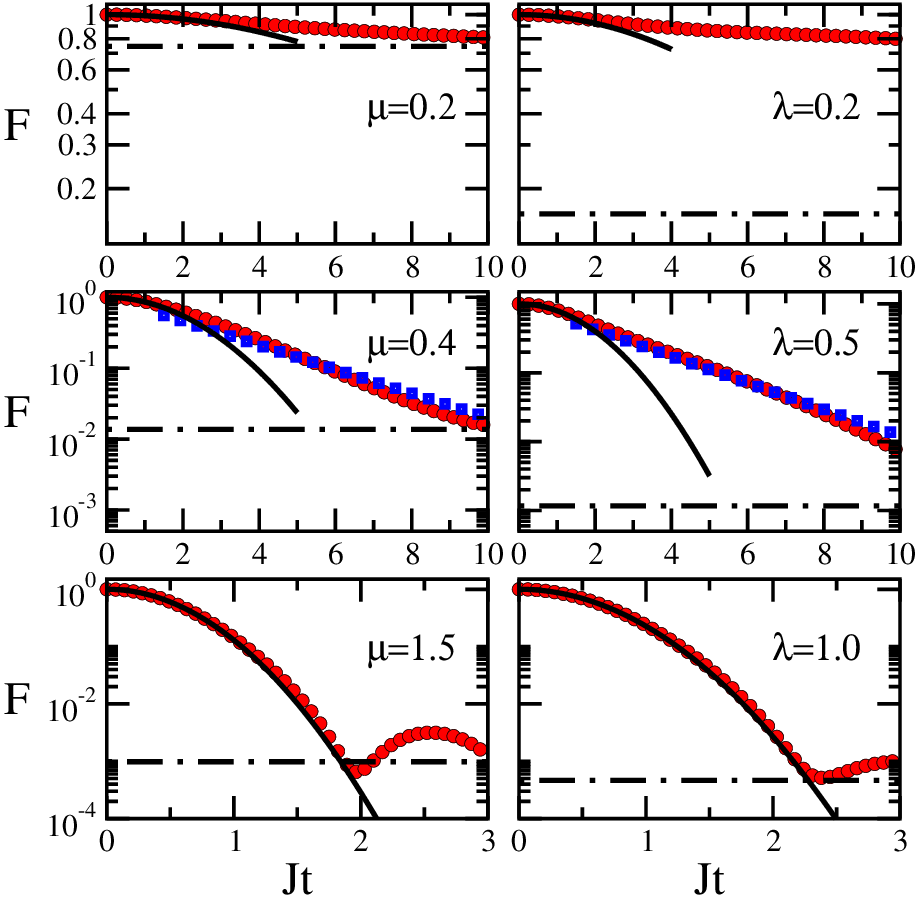}
\caption{(Color online)
Survival probability for Model 1 (left) and Model 2 (right). Numerical results are represented by circles; the analytical Gaussian decay, see Eq.~(\ref{67}), is given by solid curves; the exponentials in the middle panels correspond to squares. The infinite time average is given by dot-dashed horizontal lines.  The energy $\tilde{E}_{k_0}$ of the initial states is away from the edges of the total energy spectrum and closest to $E_T=\sum_\alpha E_\alpha e^{-E_\alpha/T}/\sum_\alpha e^{-E_\alpha/T}$, where $T=4.4J$. This choice and the parameters are the same as in Fig.~6;  $L=18$ and
${\cal S}^z_{tot}=-3$ (after \cite{Torres2014AIP}).}
\label{fig:fidelitySpin}
\end{figure}
When the perturbation is  weak ($\mu=0.2$, $\lambda=0.2$), the strength function is close to a delta function and the decay of the survival probability is very slow [top panels of Fig.~\ref{fig:fidelitySpin}]. As the perturbation increases ($\mu=0.4$, $\lambda=0.4$) and the strength function becomes Breit-Wigner, the decay at very short times is quadratic, as expected from perturbation theory, and then it switches to exponential [middle panels of Fig.~\ref{fig:fidelitySpin}]. When the regime of strong perturbation is reached ($\mu=1.5$, $\lambda=1.0$) and the strength function becomes Gaussian, the decay is also Gaussian [bottom panels of Fig.~\ref{fig:fidelitySpin}] until saturation. Fig.~\ref{fig:fidelitySpin}  confirms that the behavior of the survival probability is similar for both models, integrable and chaotic.

In the same figure, the  horizontal dot-dashed lines correspond to the infinite time average. After the saturation, the survival probability fluctuates around the value
\begin{equation}
\overline{W_0}=\sum_{\alpha} |C^{\alpha}_{k_0} |^4 = M_2(k_0)    \label{72}
\end{equation}
indicated by the dot-dashed horizontal lines. Comparing with Eq.~(\ref{31}), one sees that the value of $\overline W_0$ corresponds to the inverse participation ratio $M_2$ of the unperturbed basis state $|k_0\rangle$  projected onto the total energy basis of $H$. As for the variance of the temporal fluctuations, it is given by \cite{Torres2014NJP,TorresKollmarARXIV}
\begin{eqnarray}
\sigma_{W_0}^2  &=&\mathop{\sum_{\alpha \neq \beta} }_{\gamma \neq \delta}
|C^{\alpha}_{k_0}|^2  |C^{\beta}_{k_0}|^2 |C^{\gamma}_{k_0}|^2  |C^{\delta}_{k_0}|^2
\overline{e^{i (E_{\alpha} - E_{\beta} - E_{\gamma} + E_{\delta}) t}}  \nonumber \\
&=& \left( \sum_{\alpha} |C^{\alpha}_{k_0} |^4 \right)^2 - \sum_{\alpha } |C_{\alpha}^{k_0}|^8 \approx [M_2^(k_0)]^{2}.                                                 \label{73}
\end{eqnarray}
The second equality here holds for systems without too many degeneracies  of energies and also of spacings, that is $E_{\alpha} - E_{\beta} = E_{\gamma} - E_{\delta} \Rightarrow E_{\alpha} = E_{\gamma}$ and  $E_{\beta} = E_{\delta}$. Systems that fall in this category are those without level clustering~\cite{Zangara2013}. They include chaotic models and those with the Poisson level spacing distribution, known to be a fingerprint of integrable systems, but obviously exclude systems characterized by a picket fence spectrum. Defining the time, $t_R$, as the time it takes for the survival probability to first reach the saturation value, a simple expression can be obtained for the cases where the Gaussian decay holds until $\overline{W_0}$ \cite{Torres2014PRA,Torres2014NJP},
\begin{equation}
\exp \left( - \Delta_{k_0}^2  t_R^2 \right) =M_2^{k_0}  \Rightarrow t_R=\frac{\sqrt{\ln M_2^{k_0} }}{\Delta_{k_0} } .                                  \label{74}
\end{equation}
The lifetime of the initial state is therefore determined by its level of delocalization in the energy eigenbasis and by the width $\Delta_{k_0}$ of the strength function.

\subsubsection{Experimentally accessible initial states}
\label{SubSec:ReturnSpinExp}

Above we have studied the survival probability for initial states with average energy $\tilde{E}_{k_0}$ close to the middle of the total energy spectrum.  We restricted our focus to the dependence of the dynamics on the strength of the perturbation. We now fix the strength and analyze the role of the energy $\tilde{E}_{k_0}$. Specifically, we consider as initial states the  site-basis vectors. They correspond to tensor products of the states of each site, where the spin is either pointing up or down in the $z$ direction, such as in $|\downarrow \downarrow \uparrow \downarrow \rangle $. They are configuration basis vectors and not the eigenstates of the unperturbed Hamiltonian $H_0$.  These states can be prepared experimentally with cold atoms in optical lattices~\cite{Weld2009,Trotzky2008,Koetsier2008,Mathy2012}. The perturbation strength required to evolve a site-basis vector according to the total Hamiltonian of Model 1 or 2 is strong, because we are basically quenching the parameter $\mu$ from infinity to a finite value. Thus, a Gaussian strength function is expected.

It is straightforward to calculate  the center $\tilde{E}_{k_0}$ and the width $\Delta_{k_0}$ of the strength function for such initial states~ \cite{Torres2014PRA,Torres2014NJP}. The values are given in Table~\ref{table:initial} for two examples typically investigated in magnetic systems. The first, referred to as the domain wall state, $|\rm{DW}\rangle$, has two well separated regions, in the first half of the chain all spins point up and in the second half all spins point down. The second state, called the N\'eel state, $|\rm{NS}\rangle$, occurs in materials exhibiting antiferromagnetism. Note that the width $\Delta_{k_0}$ does not depend on the anisotropy parameter $\mu$.
\begin{table}[h]
\caption{Energy of $|k_0\rangle$ and width of its strength function.}
\begin{center}
\begin{tabular}{ccc}
\hline
\hline
  &    $\tilde{E}_{k_0}$  & $\Delta_{k_0}$ \\ [0.1 cm]
$|\rm{DW}\rangle =| \uparrow \uparrow \uparrow \ldots \downarrow \downarrow \downarrow \rangle $
   & \hspace{0.2 cm} $\frac{\displaystyle J\mu}{\displaystyle 4} [(L-3) + (L-6)\lambda]$ & $ \frac{\displaystyle J}{\displaystyle 2} \sqrt{1+2\lambda^2}$    \\ [0.2 cm]

$ |\rm{NS}\rangle = |\downarrow \uparrow \downarrow \uparrow  \ldots  \downarrow \uparrow \downarrow  \uparrow \rangle$ & \hspace{0.2 cm} $
 \frac{\displaystyle J\mu}{\displaystyle 4}  [ -(L-1) + (L-2)\lambda ]$ &   $\frac{\displaystyle J}{\displaystyle 2} \sqrt{L-1}$  \\  [0.2 cm]

\hline
\hline
\end{tabular}
\end{center}
\label{table:initial}
\end{table}

The two states of Table 1 show different dependence on $\mu$, $\lambda$, and $L$, with consequences further explored in the figures below.  For example, contrary to
$|\rm{DW}\rangle$, the width for the N\'eel state is independent of $\lambda$ being the same for the integrable Model 1 and the chaotic Model 2. For the domain wall, $\Delta_{k_0}$ is independent of the system size. As $\lambda$ increases ($0\leq \lambda \leq 1$),
the energy of the $ |\rm{NS}\rangle$ state approaches the middle of the spectrum being zero when $\lambda = (L-1)/(L-2)$, whereas, for the $|\rm{DW}\rangle$ state, $\tilde{E}_{k_0}$ moves away from the middle.

In Fig.~\ref{fig:Neel}, the left panels show the Gaussian strength function for the N\'eel state projected onto the integrable Model 1 and chaotic Model 2, and on the right panel the corresponding  behavior of $W_{|\text{NS}\rangle}(t)$. The energy shell is overall well filled, leading to the Gaussian decay up to the saturation line for both models. The curves fall on top of each other, since $\Delta_{|\text{NS}\rangle}$ is the same for both systems, and they agree  well with the analytical expression. The saturation point, indicated with the horizontal dashed line in Fig.~\ref{fig:Neel}, depends on the filling of the energy shell, which improves as $\tilde{E}_{|\text{NS}\rangle}$ approaches the middle of the spectrum.

\begin{figure}[htb]
\centering
\includegraphics*[width=4.in]{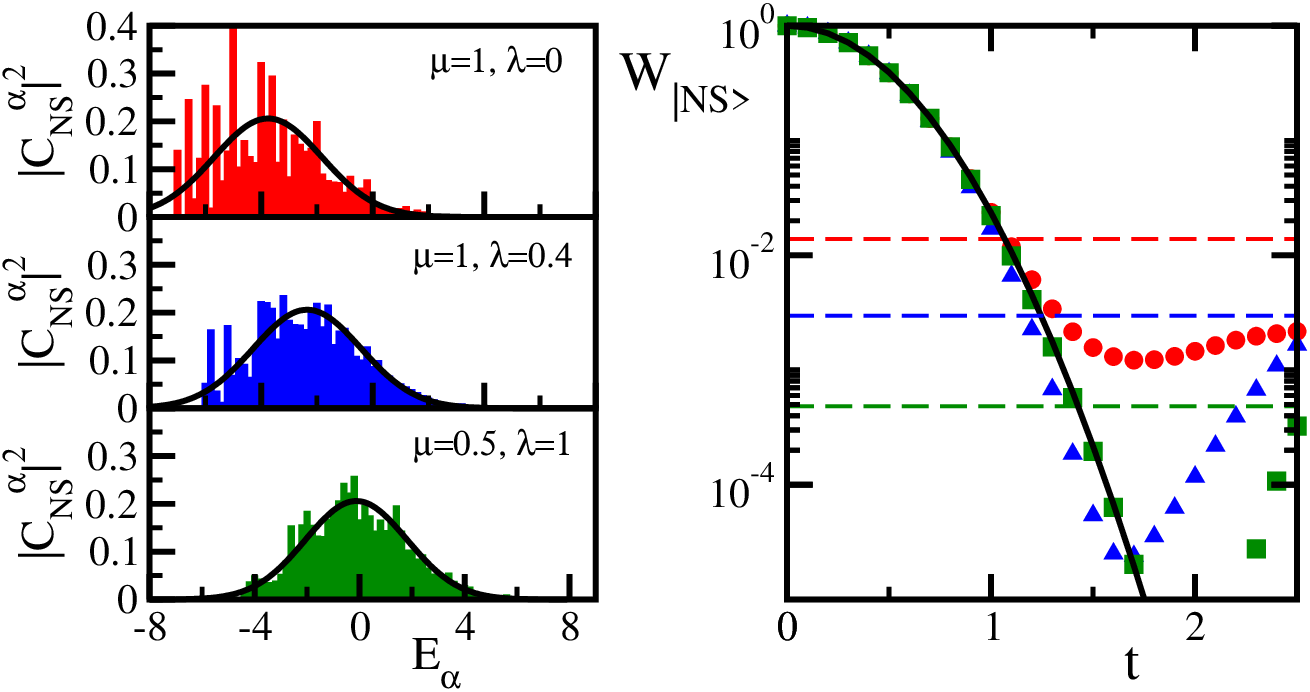}
\caption{(Color online) Strength function for the N\'eel state (left) and the corresponding survival probability decay (right) for $\mu=1,\lambda=0$ (circles), $\mu=1,\lambda=0.4$ (triangles), and  $\mu=0.5,\lambda=1$ (squares); $L=16$, $L/2$ up-spins.  Solid lines: Gaussian envelope of the strength function (left) and Gaussian decay (right) for $\Delta_{|\text{NS}\rangle}$ from Table~\ref{table:initial}. Dashed horizontal line from top to bottom: saturation point for the integrable, weakly chaotic, and strongly chaotic case (after \cite{Torres2014PRA,Torres2014NJP}).}
\label{fig:Neel}
\end{figure}

\begin{figure}[htb]
\centering
\includegraphics*[width=4.in]{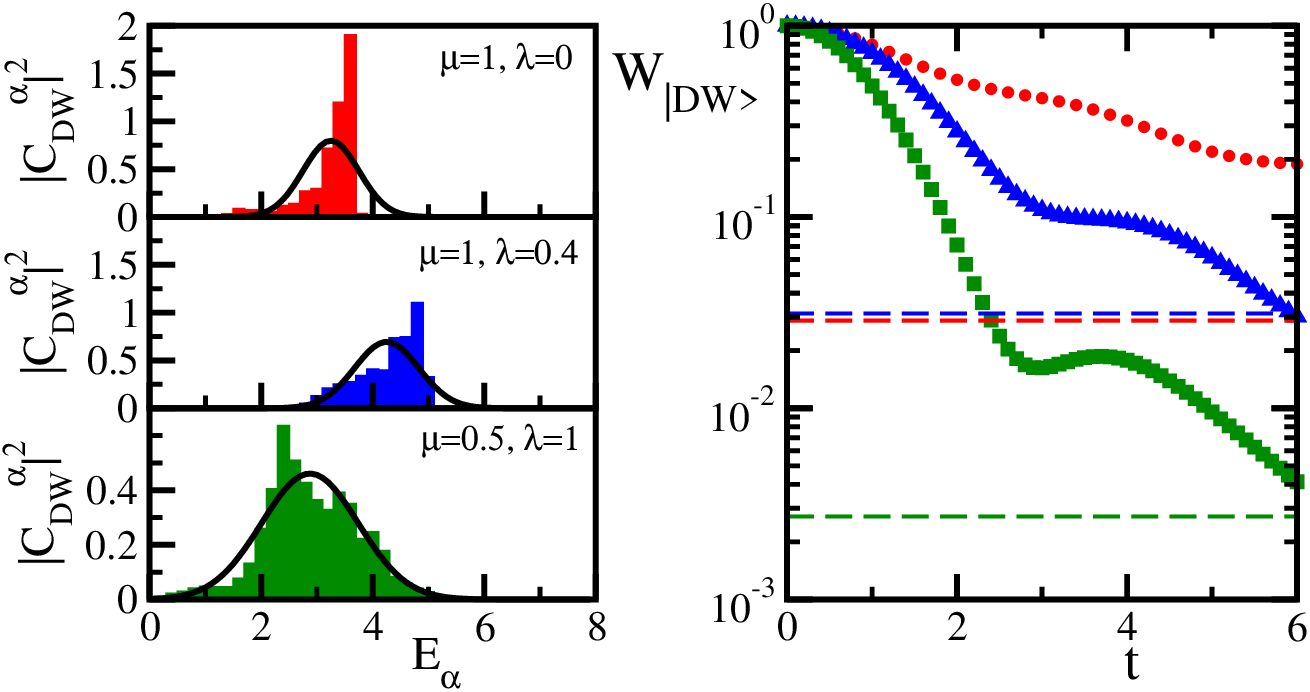}
\caption{(Color online) Strength function for the domain wall state (left) and the corresponding survival probability decay (right) for $\mu=1,\lambda=0$ (circles), $\mu=1,\lambda=0.4$ (triangles), and  $\mu=0.5,\lambda=1$ (squares); $L=16$, ${\cal S}_z=0$.  Solid lines (left): Gaussian envelope of the strength function for  $\Delta_{|\text{DW}\rangle}$ from Table~\ref{table:initial}. Dashed horizontal line from top to bottom: saturation point for the weakly chaotic, integrable, and strongly chaotic case (after \cite{Torres2014PRA,Torres2014NJP}).}
\label{fig:DW}
\end{figure}

Figure~\ref{fig:DW} shows the strength functions and survival probability  for the sharp domain wall. The same Hamiltonians as in  Fig.~\ref{fig:Neel} are considered, but the results are quite different. In comparison with the N\'eel state, the decay for $|\rm{DW}\rangle$ is much slower due to its low connectivity, and therefore narrow and poorly filled energy shell.  Apart from short time, the behavior of $W_{\text{DW}}(t)$ is neither Gaussian nor exponential. The nonmonotonous behavior of $W$ with the apparent transition to another exponent indicates
the interference of two competing  relaxation processes taking place with the increase of $\lambda$; similar effects are seen in the theory of radioactive decay with several interfering
decay chains \cite{peshkin14}.

\subsubsection{Decays faster than Gaussian}

Decays even faster than Gaussian are expected in at least two situations: (i) in the limit of very strong interaction not necessarily limited to two-body or (ii) when the strength function involves more than a single peak~\cite{Torres2014PRAb}. The first case is less realistic, but serves to provide a lower bound. The extreme scenario occurs when the final Hamiltonian coincides with a full random matrix. Here, not only the density of states, but also the strength function approaches the semicircular shape, as shown in Fig.~\ref{fig:RMTCos} (a). This leads to~\cite{Torres2014PRA,Torres2014NJP,Torres2014PRAb}
\begin{equation}
W_0(t) = \frac{ [{\cal J}_1( 2 \Delta_{k_0} t)]^2}{\Delta_{k_0}^2 t^2},   \label{75}
\end{equation}
where  ${\cal J}_1$ is the Bessel function of the first kind. The agreement between this expression and the numerical results in Fig.~\ref{fig:RMTCos} (b) is excellent.

\begin{figure}[htb]
\centering
\includegraphics*[width=4.in]{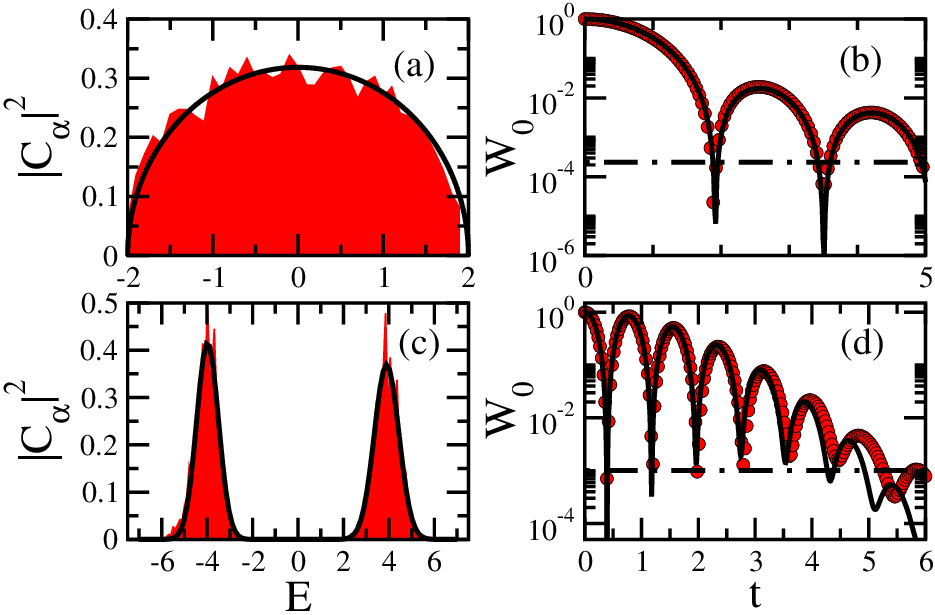}
\caption{(Color online) Strength function (left) and the corresponding survival probability decay (right). Top: Initial state from a GOE matrix projected onto another GOE full random matrix, ${\cal N} = 12\,870$.  The random numbers are normalized so that $\Delta_{k_0}= 1$. Bottom: Quench from the XXZ model to a Hamiltonian with an excessive on-site energy $d=8.0$ on site $L/2$;  $\mu=0.48$, $L=16$, 8 up-spins. The initial state is in the middle of the spectrum. The centroid and width of the two Gaussians are respectively  $E_1=-3.98$, $E_2=3.90$, $\xi_1=0.48$, and $\xi_2=0.54$. Numerical results are shown by shaded area (left) and circles (right). Solid lines correspond to the analytical expressions. Dot-dashed horizontal lines indicate the saturation point (after \cite{Torres2014NJP,Torres2014PRAb}).}
\label{fig:RMTCos}
\end{figure}

Since for full random matrices, $M_2^{k_0} \sim 3/{\cal N}$, where ${\cal N}$ is the dimension of the matrix, $t_R$ may be obtained from the relation,
\begin{equation}
\frac{   [{\cal J}_1( 2 \Delta_{k_0} t_R)]^2   }{  \Delta_{k_0}^2 t_R^2   } = \frac{3}{{\cal N}} .
                                                                                                 \label{76}
\end{equation}
After reaching the saturation point for the first time, the survival probability still shows some small oscillations decaying as $t^{-3}$ \cite{TorresKollmarARXIV}.

The second case of faster than Gaussian decay occurs if  the strength function is bimodal (or multimodal). It can happen, for example, when the system is initially prepared in an eigenstate of the total Hamiltonian of Model 1 and the evolution is started by abruptly turning on a strong static magnetic field that acts only on a single site of the chain. When the amplitude of this field is very large, inducing a large on-site Zeeman splitting, the density of states and consequently also the strength function splits in two separated Gaussian peaks~\cite{Torres2014PRAb}. An illustration is provided in Fig.~\ref{fig:RMTCos} (c). Another case of the double structure was discussed in relation to Eq.~(3.16).

For initial states with energy close to the middle of the spectrum, both peaks have similar width $\xi$, and the survival probability is approximately given by~\cite{Torres2014PRAb},
\begin{equation}
W_0(t) = \cos^2\left(\gamma t \right)\exp(-\xi^2 t^2),                            \label{77}
\end{equation}
where $\gamma$ is half the distance between the two peaks. A good agreement between this expression and the numerical results is seen in Fig.~\ref{fig:RMTCos} (d). Note that for $t<\pi/(2\gamma)$ the probability decay coincides with the ultimate bound established by the energy-time uncertainty relation, $W_0(t)  \geq \cos^2 (\gamma t)$ \cite{Mandelstam1945,Fleming1973,Bhattacharyya1983,Pfeifer1993,GiovannettiPRA2003}. After reaching the saturation point for the first time,  $W_0(t) $ still shows large oscillations that decrease according to the Gaussian envelope, $W_0(t) \sim \exp(-\xi^2 t^2)$.

\subsection{Relaxation in the energy shell}
\label{SubSec:EntropyDynamics}

The knowledge of the strength function, and, therefore, the survival probability, allows one to
estimate the time evolution of observables that can be directly associated with experimental data. In the approach developed in Ref.~\cite{FI01b}, it was suggested to consider the dynamics of wave packets in the TBRI-model in the following way. The projection of the wave function $\Psi(t)$ onto the state $|k_0\rangle$ is given by Eqs.~(\ref{59},\ref{60}). By writing $ w_{kk_0}(t)=w_{kk_0}^s+w_{kk_0}^{{\rm fluct}}(t)$ and assuming, as always, that in the long-time limit the fluctuating term $w_{kk0}^{{\rm fluct}}(t)$ vanishes, one obtains
\begin{equation}
w_{kk_0}^s =\sum\limits_\alpha|C_{k_0}^{(\alpha)}|^2|C_k^{(\alpha)}|^2 \simeq
\int \frac{dE}{\rho(E)} F_{k_0}(E) F_k(E) .                 \label{78}
\end{equation}
The ingredients here are the two strength functions, $F_{k_0}(E)$ and $F_k(E)$, for the initial and final basis state, respectively. The integral in Eq.~(\ref{78}) can be easily evaluated when the strength functions have either the BW or Gaussian shape. The similar expression for $W_0$ contains $|C_{k_0}^{(\alpha)}|^4$. For Gaussian fluctuations of the components we obtain $ \overline{ |C_{k_0}^{(\alpha)}|^4}$ = 3 $(\overline{|C_{k_0}^{(\alpha)}|^2})^2$. Therefore, if the number of principal components $N_{{\rm pc}}$ of the eigenstates is very large, the probability of return to the initial state $\left| k_0\right\rangle $ in the long-time limit is, at least, three times larger than the probability to find the system in any other state $\left| k\right\rangle $, see Fig.~\ref{Wfinal}. This is the same effect as the one  known as the {\sl elastic enhancement factor} in the ratio of fluctuational cross sections in elastic and inelastic channels in the processes going through the complex compound systems \cite{CIZB07}.

\begin{figure}[ht]
\centering
\subfigure{\includegraphics[angle=-90,scale=0.25]{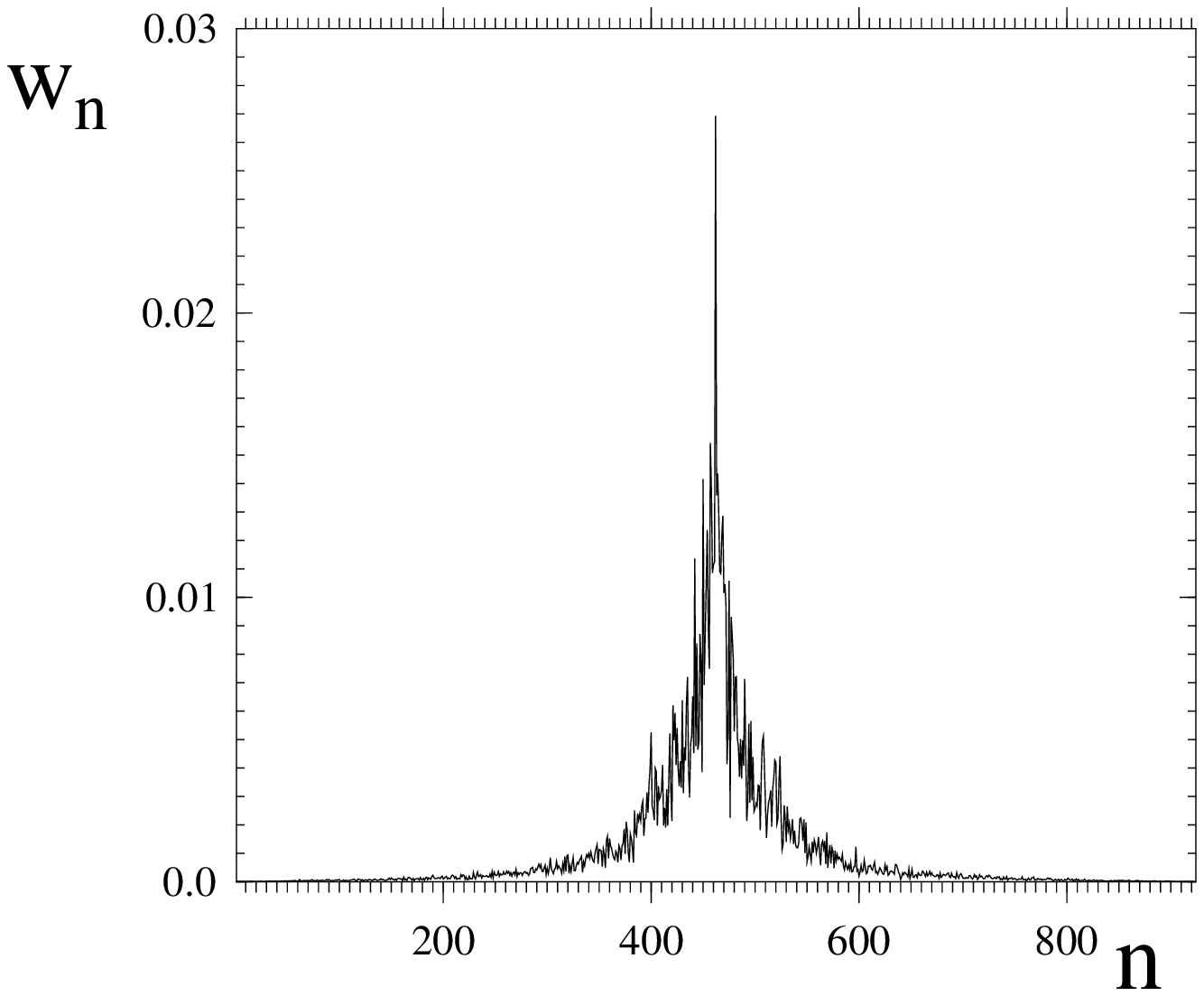}}
\hspace{1cm}
\subfigure{\includegraphics[angle=-90,scale=0.25]{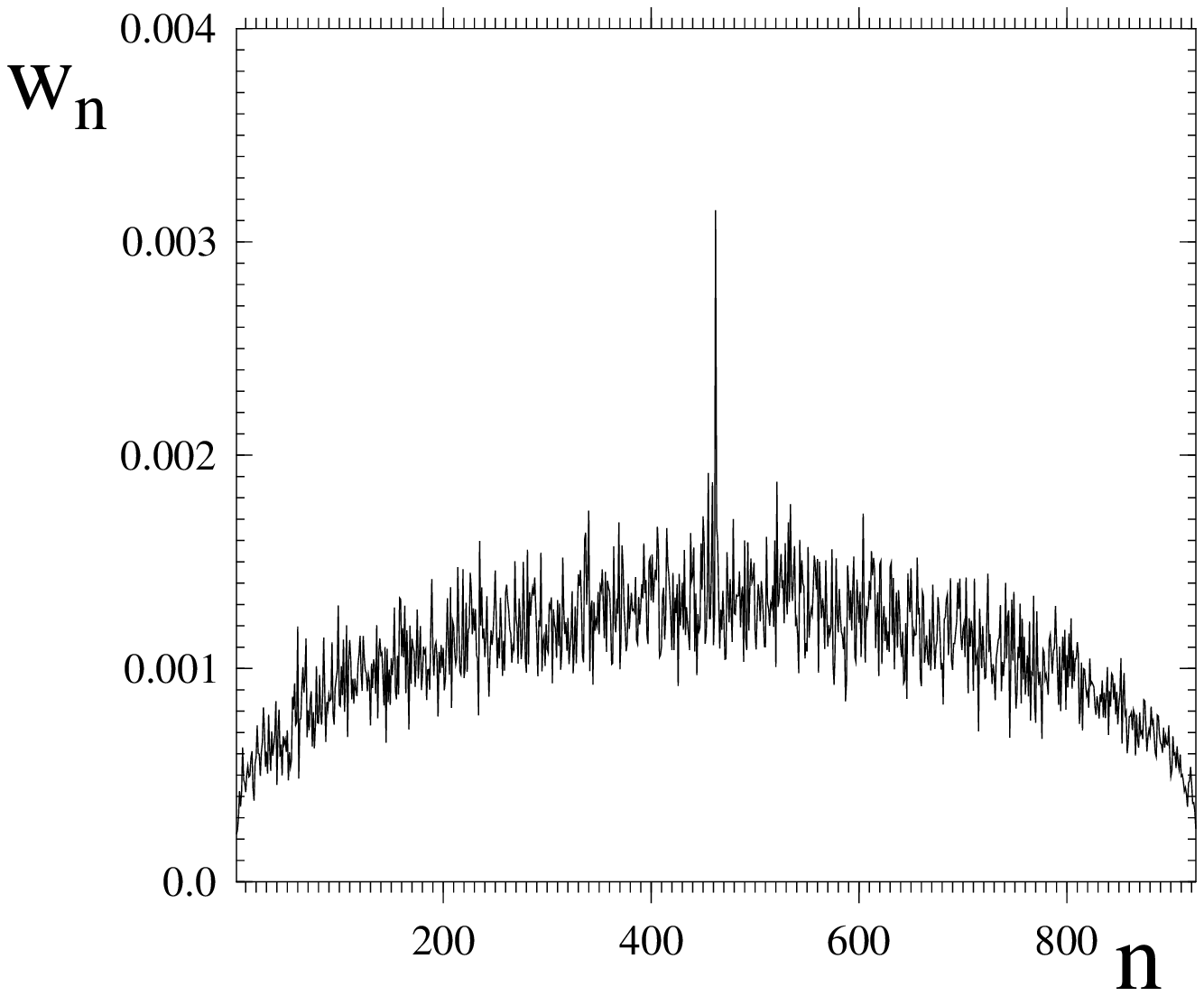}}
\caption{Asymptotic distribution $w_n = w_{nk_0}$ for the TBRI-model with 6 particles and 12 single-particle levels. As initial state we choose $k_0=462$. Left: $v_0^2 \approx
0.003, \, \Gamma_0 \approx 0.50,\,\Delta_{k_0} \approx 1.16 $, and the strength function is of the Breit-Wigner form; the average is performed over $10$ matrices with different realization of
disorder. Right: $v_0^2\approx 0.083 \,, \Gamma_0\approx 10.5 \,, \Delta_{k_0} \approx 5.8$, and the strength function is close to a Gaussian. The average is taken over $50$ matrices (after
\cite{FI01b}).} \label{Wfinal}
\end{figure}

In Fig. 26 the distribution of probabilities $w_k$ in the TBRI-model is shown after a very long evolution time, $t=40$, for two different strengths of interaction; the time $t$ is measured in units $\hbar/d_0$, where $d_0$ is the average spacing. Initially, only one basis state $k_0=462$ was populated at the center of the energy spectrum in order to avoid the asymmetry of the distribution in the basis representation. The two characteristic values of $v_0$ are chosen in such a way that the strength function have the BW or the Gaussian shape in the energy representation.
The data shown in Figure in the basis representation demonstrate a strong dependence of the dynamics on the shape of the strength function. The transition between the shapes is fast compared to the small change of the interaction strength $v_0$.

In order to study how the survival probability decays due to the perturbation $V$, it is convenient to introduce sub-classes for all basis states in the following way. The {\it first class} contains $N_1$ basis states directly coupled to the initial state by the matrix elements $H_{kk_0 }$. The {\it second class} consists of $N_{2}$ basis states coupled with the initial one in the second order  of the perturbation and so on. For large times, $t\gg \Gamma_0/(\Delta E)^2$, assuming the BW shape of the strength function, the time dependence of
occupancies of the states in different classes can be effectively described by kinetic equations, similar to what is used in the theory of compound nuclear reactions going through the hierarchy of intermediate states of various degree of complexity \cite{agassi},
\begin{equation}
\frac{dW_0}{dt}=-\Gamma_0 W_0,\,\,\,\,\, \frac{dW_1}{dt}=\Gamma_0 W_0-\Gamma_0
 W_1,\,\,\,\,\, ... \,\,\, ,\, \frac{dW_j}{dt}=\Gamma_0 W_{j-1}-\Gamma_0 W_j,\, ...    , \label{79}
\end{equation}
where $W_j$ is the probability for the system to be in a state of class $j$. Here we consider a system far from the equilibrium. If the system is at the equilibrium, the probabilities of all states within the energy window defined by $\left| E_k-E_{k_0}\right| \leq \Gamma_0$, where $ E_{k_0}$ is the energy of the initial basis state, are of the same order, $w_k\simeq N_{{\rm pc}}^{-1}$, with $N_{{\rm pc}}$ estimated as the total number of states inside the energy shell. In order to neglect the return flux, one needs the condition $w_k \simeq W_j/N_j\gg 1/N_{{\rm pc}}$ to be fulfilled.

The solution of the infinite set of equations (\ref{79}) is given by the Poisson distribution,
\begin{equation}
W_0=\exp (-\Gamma_0 t\,)\,; \,\,\,\,\, W_j=\frac{(\Gamma_0 t)^j}{j!}\exp
\left( -\Gamma_0 t\right)
=\frac{(\Gamma_0 t)^j}{j!} W_0.                           \label{80}
\end{equation}
The maximum probability $W_j=(j^{j}/j!)\exp (-j)\approx 1/\sqrt{2\pi j}$ to be in the $j$-th class is determined by the condition $dW_j/dt=0$ and occurs for $t=j/\Gamma_0$. The solution (\ref{80}) can be considered as a {\it cascade} in the population dynamics  of different classes. At small times, $t\ll \tau \equiv 1/\Gamma_0$, the system is practically in the initial state; at times $t\approx \tau \,$ the flow spreads into the first class, for $t=j\tau $ it spreads into the $j$th class, etc. For an infinite chain, the normalization condition $\sum\limits_{j=0}^\infty W_j=1$ remains valid.

A quantity of special interest is the Shannon entropy of wave packets in the many-body basis,
\begin{equation}
S(t)=-\sum\limits_k w_k\ln w_k \approx -\sum \limits_{j=0}^\infty W_j \ln \frac{W_j}{N_j},  \label{81}
\end{equation}
where $w_k\approx W_j/N_j$ stands for the population of basis states of the class $j$ with $N_j$ states in this class. In fact, for $t \sim j \tau$ one needs to count only the states inside the energy shell since the population of the states outside the energy interval $\left| E_k-E_{k_0}\right| > \Gamma_0$ is small \cite{FI01b}. Using $N_j=K^j$ (where $K$ is the number of basis states directly coupled by the perturbation) and $\sum\limits_{j=0}^\infty j(\Gamma_0 t)^j/j!= \Gamma_0 t\exp (\Gamma_0 t)$, one comes to the following expression,
\begin{equation}
S(t)\approx \Gamma_0 t \ln K + \Gamma_0 t -e^{-\Gamma_0 t} \sum\limits_{j=0}^\infty
\frac {(\Gamma_0 t)^j}{j!} \ln \frac {(\Gamma_0 t)^j}{j!}.         \label{82}
\end{equation}
The two last terms in the right hand side of Eq.~(\ref{82}) turn out to be smaller than the first one, therefore,
\begin{equation}
S(t)\approx \Gamma_0 t \ln K [1 + f(t)],     \label{83}
\end{equation}
where the function $f(t) \ll 1$  depends weakly on time.

Neglecting the second term in Eq.~(\ref{83}) we obtain the linear increase of entropy which means that the number of principal components $N_{{\rm pc}}(t)$ increases exponentially  with time. This behavior can be compared with a linear increase of the dynamical entropy $S_{{\rm cl}}(t)$ in classically chaotic systems where $S_{{\rm cl}}(t)$ is known to be related to the exponential divergence of close trajectories in the phase space: $S_{cl}(t)\propto \lambda t$ where $\lambda $ is the Lyapunov exponent, see, for example \cite{LB99,P99,BLR01,GPP00}. The non-trivial point is that the linear increase of entropy also occurs for systems without classical limit \cite{GPP00}, therefore, the product $\Gamma_0 \ln K$ may be treated as describing the ``quantum Lyapunov exponent".   It should be stressed that for short times,
$t\ll \Gamma_0 /\Delta_{k_0}^2\,$, the entropy increases quadratically in time, $S(t) \approx \Delta_{k_0}^2 t^2$, which is an universal result valid for any shape of the strength function.

The cascade model assumes an infinite number of ``classes". In systems with a finite number of many-body states, any basis state can be reached dynamically in several ``interaction steps''. For example, in a system of $N_p=6$ particles and $N_s=12$ single-particle levels, three steps are needed in order to have all 924 basis states effectively populated if all eigenstates are fully delocalized. If the number of classes is finite, the dynamics saturates and one can expect the emergence of a steady-state distribution of population in the unperturbed basis. A simple expression for $S(t)$ was suggested for the systems with a small number of classes \cite{FI01b},
\begin{equation}
S(t)=-W_0(t)\ln W_0(t)\,-(1-W_0(t))\ln \left( \frac{(1-W_0(t))}{N_{{\rm pc}}}\right).  \label{84}
\end{equation}
This expression takes into account the normalization condition $\sum\limits_{k\neq k_{0}}w_k = 1-W_0$, and has a reasonable behavior for both small and large times, except for very short times when the quadratic dependence of $S(t)$  on time dominates. Since $W_0(t)$ is entirely defined by the strength function, the key parameter in Eq.~(\ref{84}) is the number $N_{\rm pc}$ of principal components, or the degree of delocalization of eigenstates in the unperturbed basis.

We can now compare the obtained analytical expressions with numerical data for the TBRI model (\ref{3}). For the strength function of  BW shape, the time dependence of the entropy is shown in Fig.~\ref{Shannon-BW}.
\begin{figure}[htb]
\centering
\includegraphics[width=0.5\textwidth]{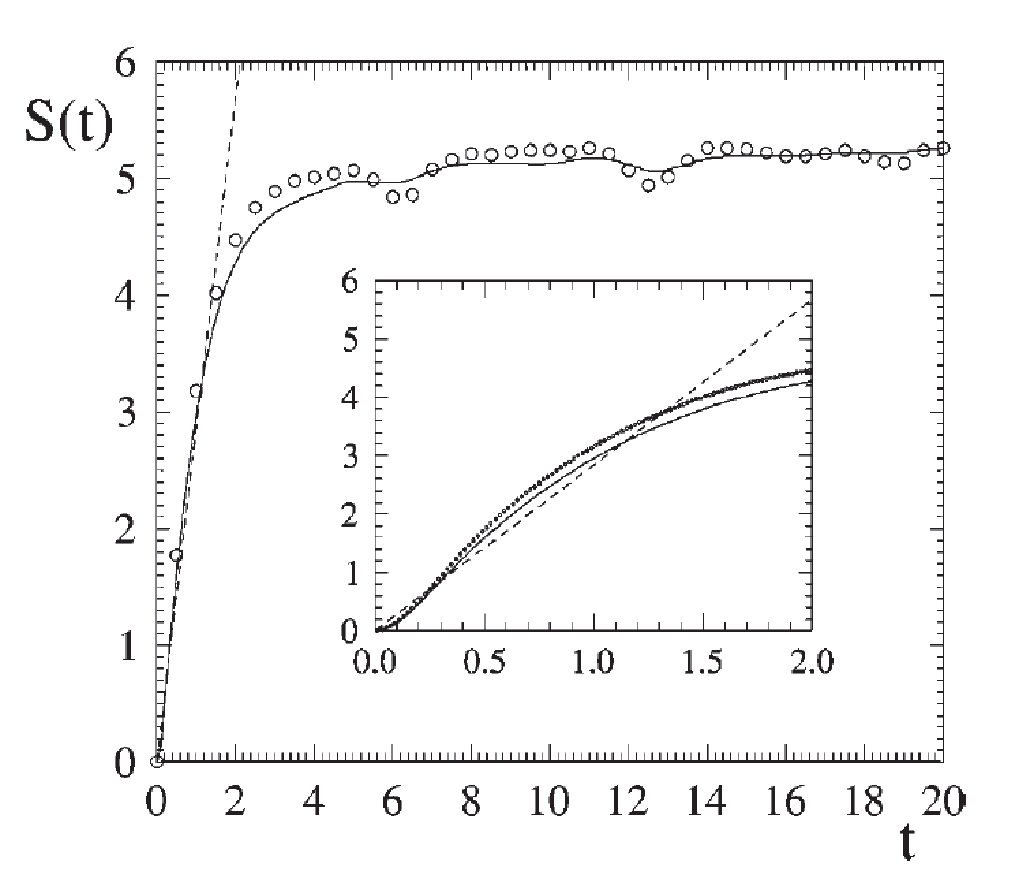}
\caption{Entropy versus time for the TBRI model with parameters $N_p=6,\, N_s=12,\ v_0^2 \approx 0.003, \Gamma_0 \approx 0.50,\,\Delta_{k_0} \approx 1.16$, when the SF has the Breit-Wigner shape. Circles stand for numerical data, solid curve is Eq.~(\ref{84}), and dashed line is the linear dependence (\ref{86}). In the inset the same is shown for a smaller time scale (after \cite{FI01b}).}
\label{Shannon-BW}
\end{figure}
The number $K$ of the basis states directly coupled by the random two-body interaction, is determined \cite{Flambaum1996a} by
\begin{equation}
K=N_p(N_s-N_p)+\frac{N_p(N_p-1)(N_s-N_p)(N_s-N_p-1)}4,               \label{85}
\end{equation}
where the first term gives the number of one-particle transitions, and the second stands for two-particles transitions. In the case of $N_p=6$ particles and $N_s=12$ orbitals, considered above, the total number of basis states is ${\cal N}=924$ and $K=261$. The {\it effective number} $n_{c}$ of classes in the cascade model can be determined from the relation $K^{n_c}=
{\cal N}$. This gives $n_c=\ln {\cal N}/\ln K\approx 1.2$. Thus, we can use the simple expression (\ref{84}) to describe the time dependence of the entropy. The data in Fig.~\ref{Shannon-BW} demonstrate an excellent agreement between the numerical results and Eq.~(\ref{84}).

A good approximate description of data on the whole time scale depends on the effective number $N_{{\rm pc}}$ of principal components in the stationary distribution $w_k(t\rightarrow \infty)$ relating to the limiting value of the entropy, $N_{{\rm pc}}=\ln S(t\rightarrow \infty)$. This number can be estimated as discussed above from the width of the energy shell; when plotting the solid curves in Fig.~\ref{Shannon-BW} we have used the value of $N_{{\rm pc}}$ found from  numerical data. The {\it actual number} of classes in the case of $N_p=6$ particles and $N_s=12$ orbitals is equal to $3$ so that all basis states can be populated in the third step by the two-body interaction. However, the amount of states in the second, $k=2$, and third, $k=3$, classes is much smaller than what follows from the exponential relation $N_k=K^k$ (in practice, this relation may be justified for a large number of particles only). For this reason the one-class formula, Eq. (\ref{84}) works well. It is also instructive to compare the entropy with the linear time dependence,
\begin{equation}
S(t)=\Gamma_0 \,t\,\ln K                 \label{86}
\end{equation}
that stems from Eq.~(\ref{77}) if the only first term is left. On some time scale, the data in Fig.~\ref{Shannon-BW} roughly correspond to Eq.~(\ref{80}), but a clear difference of $S(t)$ from the linear increase is seen in the inset. As we already noted, at short time the universal dependence is quadratic.
\begin{figure}[htb]
\centering
\includegraphics[width=0.5\textwidth]{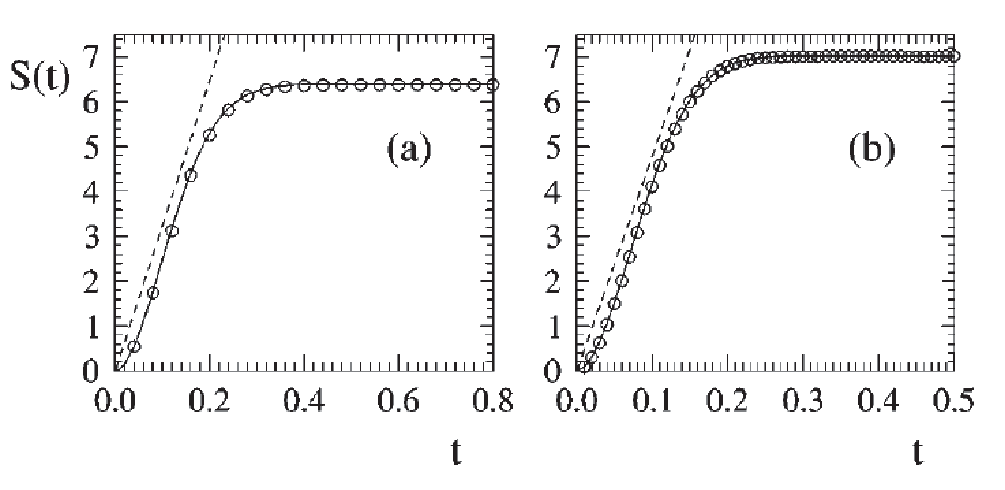}
\caption{Time dependence of entropy for the TBRI model when the strength function is  Gaussian:  (a) $N_p=6, N_s=12, v_0\approx 0.083, \Gamma_0\approx 10.5, \Delta_{k_0}\approx 5.8$, and (b) $N_p=7, N_s=13, v_0\approx 0.12, \Gamma_0\approx 14.6, \Delta_{k_0} \approx 8.13$. Circles are numerical data, solid curves stand for the approximate expression (\ref{78}), and dashed lines represent the linear dependence (\ref{80}) (after \cite{FI01b}).} \label{SG}
\end{figure}

Results for the strong interacting case, when the SF is close to a Gaussian, are reported in Fig.~\ref{SG}. The interaction strength is chosen to have the same ratio
$\Gamma_0/\Delta_{k_0} \approx 1.8$, as in Fig.~\ref{Shannon-BW}. For such a strong interaction, the half-width of the strength function is determined by its second moment rather than by $\Gamma_0$. Therefore, $\Gamma_0$ in the expressions (\ref{83}) and (\ref{84}) has to be substituted by the width $\Delta_{k_0} \approx \sigma_{k_0}$. In both cases shown in Fig.~\ref{SG} the numerical data give strong evidence of the linear entropy increase,
\begin{equation}
S(t)=\Delta_{k_0} \, t \, \ln K,    \label{87}
\end{equation}
before the saturation. This estimate gives a correct value for the slope of $S(t)$. The linear dependence for $S(t)$ has been shifted in the figure for a better comparison with numerical data
outside  the initial time scale where the time dependence is quadratic.

The linear dependence of $S(t)$ in Fig.~\ref{SG} is much more pronounced than in the BW-region, see Fig.~\ref{Shannon-BW}. Thus, we confirm the difference between the two cases (Breit-Wigner and Gaussian shape of the SF). This point is supported by the recent results \cite{CIK00,CK00,CK01} that for not very strong interaction, $\Gamma_0<\Delta_{k_0}$, resulting in the Breit-Wigner shape of the SF, there is no detailed quantum-classical correspondence for the evolution of wave packets in the energy space. On the other hand, for the Gaussian shape of the SF, this correspondence is observed. The strength function in the energy representation has a clear classical analog in the cases when the Hamiltonian has its well defined classical limit. The results presented in Refs.~\cite{WIC98,BGI98,MIU00,LMI00,BISS00,I01,LMI01,BCIC02,LMI02,BFH03} give evidence for a quantum-classical correspondence of the strength function. In the classical limit, the meaning of the strength function is just a projection of the energy surface of $H_0$ onto that of $H$, the fact that simplifies the analysis of quantum systems having a well defined classical limit. In such a case, the strength function can be found directly from classical equations of motion, at least numerically.

The above analysis, applied to the Shannon entropy, can be easily extended to other dynamical observables. For example, the expression for the inverse participation ratio $M_2(t)$ that can be associated with an effective number of components in the wave packet $\Psi(t)$ related to the initial state $|k_0\rangle$, becomes
\begin{equation}
M_2(t) = W_0^2 I_0 \left( \frac{2\ln (W_0^{-1})}{\sqrt{K}} \right) ,      \label{88}
\end{equation}
where $I_0(z)$ stands for the modified Bessel function. Numerical data confirm well this estimate, see details in Ref.~\cite{FI01b}.

With the approach we discuss here, one can also describe the evolution of occupation numbers $n_s(t)$ towards the steady-state distribution emerging after a long time \cite{FI01c}.
Assume that in the TBRI model (\ref{3}) the system at $t=0$ is in a basis state $\left| k\right\rangle $ with the occupation numbers $n_s (0)$ equal to $0$ or $1$. In the Breit-Wigner regime, the first class is populated within the time $\tau \sim 1/\Gamma_0$ \cite{FI01b}, or, $\tau \sim 1/\Delta_{k_0}$ in the Gaussian regime. At that time the occupation numbers $n_s$ already strongly deviate from their initial values since the two-body interaction can move any  two particles to new single-particles levels. Thus, the characteristic time for an initial stage of thermalization is determined by the population time $\tau$ for the first class. Note that this time is also a characteristic time for the ``decay" of the probability $W_0(t)$. The population of all $n_c$ classes in a particular system requires a longer time $\tau_{n_c} \sim n_c \tau$ \cite{FI01b}. Thus, in the case of $n_c \gg 1$ (e.g. in a mesoscopic system) the thermalization of the occupation numbers may roughly occur on a time scale smaller
than the onset of a complete statistical equilibrium.

This suggests a simple derivation of the time dependence $n_s(t)$ for occupation numbers. From the normalization condition $\sum_{j=0}^{n_c} W_j =1$ one can find the population of all classes with $j\neq 0$ as $\sum_{j=1}^{n_c} W_j = 1-W_0$. Now we assume that the thermalization of the occupation numbers occurs on the time scale $\tau$. This assumption leads to the simple expression \cite{FI01c},
\begin{equation}
n_s(t) = n_s(0) W_0(t) + n_s (\infty) \left (1-W_0(t) \right),             \label{89}
\end{equation}
where the occupancies $n_s(\infty)$ determine an equilibrium distribution after a long evolution.

In Fig.~\ref{time-dep} we compare numerical data for $n_s(t)$ with the estimate (\ref{89}). Two situations are studied, corresponding to a strength function close  to a Breit-Wigner, $\Gamma_0 \ll \Delta_{k_0}$, and  close to a Gaussian,  $\Gamma_0 \approx \Delta_{k_0}$. For numerical simulation, the TBRI model has been used with six fermions occupying twelve single-particle levels. Since the analytical expression for $W_0(t)$ in general is quite complicated,  the numerical values of $W_0(t)$ have been used in simulations.

Overall, there is a good agreement between the analytical estimate and numerical data, apart from fluctuations that are neglected in the theory. In order to simplify the analysis, the initial basis state $|k_0\rangle$ was taken from the middle of the many-body spectrum of $924$ levels. In this case the final values of the occupation numbers are expected to be $n_s(\infty)=1/2$ corresponding
to an infinite temperature $T$.
\begin{figure}[ht]
\centering
\hspace{-1.0cm}
\subfigure{\includegraphics[angle=-90,scale=0.30]{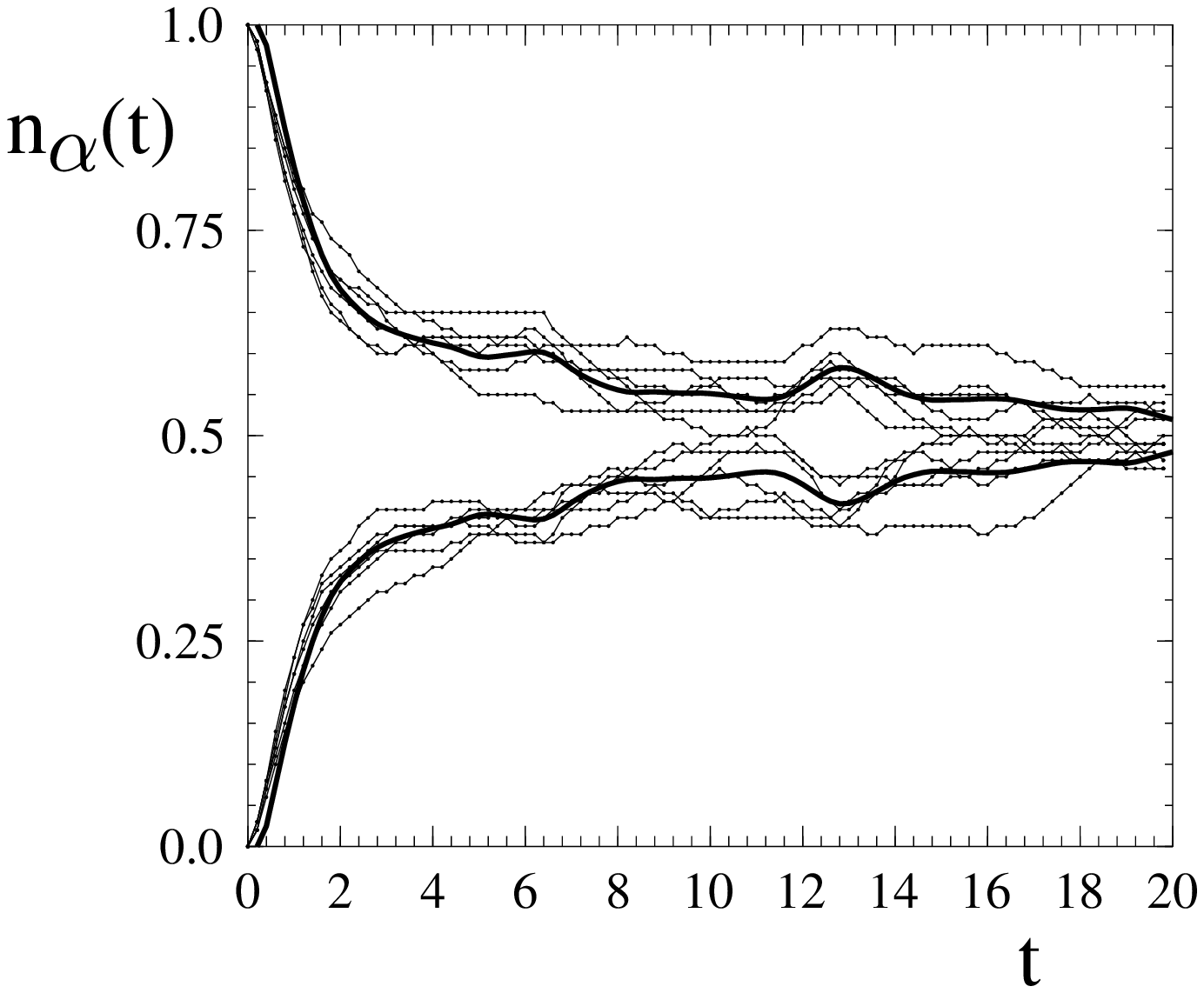}}
\hspace{-1.25cm}
\subfigure{\includegraphics[angle=-90,scale=0.30]{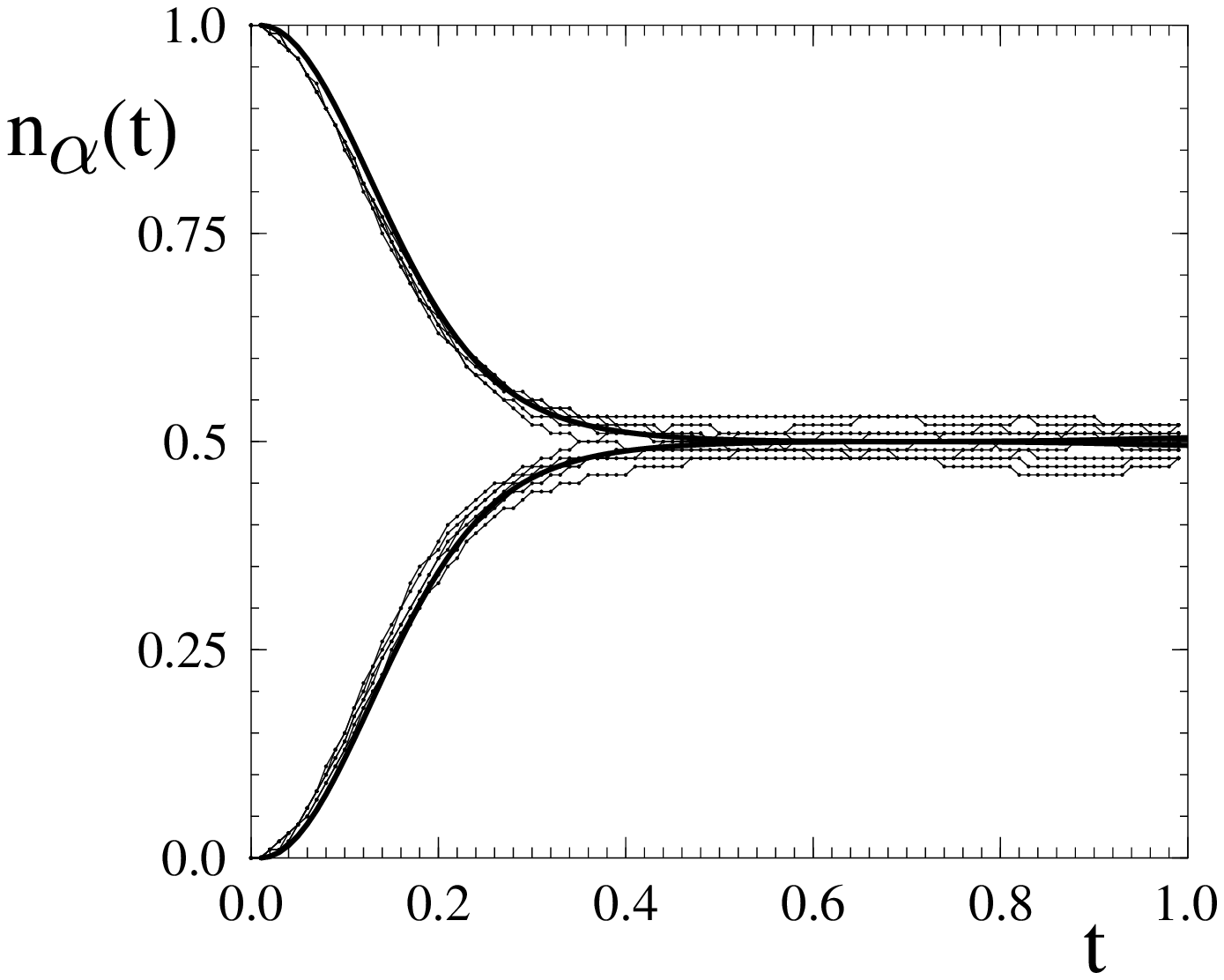}}
\caption{LEFT: Time dependence of occupation numbers $n_s(t)$ for the Breat-Wigner regime (weak interaction). Thin curves with dots present numerical data, thick smooth curves correspond to the analytical expression (\ref{89}), see the text. Computations are made for the model with random two-body interaction, with $N_p=6, N_s=12, \Gamma_0 \approx 0.50$, $\Delta_{k_0} \approx 1.16$. Right: Time dependence of occupation numbers $n_s(t)$ for the Gaussian shape of the strength function with $ \Gamma_0 \approx 10.5, \Delta_{k_0} \approx 5.8$ (after \cite{FI01c}).} \label{time-dep}
\end{figure}

The data reveal a difference between the cases of weak and strong interparticle
interaction.
For a weak interaction (left panel), the transition to equilibrium values of $n_s$ has a character of damped oscillations. As the number of principal components $N_{pc} \sim \Gamma_0 \rho^{-1}$ in this case is not very large, there are considerable fluctuations in $n_s(t)$ even at the equilibrium.
On the other hand, for strong interaction (right panel) $n_s$ shows fast and monotonic decrease to
thermal values $n_s(\infty)$  with relatively small fluctuations, see also \cite{FI01b}.
These results also indicate that one can speak of two time scales in the onset of thermalization. The first one is determined by $\tau$. It characterizes an ``initial thermalization'' (or {\it pre-thermalization}), and  allows one to use Eq.~(\ref{89}) for the time dependence of occupation numbers. For larger times, damped quantum oscillations (with period $T\sim n_c \tau$) may occur in the transition to the complete equilibrium \cite{FI01b}.

\subsection{Linear entropy increase as an indicator of thermalization}
\label{SubSec:EntropyRevivals}

In this Section we apply the analytical estimates derived for the Shannon entropy to the
fully deterministic spin Models 1 and 2 (see Eqs.~(\ref{5}) and (\ref{6})), as well as to the model of interacting bosons in a 1D trap \cite{BBIS04}. Fig.~\ref{fig:ShannonSpin} serves as an illustration of the linear increase of Shannon entropy for deterministic spin systems. Here the initial state has been chosen far from the edges of the energy spectrum, for three different values of the perturbation strength. Similarly to what has been observed for the survival probability, we verify again an analogous behavior for the integrable Model 1 and the chaotic Model 2.
\begin{figure}[htb]
\centering
\includegraphics*[width=4.in]{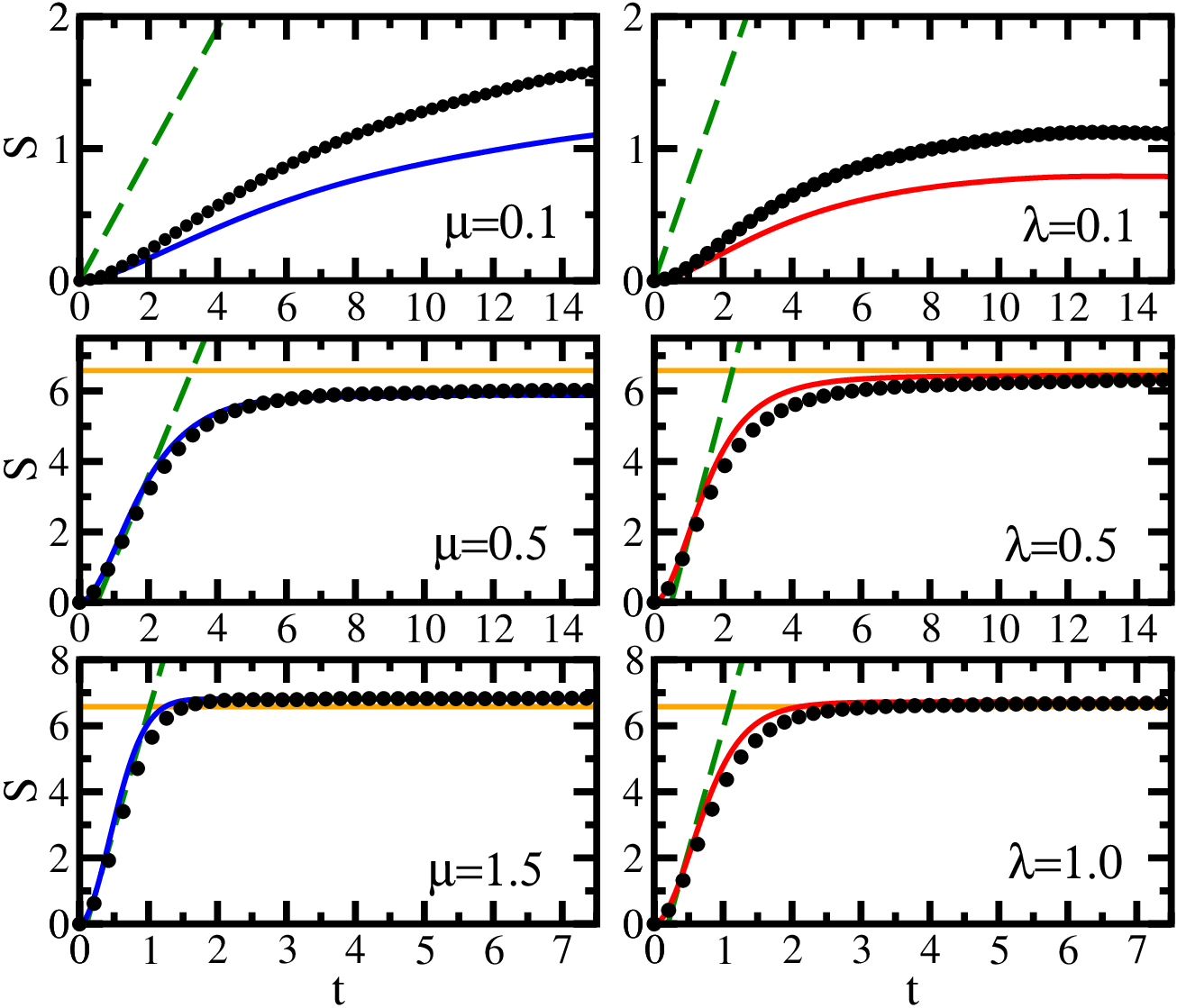}
\caption{(Color online)
Shannon entropy vs time for Model 1 (left) and Model 2 (right). Circles stand for numerical data, dashed lines show the linear dependence [Eqs.~(\ref{86}) and (\ref{87})], and solid curves correspond to Eq.~(\ref{84}) with the average  performed after the saturation for $t \in [100,200]$ in order to obtain $N_{pc} = \langle e^S \rangle$. The horizontal solid lines represent the value $S_{\text{GOE}} \approx 6.58$ of the Shannon entropy for a GOE (after \cite{Santos2012PRE}).}
\label{fig:ShannonSpin}
\end{figure}

In the top panels of Fig.~\ref{fig:ShannonSpin} (weak perturbation), the growth of $S(t)$ is slow, and it is not described by the analytical expressions given in  Eqs.~(\ref{84}), (\ref{86}), and (\ref{87}). In the middle panels, where the values of the perturbation strength give rise to the Breit-Wigner strength functions for both models~\cite{Santos2012PRL,Santos2012PRE}, the entropy increase, after the short-time quadratic behavior, is linear and well described by Eq.~(\ref{86}). In addition, the entire dynamics, from short  times all the way to saturation, show good agreement with the semi-analytical Eq.~(\ref{84}). A similar scenario emerges for the parameters in the bottom panels which induce Gaussian strength functions. In this case the dynamics agrees with Eq.~(\ref{84}) and the linear increase is captured by Eq.~(\ref{87}). Notice that Eq.~(\ref{81}) depends only on the quantities $K$ and $\Delta_{k_0}$ obtained from the elements of the final Hamiltonian matrix written in the basis of $H_0$, so the dynamics can be anticipated prior to the diagonalization of the total Hamiltonian $H$.

We stress that the expressions in Eqs.~(\ref{84}), (\ref{86}), and (\ref{87}) apply only for
initial Hamiltonians $H_0$ corresponding to the unperturbed part of $H$. Also, the initial states are taken far from the edges of the spectrum where the eigenstates of $H_0$ are known to be non-chaotic. For comparison let us notice that the initial states discussed in Section~\ref{SubSec:ReturnSpinExp} and in Ref.~\cite{Torres2014NJP} are not the eigenstates of $H_0$ and the analytical predictions for the increase of Shannon entropy are not known.

The time dependence of the Shannon entropy can serve as a good indicator for the onset of statistical relaxation in realistic systems. Below we follow Ref.~\cite{BBIS04} where the relatively simple model of interacting Bose-particles has been considered in view of a possible experimental realization. The model is specified by the deterministic Hamiltonian,
\begin{equation}
\displaystyle \hat{H} = \sum_{m} \epsilon_m \hat{n}_m
+\frac{g}{2L} \sum_{m,q,p,r} \hat{a}_m^\dagger \hat{a}_q^\dagger
\hat{a}_p \hat{a}_r \delta(m+q-p-r).                                                \label{90}
\end{equation}
The bosons are confined to a one-dimensional ring of length $L$, and the single-particle energy levels $\epsilon_m$ are defined by the standard quantization,
$\epsilon_m = 4\pi^2 m^2/L^2$. The {\it single-particle states} $|m\rangle$ are labeled by
the angular momentum numbers $m=0, \pm 1 , \pm 2... $. The interaction between bosons is defined by  point-like forces characterized by the parameter $g$ that is inversely proportional to the 1D interatomic scattering length \cite{O98} and can be experimentally controlled. The 1D regime can be achieved in optical/magnetic traps when the radial degrees of freedom are frozen by the tight transverse confinement.

Experimental achievements in effective one-dimensional harmonically confined quantum systems (see, for example, \cite{Go01,So1a,Go01a}) have stimulated several attempts to understand their main properties \cite{GWT01,AG02}. The Hamiltonian (\ref{90}) corresponds to the Lieb-Liniger model; its thermodynamical properties and the excitation spectra have been calculated analytically in Refs. \cite{LL63,LL63a}. An unexpected feature, predicted by Girardeau \cite{G60,G65}, is the onset of fermionization when $n/g \to 0$ where $n$ is the particle density. In this Tonks-Girardeau (TG) regime, the {\rm density} of the interacting bosons becomes identical to that of non-interacting fermions, while, of course, the wave-function keeps the bosonic symmetry. On the other hand, in the opposite limit, $n/g \to \infty$, the system is described in the mean-field (MF) approximation as a weakly interacting Bose gas. The crossover between these two regimes occurs near $n / g \sim 1$. In this region the MF approach breaks down and more complicated two-body correlations become crucial. Below we show that the crossover between the two regimes is signaled by the presence or absence of a linear time evolution of the  Shannon entropy for  wave packets that evolves linearly in time.

Let all bosons initially occupy the single-particle level with angular momentum $m=0$. Thus, at $t=0$ the system is in the unperturbed ($g=0$) ground state $|\Psi_0\rangle$; note that the total angular momentum is conserved in time. Our main interest is in the evolution of the system for
different values of the control parameter $n / g$. {\it Many-body} basis states $|k\rangle$ are defined by the occupation numbers $\{n^k_m\}$ of single-particle levels $|m\rangle$.

Numerically, the finite number of particles $N_p$ occupying $N_s$ single-particle states cannot be taken very large. The numbers $N_p$ and $N_s$ should be chosen in a consistent way in order to have the possibility to extrapolate the results to a large number of atoms. In a 1D geometry on a ring, $N_p$ particles define the smallest spacing, which corresponds to the largest value of the momentum $m \approx N_p$; this relation was satisfied in the numerical study \cite{BBIS04} when changing $N_p$ and $N_s=2m+1$.

The crossover from the MF to the TG regime can be understood using the following arguments. Having all particles initially in the lowest state with $m=0$, we can estimate the strength of the
interaction necessary to move two particles from the unperturbed ground state to the upper (and lower) single particle levels $\pm m$. This interaction will result in an ergodic filling (in time) of all single-particle states. The energy required for this excitation is approximately $m^2/L^2 \approx N_p^2/L^2$, and the matrix element of the interaction between the corresponding states is  $V_{kk'}\sim g\sqrt{N_p(N_p-1)}/L$. Equating these values we obtain $n=N_p/L \sim g$ that is associated with the crossover from the bosonic to fermionic regime. Dynamically this crossover is reflected by a rapid depletion of the occupancy of the single-particle state with $m=0$. Note that at $n / g \sim 1$, the ratio $N_0 / N_p \sim 1/2$, with $N_0 = \langle \hat{a_0}^\dagger
\hat{a_0} \rangle$.

We define the Shannon entropy $S(t) =-\sum_k |\Psi_k (t)|^2 \ln |\Psi_k (t) |^2$ of the wave packet, where $\Psi_k(t) = \langle k | \Psi(t)\rangle$ is the projection of the wave function
onto the noninteracting many-body basis. After switching on the interaction, the wave function evolves, $|\Psi(t) \rangle  = e^ {-iHt} |\Psi (0)\rangle $, and spreads over the unperturbed basis. This spread is shown in Fig.~\ref{becf3} for different values of the control parameter $n/g \gg 1$. The numerical data clearly manifest that for $n/g \gg 1$ the entropy oscillates in time, while for $n/g \ll 1$ there is a generic linear increase of $S(t)$ followed by a saturation.
On a short time scale, the time dependence of $S(t)$ is quadratic  rather than linear. As discussed above, the quadratic growth of the Shannon entropy is a generic property for any system, and out of our interest.
\begin{figure}
\centering
\includegraphics[scale=0.45]{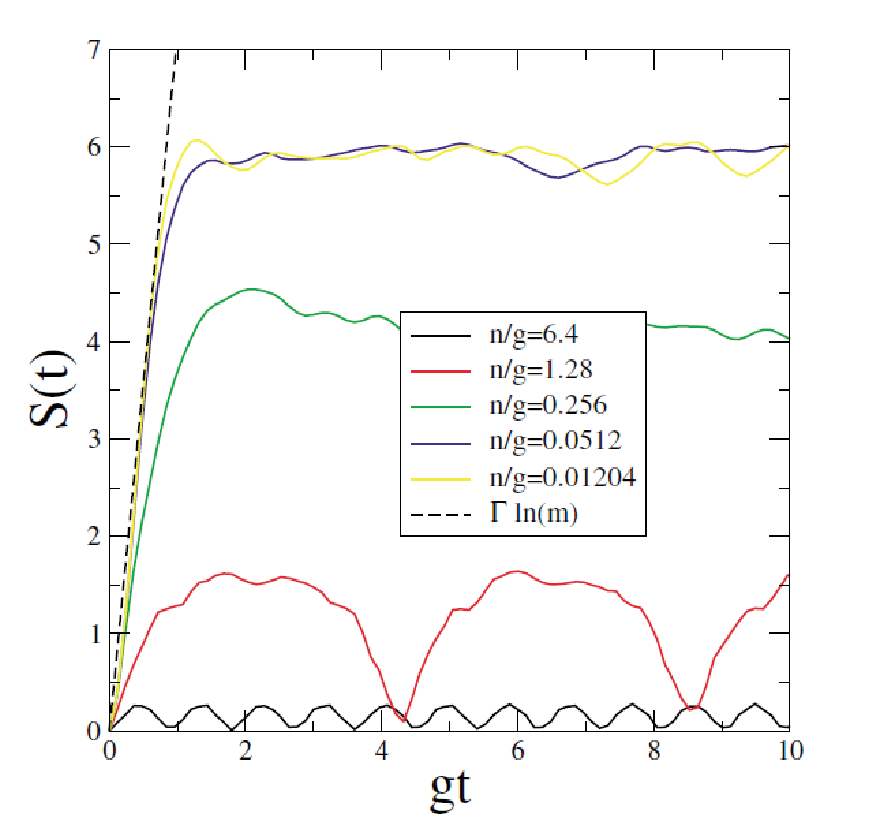}
\caption{ Entropy as a function of the rescaled time $gt$ for fixed $m=6$ and different values of $n/g$. Data are given for $N_p=6$ number of the bosons. The crossover from the oscillating saturation of $S(t)$ for small values of $n/g$ manifests the onset of the Tonks-Girardeau regime.
The dashed line corresponds to the theoretical prediction (\ref{86}) (after \cite{BBIS04}). } \label{becf3}
\end{figure}

The onset of a linear increase of Shannon entropy can be used to mark the crossover to the Tonks-Girardeau regime of statistical relaxation to a steady state momentum distribution. This transition can be observed experimentally by studying the interference fringes obtained after releasing the trap and letting the boson system expand ballistically (for details, see Ref.~\cite{BBIS04}). Similar crossover for the Shannon entropy has been also numerically observed for the case of effectively attractive interaction between bosons \cite{LXZY05}.


\subsection{Relaxation of Few-Body Observables}
\label{SubSec:ObservableDynamics}


Experimentally, information about the dynamics of the system is obtained via observables $O$ such as magnetization, spin-spin correlations, and the number of atoms on a site or a region in space. The equation for the evolution of observables contains explicitly the survival probability as
\begin{equation}
O(t) = W_0(t) O(0)
+\sum_{k,k'} \langle k_0 | e^{i H_F t} | k \rangle O_{kk'} \langle k' | e^{- i H_F t} | k_0 \rangle  ,
\label{eq:TotObs}
\end{equation}
where $O_{k,k'} = \langle  k | O | k'\rangle$, $|k\rangle $ are the eigenstates of the Hamiltonian that defines the initial state, and the sum involves the states where $k$ and $k'$ or both are different from $k_0$.

In general, the dynamics of the observables is very fast. After a transient time, they equilibrate in a probabilistic sense, fluctuating around the infinite time average,
\begin{equation}
\overline{O}=\sum_\alpha |C_{k_{0}}^{\alpha}|^2 O^{\alpha\alpha},
\end{equation}
where $O^{\alpha\alpha} = \langle  \alpha | O | \alpha \rangle$. In systems without an excessive number of degeneracies and for initial states away from the edges of the spectrum, in other words when $| k_0 \rangle$ has a chaotic structure, the temporal fluctuations decrease exponentially with the system size~\cite{Zangara2013}. It has been shown that this holds not only for models with a Wigner-Dyson level spacing distribution, but also for integrable models with interaction, such as Model 1.

The size of the temporal fluctuations of observables in isolated finite systems has played an important role in the studies of relaxation. Earlier semiclassical arguments based on full random matrices were used to describe the exponential decay of the fluctuations~\cite{Feingold1986,Deutsch1991,Prosen1994,SrednickiARXIV,Srednicki1996,Srednicki1999}. More recently, analytical upper bounds for
the fluctuations were derived in \cite{Short2011,Short2012}. Analytical~\cite{Venuti2013} and numerical~\cite{Cassidy2011,Gramsch2012,HeSantos2013} studies also exist for noninteracting integrable models, where the size of the fluctuations decreases much slower, as $1/\sqrt{L}$.

\section{Thermal Equilibrium}

\label{Sec:Equilibration}
Once verified that the observables equilibrate, one can ask whether a statistical approach can be used to obtain the same result as the infinite time average. In particular, the question is about the agreement between the infinite time average and the standard statistical ensembles of conventional thermal equilibrium. Here, we discuss under what conditions this scenario can hold.

The question to be addressed is represented by the following equation (see, for example, \cite{Rigol2008,Deutsch1991}),
\begin{equation}
\overline{O}=\sum_\alpha |C_{k_{0}}^{\alpha}|^2 O^{\alpha\alpha}
\overbrace{\sim}^{? }
O_\text{ME}  \equiv  \frac{1}{{\cal{N}}_{\tilde{E}_{k_0},\delta  E}} \sum_{\substack{\alpha \\ |\tilde{E}_{k_0}-E^\alpha|<\delta E}}O^{\alpha\alpha} .                                        \label{91}
\end{equation}
The infinite time average, on the left hand side, depends on the initial state via
$|C_{k_{0}}^{\alpha}|^2$, which is made explicit with the subscript $k_0$. The right hand side corresponds to the microcanonical ensemble (ME), where ${\cal{N}}_{\tilde{E}_{k_0},\delta E}$ stands for the number of  eigenstates in a small energy window $\delta E$ centered  at $\tilde{E}_{k_0}= \langle k_0 | H |k_0\rangle$. We would like to specify when, for a finite system, the relation $\overline{O} \approx O_\text{ME}$ holds, and  what guarantees that the difference between the two averages goes to zero in the thermodynamic limit.

Recently, it became a widespread trend to relate the conventional thermalization with the so-called eigenstate thermalization hypothesis (ETH) \cite{Deutsch1991,Srednicki1994,SrednickiARXIV,Rigol2008}. The statement is that for an initial state $|k_0\rangle$ covering a sufficiently small energy window, the value of $\overline{O} $ will be independent of $|C_{k_{0}}^{\alpha}|^2$ if the eigenstate expectation values of the observable, $O^{\alpha \alpha} $, is a smooth function of energy \cite{SrednickiARXIV}. In this case, a single eigenstate inside the microcanonical window suffices to compute $\overline{O} $ and the result agrees with the microcanonical average.

As we have already mentioned, the equivalence between an observable corresponding to an individual state and the statistical average over a small energy window is basically a restatement of the fundamental principle of statistical mechanics \cite{LL58}. The ETH {\em per se} does not clarify {\it when} this equivalence should hold. Below, we discuss the conditions for the proximity between $\overline{O}$\ and $O_\text{ME}$ in realistic systems using our approach based on the notion of energy shell. These conditions are intimately connected with the concept of quantum chaos, but not chaos associated uniquely with level repulsion; instead the focus should be on the existence and properties of chaotic eigenstates.

In the extreme case of full random matrices, the eigenstates are completely delocalized in the whole basis, that is $C_{k_{0}}^{\alpha}$ are simply random numbers. For these (pseudo-)random vectors, the results for $O^{\alpha \alpha} $ (for any observable) are obviously independent of the particularly chosen energy eigenstate $| \alpha  \rangle $. In this case, more than just being a smooth function of energy, $O^{\alpha \alpha} $ is actually constant throughout the spectrum apart from small fluctuations that decrease exponentially with the dimension of the Hamiltonian matrix. But full random matrices do not describe isolated systems of interacting particles; typically, an exact eigenstate occupies only a fraction of the mean-field basis.

As shown above, in realistic systems strongly chaotic eigenstates have their widths comparable to the widths $\sigma_{k_0}$ of the strength function, defined as $\sigma_{k_0} = (\sum_{k\ne k_0}  |H_{kk_0}|^2)^{1/2}$ and of the same order as the  width of the energy shell $\Delta_{k_0}$. We remind that the strength function is defined by the projection of the eigenstates of the initial Hamiltonian onto those of the final one. Thus, we may expect that only for those chaotic eigenstates of the final Hamiltonian which belong to this energy window, $\sigma_{k_0} \simeq \Delta_{k_0}$, their expectation values $O^{\alpha \alpha} $ become nearly constant so that they can be taken out of the sum of the left hand side of Eq.~(\ref{91}). This part of the equation will then have a single value of $O^{\alpha \alpha}$ multiplied by $\sum_{\alpha} |C_{k_0}^{\alpha}|^2=1$. When this happens, we will have an approximate equality with the right hand side of the same Eq.~(\ref{91}) only if the chosen small energy window $\delta E$ is smaller than the energy shell, $\delta E \leq \Delta_{k_0}$. In this case the microcanonical average will approximately agree with a single value of $O^{\alpha \alpha} $, just as the left hand side. As one can see from Eq.~(\ref{91}), this equality holds only if
$|C_{k_0}^\alpha|^2 = 1/{\cal N}$, namely, it is constant and does not depend on the total energy $E^\alpha$. Since this is never true, another possibility could be that $O^{\alpha \alpha}$ does not depend on $\alpha$. In this second case the equality is always satisfied since it becomes trivial due to the normalization $\sum_\alpha |C_{k_0}^\alpha|^2=1$.

In summary, the prerequisites for conventional thermalization are:

(1) there is a region of the energy spectrum where the eigenstates are chaotic which is given by the equality of the width of the strength function and the width of the energy shell, $\sigma_{k_0} \simeq \Delta_{k_0}$, both being dependent on the initially chosen basis. This results in the existence of observables for which $O^{\alpha \alpha} $ is nearly constant and therefore uncorrelated with $|C_{k_{0}}^{\alpha}|^2$;

(2) the width of the energy window for the microcanonical averaging should be smaller than the energy shell, $\delta E \leq \sigma_{k_0} \sim \Delta_{k_0}$.

These conditions clarify that the approximate validity of Eq.~(\ref{91}) is strictly dependent on the following parameters: the width of the strength function in connection with that of the energy shell and the microcanonical energy window. They also make evident that the viability of thermalization depends on the chosen initial state~\cite{He2012,He2013,Torres2013,Torres2014PRE} (specifically, close to the edges of the energy band this approach might be  not valid). Conventional thermalization is therefore not expected to occur for initial states whose widths of strength functions are smaller than the energy shell. Note that an ergodic filling of the energy shell by eigenstates can emerge independently of the character of the  level spacing distribution (Wigner-Dyson or Poisson).

The analysis of how the energy of the initial state affects the viability of conventional thermalization was performed in \cite{He2012,He2013,Torres2013,Torres2014PRE}. In particular, it has been shown that thermalization may happen even in quenches where the final Hamiltonian is integrable provided the initial state spans over chaotic-like states of the total Hamiltonian.
In this case the initial state samples at random different symmetry sectors of the integrable system, allowing for thermal features to emerge. One can observe  that, for isolated integrable systems of a finite number of particles , the energy of the initial state needs to be closer to the middle of the spectrum than for chaotic systems to ensure the proximity of $\overline{O}$ to $O_{\text{ME}}$.

Several studies have shown that quenches to an integrable final Hamiltonian  lead to an equilibrium described by the generalized Gibbs ensemble (GGE) \cite{Rigol2007} or generalized microcanonical ensemble (GME) \cite{Cassidy2011}. Very close to the middle of the spectrum  both ensembles give approximately the same values for average observables, but away from the center deviations have been seen even in the thermodynamic limit \cite{Rigol2014}. The generalized ensembles take into account the symmetries associated with the integrability of the model. While for  integrable systems composed of non-interacting particles it is clear which conserved quantities should be taken into account, for integrable systems with interacting particles, such as the XXZ model, this is still an open question~\cite{Sotiriadis2014,Fagotti2014,Pozsgay2014,Wouters2014,PozsgayARXIV,GoldsteinARXIV}.

\section{Concluding remarks}
In this review we presented a summary of our approach to the problem of the onset of chaos and
thermalization in isolated systems of interacting quantum particles. The approach was
developed by the authors and their collaborators during the last two decades. For a long time, the generally accepted viewpoint was that an isolated system of a finite number of particles cannot be
treated by conventional statistical mechanics if this system was deterministic, especially if the intrinsic dynamics was integrable. Indeed, in contrast to classical mechanics that allows
the emergence of deterministic chaos caused by local instability of motion, in quantum systems
this mechanism of chaos is absent due to the linearity of the equations of motion. This point is
also reflected by the  fact that the energy spectrum of a bounded isolated quantum system is discrete, thus indicating that dynamics is periodic or quasi-periodic. This
is in contrast with classical mechanics where the spectrum of the motion can be either
discrete or continuous depending whether the motion is regular or chaotic. Yet,
as our approach shows, chaos and thermalization can take place in isolated interacting quantum systems.

At the early stage of the studies of quantum systems that were strongly chaotic in the classical limit, it was understood that quantum chaos could be essentially quantified by specific
properties of energy spectra. In Ref.~\cite{casati80} it was conjectured that quantum chaos
should be characterized in terms of the statistical theory of spectra developed originally
for the description of billiard-like systems and compound nuclear reactions. Since the mathematical tools of this
theory were related to random matrices, it was claimed that the properties of quantum chaos
had much in common with those of random matrices. This conjecture was  numerically
confirmed in Ref.~\cite{Bohigas84} and since then it has been accepted that the strongest
properties of quantum chaos are manifested by local fluctuations of the energy spectra and by the chaotic structure of the eigenstates, as predicted by random matrix theories.

For a long time, the
Wigner-Dyson distribution of spacings between nearest energy levels was the main tool for
detecting quantum chaos. In this way, one-body chaos was found, both numerically and
experimentally, for a hydrogen atom  in a strong magnetic field or for an electron  in a
quantum dot. For many-body systems, such as Bose-particles in optical traps or
Fermi-particles on a lattice, chaotic properties emerge due to the many-body effects related
to inter-particle interactions. Thus a proper  characterization of
many-body quantum chaos is in terms of the chaotic (exceedingly complicated) structure of many-body
eigenstates. This structure can be properly quantified and used as a powerful instrument in
theoretical studies and in the analysis of experimental data. A particular problem is the
many-body localization, currently a subject of  extensive studies (see, for example, Refs.
\cite{Polkovnikov2011RMP,Huse2015} and references therein).

As shown in Ref.~\cite{Srednicki1994}, the chaotic structure of individual many-body eigenstates is directly
related to the conventional statistical distributions (Fermi-Dirac, Bose-Einstein and
Boltzman).  For isolated systems, an impressive demonstration of the emergence of the
Fermi-Dirac distribution for an isolated atom was reported in Ref. \cite{GGF99}. This
fact is in a correspondence with the remark given in the book by Landau and Lifshits \cite{LL58}, stating that conventional statistical mechanics can appear not only due to
ensemble average but also with the use of a single typical wave function.

In this review we demonstrated how the thermalization mechanism is related to the chaotic
structure of many-body eigenstates. The crucial point here is a proper choice of the basis
used for evaluating the complexity of the eigenstate structure. For example, a discussion of
the Anderson localization of electrons in a disordered potential always assumes the
configuration basis. For many-body systems in atomic and nuclear physics  the mean field
basis that defines interacting constituents, particles or quasi-particles, naturally
separates regular features of dynamics from incoherent collision-like interactions
responsible for quantum chaos. In specific models, it is sometimes  useful to go from one mean field representation to another in order to
better understand the role of inter-particle interaction [272,273]. In many-electron atoms or heavy nuclei,
the residual interaction  written in the basis of non-interacting particles is typically
quite complex. In our review, we also discussed the analogous picture for the case of lattice models,
both integrable and non-integrable. The emergence of chaotic features depends on the strength of interaction and
the energy of the considered eigenstates. The conditions for the crossover from non-chaotic to
chaotic eigenstates have been obtained by means of the concepts of energy shell and
strength function. An instructive example of a proper choice for a mean field is given in Refs.~\cite{berman1,berman2,mark} where two models of quantum computers have been studied in view of the onset of chaos and many-body localization emerging due to interaction between {1/2-}spins. 

The notion of energy shell was discussed for the first time in terms of band random
matrices in Refs.~\cite{Casati1993,Casati1996}. On increasing the interaction strength, the
eigenstates of the total Hamiltonian, presented as the sum $H=H_0+V$ of the regular
(mean-field) and residual part, begin occupying a region of the basis of $H_0$, the size of
which depends on the form and strength of the interaction term $V$. The maximal energy range
that can be filled by the exact eigenstates is defined by the width of the energy shell. The latter can
be estimated from the structure of the Hamiltonian matrix  in the unperturbed basis without
any need to diagonalize the total Hamiltonian $H$.

The strength function is a quantity of special interest. It  is the expansion of an
eigenstate of $H_0$ in the eigenbasis of the total Hamiltonian $H$, written in the energy
representation. It gives the energy spreading of the chosen mean-field basis vector. At weak interaction, when perturbation theory is valid,  the SF is
a narrow peak around the unperturbed energy, with  small admixtures of other basis states due to the interaction $V$. With an increase of interaction, the Breit-Wigner form of the SF
emerges, the width of which is defined by the standard Fermi golden rule. This fact has been known since
early times of application of random matrices to the statistical description of energy
spectra of heavy nuclei \cite{wigner1955,wigner1957}. Unlike chaotic
billiard-type problem that can be modeled  by full random matrices, isolated many-body
systems with finite interparticle interaction are described by banded random matrices. This was first understood by Wigner, who introduced an ensemble of such matrices
(WBRM). This allowed to find an analytical form of the SF depending on the model
parameters. As  shown in many studies for interacting particles in isolated systems, as the interaction strength increases, the SF typically shows a crossover from the Breit-Wigner form to a Gaussian-like. In our review we show  that this
crossover can be used to identify the emergence of chaotic eigenstates filling the energy shell.

The knowledge of the form of the SF is crucial for the description of quench dynamics. The Fourier transform of the SF determines the
return probability $W_0(t)$ for finding the system in its initial state. For the Breit-Wigner SF, $W_0(t)$ decays exponentially, apart from a short time scale where the standard perturbation theory predicts the Gaussian decay. On increasing the interaction strength, the
form of the SF changes and for the Gaussian SF, the Gaussian decay of $W_0(t)$ lasts for a
long time. The tails of the SF cannot follow the exponential decay that would correspond to
 a diverging second moment of the SF. There is always a characteristic time after which the dynamics of a system is non-standard, and the decay of $W_{0}(t)$ becomes algebraic. This time scale is associated with the lower bound of the spectrum and is not discussed in our review.

With the form of the SF  known, one can predict the dynamic properties of the wave packet in the mean-field representation. The appropriate analytical method to do this is
the so-called ``cascade model" of proliferation of excitations developed in
Ref.~\cite{FI01b}. Relatively simple expressions allow one to find the time dependence for
such important quantities as the Shannon entropy associated with the packets and and the
inverse participation ratio (the second moment) of wave packets in the unperturbed basis.
Obtained analytically and confirmed numerically is the linear
increase of the Shannon entropy for the Gaussian SF. This means that in the region of strong
quantum chaos the number of excited many-body states increases exponentially in time. This
may be treated as another fingerprint of quantum chaos and  compared with the exponential
instability of motion in classical systems with strong deterministic chaos. As  shown in Ref.~\cite{BBIS04}, the crossover from a time-periodic behavior of the Shannon entropy to the linear increase for interacting bosons in one-dimensional optical traps corresponds to the crossover between the mean field and the Tonks-Girardeau regimes \cite{G60,GWT01}. A similar
effect, namely an exponential increase of the number of harmonics in the time-dependent Wigner
function of the Ising model, has been found in Ref.~\cite{Balachandran2010}.

We considered  isolated systems with a  relatively small number $N$  of interacting particles whose dynamics was analyzed on a finite time scale. We did not discuss the problem of thermalization in the thermodynamic limit, $N \rightarrow \infty$.
However, few remarks are worthwhile here. When the number of particles is infinite,
the system can  be, in essence, treated as a heat bath itself. According to
Ref.~\cite{Tatarskii}, in such a case the behavior of a particle follows the standard
statistical predictions even if the system is integrable. It is known that the mechanism for
the onset of the statistical behavior in the thermodynamic limit is related to an infinite number of
non-commensurate frequencies in the system dynamics and random phases of the density matrix. As shown by Bogoliubov
\cite{B45,Bogolubov70} (see also the discussion in Ref.~\cite{C86}),  conventional
statistical properties emerge even in an integrable system of linearly coupled oscillators
under quite modest mathematical assumptions. The energy spectrum of such a system, being
discrete  for finite $N$, becomes continuous in the thermodynamic limit. It was
demonstrated in Ref.~\cite{Tatarskii} that  the time scale on which there is a clear
exponential relaxation to the equilibrium, strongly increases with the number of particles
and tends to infinity in the thermodynamic limit.

As argued by Chirikov \cite{C86,c97}, quantum chaos is, in essence, chaos on a finite
time scale. However, this time scale can be extremely large, thus allowing to treat the
dynamics as  occurring in a continuous spectrum. The situation is somewhat similar to that
in classical mechanics where, before the development of the concept of chaos, the old known mechanism for the onset of standard statistical
properties was the thermodynamic limit itself. Then it is practically  irrelevant
whether the considered system is integrable or not. For instance, the computation of the
Lyapunov exponent on a very long time scale in the integrable Toda lattice with a large number of particles gives the same non-zero result as for the chaotic
Fermi-Pasta-Ulam model,  Ref. \cite{Casetti}. Along similar lines, we can think about computers. They make computations with finite precision which means that the effective phase space is
discrete, therefore all trajectories are periodic. Strictly speaking there are no chaotic
trajectories in computation. However, the time scale on which one can detect  the
periodicity of the motion is extremely long. That is why any numerical manifestation of
classical chaos is nothing but chaos in the discrete spectrum, observed on a finite time
scale. This very fact can be compared with the meaning of quantum chaos.

\section{Acknowledgements}
The authors are grateful to L.~D'Alessio, B.A. Brown, B.~Fine, V.V. Flambaum, M. Horoi, M.~Olshanii, G.~Ortiz, A.~Polkovnikov, M.~Rigol, A. Volya, and V.~Yurovsky for fruitful discussions.
The authors gratefully acknowledge support from the CONACyT Grant No. N-133375 and from the VIEP-BUAP Grant IZF-EXC13-G (F.M.I.); from the NSF grant DMR-1147430 (L.F.S.);  from the NSF grants PHY-1068217 and PHY-1404442 (V.Z.).


\section{Further reading:} 
These are relevant papers that came to our knowledge after the acceptance of this review article.

[1] S.K. Haidar, N.~D. Chavda, M. Vyas, V.~K.~B. Kota, Fidelity decay and entropy production in many-particle systems after random interaction quench,
arXiv:1509.01392.

[2] M. Tavora, E.~J. Torres-Herrera, L.~F. Santos, Powerlaw decay exponents as predictors of thermalization in many-body quantum systems,
arXiv:1601.05807.

[3] E. Ilievski, M. Medenjak, T. Prosen, L. Zadnik, Quasilocal charges in integrable lattice systems, arXiv:1603.00440 (integrals of motion of the XXZ model).

\section*{References}


\end{document}